\documentclass[aos,preprint]{imsart}

\RequirePackage[OT1]{fontenc}
\RequirePackage[numbers]{natbib}
\RequirePackage{amsthm,amsmath,amsfonts}
\RequirePackage[colorlinks,citecolor=blue,urlcolor=blue]{hyperref}
\RequirePackage{multirow}
\RequirePackage{multicol}
\RequirePackage{graphics}
\RequirePackage{graphicx}
\RequirePackage{booktabs}
\RequirePackage{epstopdf}
\RequirePackage{threeparttable}
\RequirePackage{longtable}
\RequirePackage{textcomp}

\startlocaldefs

\newtheorem{corollary}{Corollary}
\newtheorem{proposition}{Proposition}
\theoremstyle{definition}

\newtheorem{example}{Example}
\newtheorem{remark}{Remark}
\numberwithin{equation}{section}
\theoremstyle{plain}
\newtheorem{thm}{Theorem}[section]
\numberwithin{equation}{section}
\theoremstyle{plain}
\endlocaldefs
\begin{document}

\begin{frontmatter}
\title{Unified statistical inference for a  novel nonlinear dynamic functional/longitudinal data model}
\runtitle{Unified Statistical Inference}

\begin{aug}
\author{\fnms{Lixia} \snm{Hu}\thanksref{m1}
\ead[label=e1]{hulx18sufe@163.com}},
\author{\fnms{Tao} \snm{Huang}\thanksref{m2}
\ead[label=e2]{huang.tao@mail.shufe.edu.cn}}
\and
\author{\fnms{Jinhong} \snm{You}\thanksref{m2}
\ead[label=e3]{johnyou07@163.com}
}

\runauthor{Lixia Hu, Tao Huang and Jinhong You}

\affiliation{Shanghai Lixin University of  Accounting and Finance\thanksmark{m1}
and \\  Shanghai University of Finance and Economics\thanksmark{m2}
}
\address{
School of Statistics and Mathematics \\
Shanghai Lixin University of\\
Accounting and Finance\\
Shanghai, 201209 
China\\
\printead{e1}\\}

\address{
School of Statistics and Management\\
Shanghai University of Finance and\\
Economics,  Shanghai, 200433
China\\
\printead{e2}\\
\phantom{E-mail:\ }\printead*{e3}}
\end{aug}

\begin{abstract}

In light of recent work studying massive functional/longitudinal data, such as the resulting data from the COVID-19 pandemic, we propose a novel functional/longitudinal data model which is a combination of the popular varying coefficient (VC) model and additive model. We call it Semi-VCAM in which the response could be a functional/longitudinal variable, and the explanatory variables could be a mixture of functional/longitudinal and scalar variables. Notably some of the scalar variables could be categorical variables as well. The Semi-VCAM simultaneously allows for both substantial flexibility and the maintaining of one-dimensional rates of convergence. A local linear  smoothing with the aid of an initial B spline series approximation is developed to estimate the unknown functional effects in the model. To avoid the subjective choice between the sparse and dense cases of the data, we establish the asymptotic theories of the resultant Pilot Estimation Based Local Linear Estimators (PEBLLE) on a unified framework of sparse, dense and ultra-dense cases of the data. Moreover, we construct unified consistent tests to justify whether a parsimony submodel is sufficient or not. These test methods also avoid the subjective choice between the sparse, dense and ultra dense cases of the data. Extensive Monte Carlo simulation studies investigating the finite sample performance of the proposed methodologies confirm our asymptotic results. We further illustrate our methodologies via analyzing the COVID-19 data from China and the CD4 data.
\end{abstract}



\end{frontmatter}

\section{Introduction}

\label{sec:Intro}

Increasingly, data is recorded continuously over an interval of time (spatial location, or wavelength and so on) or intermittently at several discrete points in time due to progress in modern computation technology. As a result, the data in which each individual has multiple observations becomes more and more common in almost all scientific, societal and economic fields. Obviously, a recent example of this kind of data is the COVID-19 data: the daily confirmed diagnoses, death toll and suspected cases of different countries are recorded and made available.
When a variable is measured or observed at different times, the variable is usually treated as a function of time. As a result, the variable is  called a functional variable, the data for the variable are called functional data and the related statistical analysis is called functional data analysis (FDA) (\cite{wang2016f}). The functional data and corresponding FDA have been successfully applied to explore the interactions and co-movements among a group of temporally evolving subjects. Several monographs by  \cite{Ram2002}, \cite{Ram2018} and \cite{fer2006} provide comprehensive discussions on the methods and applications. More recent work about FDA could refer to \cite{wang2016f}.

According to \cite{wang2016f}, usually, the functional data could be divided into two cases: sparse and dense. Sparse functional data usually occurs in longitudinal studies where subjects are measured at different time points and the number of measurements for each subject is often bounded away from infinity. Inversely, in the dense functional data the number of measurements of each subject tends towards to infinity. In theory, the difference between sparse (longitudinal) and dense function data is clear. However, due to the limitations of humans, the observations in real data sets could not be infinite and are definitely finite. Therefore, the edge of sparse (longitudinal) and dense function data in practice is vague in some scenarios, especially when the number of measurement of each subject is moderate or different subjects have different numbers of measurements.

In many functional/longitudinal studies, repeated measurements within each subject are possibly correlated with each other,
but different subjects can be supposed to be independent.
One approach to take intra-subject variation into account is  the mixed-effects model \cite{wu2002},
which decomposes   regression function into a fixed population mean and a subject-specific random trajectory
with zero mean.
For sparse and dense functional data, \cite{kim2013} considered a mixed-effects nonparametric
regression model absence of covariates,
and showed that  the asymptotic distributions  of kernel estimators are essentially different in these two situations.
Therefore, a subjective choice between sparse and dense cases may lead to erroneous conclusion.
To evade this problem, they proposed a self-normalized method,
which can deal with sparse and dense functional data in a unified framework.
Furthermore,  \cite{chen2017}  generalized the results of \cite{kim2013} to a mixed-effects VCM  presence of  covariates
with sparse or dense functional data.
Lately, \cite{zhang2016sparse} provided a comprehensive perspective that deals with
a general weighing scheme on a unified paltform for all types of sampling plan,
including sparse, dense and ultra dense case.
Motivated by a monotone relationship between gray matter volume and age in the older population,
\cite{Chen2019} considered sparse and dense cases on a unified framework under monotone constraint of
the mean function.
The research work of \cite{zhang2016sparse,Chen2019} has focused on the statistical inference
about mean function of the underlying process.
To the best of our knowledge, there exists no further development about unified inference
parallel to  \cite{zhang2016sparse} for nonparametric regression model presence of covariates,
a common case in practice.

In the analysis of longitudinal data, a varying-coefficient models (VCM) enjoying flexibility, parsimony and interpretability,
is a  widely-used nonparametric regression method.
One can refer to \cite{fan2000twostep,guo2002functional,Hastie1993Varying,huang2002bspline,
huang2004polynomial,hooer1998nonparametric,morris2006,senturk2011}. An additive model (AM) is another popular nonparametric regression method,
which has been studied by \cite{berhane1998,carroll2009additive,lin1999,luo2017functionlinear,
luo2016,qi2018,Xue2010consistent,you2007additive,scheipl2015}. Recently, \cite{hu2019est,hu2019robust,zhang2015varying,zhang2020VCAM} have investigated a novel nonparametric regression method,
named the varying-coefficient additive model (VCAM),
which can be viewed as a generalization of the VCM and AM.
Let $T_{ij}$ be the observation time when the $j$th measurement of the $i$th subject is made,
$Y_{ij}$  and $\mathbf{X}_{i}\left(T_{ij}\right):=\mathbf{X}_{ij}$  be the
response and $p$-covariates  for the $i$th subject at time $T_{ij}$, respectively.
Then $\{\left(Y_{ij},\mathbf{X}_{ij},T_{ij}\right);i=1,...,n;j=1,...,m_{i}\}$ constitutes
 a longitudinal/functional sample from $n$ randomly selected subjects with
 $m_{i}$ repeated measurements of the $i$th subject.
The VCAM for longitudinal/functional data is proposed by \cite{hu2019robust} as below
\begin{equation}\label{VCAM}
Y_{ij}=\alpha_{0}\left(T_{ij}\right)+
\sum_{k=1}^{p}\alpha_{k}\left(T_{ij}\right)\beta_{k}\left(X_{ijk}\right)+\nu_i\left(T_{ij}\right)
+\varepsilon_{ij},
\end{equation}
with the abuse of notations.
Here $\nu_{i}\left(T_{ij}\right)$  is the subject-specific random trajectory at observation time $T_{ij}$,
and $\left\{\varepsilon_{ij}\right\}$ are i.i.d. random measurement errors.
The multiplicative  factors $\alpha_{k}$($k=1,...,p$) and $\beta_k$($k=1,...,p$)
are called to be varying-coefficient component functions and additive component functions,
respectively, and $\alpha_0$ is a trend term.
Obviously, the VCAM \eqref{VCAM} reduces to an AM provided that each $\alpha_k$($k=0,...,p$)
is time-invariant, whilst it becomes a VCM if each $\beta_k$($k=1,...,p$) has a simple linear form.
Therefore,  it can be said that the VCAM is a  kind of hybird of an AM and a VCM, enjoying more flexibility,
which can greatly decrease the bias of model misspecification.
On the other hand, it is hard to address how to choose between an AM and a VCM in practice.
The general type of a VCAM provides a data-driven method to decide which model may be more suitable
for the real-life data at hand.

However, the product forms of $\alpha_k$ and $\beta_k$ in \eqref{VCAM} exclude
the discrete covariates from this model.  It will vastly limit the scope
of applications  because categorical variables are often important influence factors in the practical fields.
To accommodate both discrete and continuous covariates in regression model,  in this paper
 we consider a mixed-effects  semi varying-coefficient additive model (Semi-VCAM) to analyze longitudinal data.
Let $\mathbf{Z}_{i}\left(T_{ij}\right):=\mathbf{Z}_{ij}=\left(1,Z_{ij,1},...,Z_{ij,q}\right)^{\tau}$
be a $\left\{q+1\right\}$-vector of discrete covariates observed at time $T_{ij}$, and
 $\boldsymbol{\alpha}_0\left(t\right)=\left(\alpha_{00}\left(t\right),\alpha_{01}\left(t\right),...,
 \alpha_{0q}\left(t\right)\right)^{\tau}$ is the vector of  varying-coefficient functions for $\mathbf{Z}$
 that is i.i.d. with $\mathbf{Z}_{i}$,
 and $\alpha_{00}$ denotes the trend function.
 Then, we generalize the VCAM \eqref{VCAM} to a Semi-VCAM as below,
\begin{equation}\label{M_VCAM}
Y_{ij}=\mathbf{Z}_{ij}^{\tau}\boldsymbol{\alpha}_{0}\left(T_{ij}\right)
+\sum_{k=1}^{p}\alpha_{k}\left(T_{ij}\right)\beta_{k}\left(X_{ijk}\right)+\nu_{i}\left(T_{ij}\right)
+\sigma\left(T_{ij}\right)\varepsilon_{ij},
\end{equation}
where the subject-specific random trajectory $\nu_{i}\left(t\right)$ satisfies
 $\mathrm{E}\left[\nu_{i}\left(t\right)\right]=0$ and covariance function
$\gamma\left(t,t'\right)=\mathrm{E}\left[\nu_{i}\left(t\right)\nu_{i}(t')\right]$,
$\left\{\varepsilon_{ij}\right\}$ are random errors such that $\mathrm{E}\left(\varepsilon_{ij}\right)=0$ and $\mathrm{E}(\varepsilon_{ij}^{2})=1$,
and $\sigma\left(t\right)$ is a smooth standard deviation function of process $\varepsilon\left(t\right)$.
Note that \eqref{M_VCAM} allows a mixture of functional/longitudinal predictors and scalar covariates,
and it reduces to a partial linear additive model (PLAM),
if each varying-coefficient function is time-invariant.
Compared with the model \eqref{VCAM} studied in \cite{hu2019robust},  \eqref{M_VCAM} allows categorical
covariates and heteroscedasticity as time elapsed.
Meanwhile, in this paper we also take into account intra-subject correlation,
which was merged into random errors in \cite{hu2019robust}.
Therefore, Semi-VCAM is a more refined nonparametric model than VCAM \eqref{VCAM}
in the analysis of longitudinal data.

As a global smoothing technique, spline method is widely used to
fit a smooth nonparametric function because of its merit of cost saving.
But it usually has no asymptotic distribution due to absence of decomposition of
bias part and variance part, unless the asymptotic bias is smaller of high order than
the asymptotic variance.
All of the existing research literatures about VCAM are based upon a spline method,
\cite{zhang2015varying,zhang2020VCAM} provide no asymptotic distributions of estimators, whilst
\cite{hu2019est,hu2019robust} obtain the asymptotic distributions
under the condition that the asymptotic bias can be ignored.
Alternatively, kernel method is a local smoothing tool, based upon which
we can construct the involved asymptotic distribution presence of asymptotic bias, and make statistical inference
on certain interested function. 
Specially, local linear smoothing is popular due to its nice properties, such as design adaption, good boundary performance, and statistical efficiency in an asymptotic minimax sense, see \cite{fan1996local} for more details.

In this paper, we build a pilot estimation based local linear estimator (PEBLLE) for
varying-coefficient component functions and additive component functions, respectively.
The proposed estimation method has wide applicability,
including sparse data and dense data, and  the data presence of functional/longitudinal covariates and
scalar variables.
We have shown the consistency of PEBLLE, and as  a main contributor of this paper,
we construct the asymptotic distributions on a unified framework for sparse, dense and ultra dense data.
For the convenience of concise presentation, we only consider the same weight to each subject (SUBJ), and
our theoretical results can be viewed as a generalization of \cite{zhang2016sparse}
to nonparametric regression model presence of  covariates with SUBJ scheme.
Another intriguing question is how to judge a general Semi-VCAM or a submodel is sufficient.
 To this end, we develop two hypothesis testing to decide whether each varying-coefficient
component functions is time-invariant~(i.e., a PLAM or especially, an AM if absence of $\mathbf{Z}$ covariates),
 or whether each additive component function
has linear form~(i.e., a  VCM).
It has been shown that the proposed testing procedure is consistent on a unified framework of sparse, dense
and ultra dense case of data.

In the empirical studies, we consider the new coronavirus disease (COVID-19) breaking out in December 2019,
and apply our method to analyze the growth rate of cumulative confirmed (GRCC) cases
 in China except Hubei Province, Tibet, Macao, Taiwan and  Hong Kong.
We collect the data  from {\color{blue}{https://github.com/CSSEGISandData/COVID-19}},
and take sample period from January 22th, 2020 and April 8th, 2020.
To model GRCC,  four function covariates and one scalar covariate (population size) are chosen.
The testing procedures show that a Semi-VCAM is necessary for this dataset.
Another example  is CD4 data from the Multicenter AIDS Cohort Study (a data set in the R package ``timereg''),
which has been studied by \cite{huang2002bspline,Zhang2013}.
In this model, smoke status (1 for smoker and 0 for nonsmoker) is included.
Employing Semi-VCAM,
the testing procedure shows a VCM is sufficient, which verifies the rationality of the research results in \cite{huang2002bspline}.

The rest of this paper is organized as follows.  Section \ref{sec:est}  proposes a
pilot estimation based local linear smoothing method and Section \ref{sec:theorems} presents
a series of the asymptotic theories.
In Section \ref{sec:testing}, we propose a testing procedure to justify whether a VCM
or  a PLAM is sufficient or not, and show its asymptotic properties.
Section \ref{sec:Implement} speaks about the implementation of the proposed method.
Extensive simulation studies investing the finite-sample performance and real data applications
 illustrating our methodologies are considered in Section \ref{sec:numerical}.
Brief remarks are concluded in Section \ref{sec:remarks}.
The requirements for validity of the   asymptotic theories are presented in the Appendix,
and the main proofs  are relegated to the Supplementary Material.

\section{ Estimation Method}
\label{sec:est}

We assume that observation time $\{T_{ij}\}$ are i.i.d. copies of $T$,
which has a density function $f_{T}$  with a bounded support, say $[a,b]$.
The  vector of covariates $\mathcal{X}_i=\left(\mathbf{X}_{i1},...,\mathbf{X}_{im_i}\right)^{\tau}$
for the $i$-th subject is randomly drawn from a $p$-dimension stochastic process $\mathbf{X}(T)$,
 of which the $k$-th element $X_{k}(T)$ has a marginal density function $f_{X_{k}}$  with support $\mathbb{S}_k$.
To identify the trend term and product terms in model \eqref{M_VCAM},
we impose the conditions
$\mathrm{E}\left[\alpha_{k}\left(T\right)\right]=1$ and $\mathrm{E}\left[\beta_k\left(X_{k}\right)\right]=0$
($k=1,...,p$),  a similar practice with \cite{zhang2020VCAM,hu2019robust}.

In this section, we develop pilot estimation based local linear estimators~(PEBLLEs) for $\alpha_{k}$ and $\beta_{k}$.
Suppose that $\beta_k$'s are known, then Semi-VCAM \eqref{M_VCAM} become a VCM, and the LLE
of $\alpha_k$'s are easily obtained. Let
$\mathbf{a}_{0t}=\{a_{00}\left(t\right),...,a_{0q}\left(t\right)\}$,
$\mathbf{a}_t=\{a_{1}\left(t\right),...,a_p\left(t\right)\}^\tau$,
$\mathbf{b}_{0t}=\{b_{00}\left(t\right),...,b_{0q}\left(t\right)\}$,
$\mathbf{b}_t=\{b_{1}\left(t\right),...,b_p\left(t\right)\}$,
where $t$ is any interior point on the interval $[a,b]$.
We solve the optimization problem as below
\begin{align}\label{funII}
Q_1(\hat{a}_t,\hat{b}_t)={}&\min_{t\in (a,b)}Q\left(a_t,b_t\right)\nonumber\\
={}&\sum_{i=1}^n\frac{1}{m_i}\sum_{j=1}^{m_i}\Big[Y_{ij}-\sum_{l=0}^q
Z_{ijl}\left\{a_{0l}\left(t\right)+b_{0l}\left(t\right)\left(T_{ij}-t\right)\right\}\nonumber\\
{}&\qquad\quad-\sum_{k=1}^p\left\{a_k\left(t\right)
+b_k\left(t\right)\left(T_{ij}-t\right)\right\}\beta_{k}\left(X_{ijk}\right)
\Big]^2k_{h_{\mathrm{C}}}\left(T_{ij}-t\right),
\end{align}
where $k_{h}\left(\cdot\right)=k\left(\cdot/h\right)/h$ for certain kernel function $k$.
Then the LLE of varying-coefficient component functions are given by
$\hat{\alpha}_{0l}\left(t\right)=\hat{a}_{0l}$ for $l=0,...,q$ and
$\hat{\alpha}_k\left(t\right)=\hat{a}_k$ for $ k=1,...,p$.

On the other hand, if $\alpha_k$'s are known,  then Semi-VCAM reduces to an AM.
Suppose  that we have got estimation of additive component functions except
$\beta_k$, denoted as $\tilde{\beta}_l$ for $l\neq k$, and consider
the following minimum problem
\begin{align}\label{funIII}
{}&Q_2(\hat{a}_x,\hat{b}_x)=\min_{\mathbb{S}_k}Q_2\left(a_x,b_x\right)\nonumber\\
={}&\sum_{i=1}^{n}\frac{1}{m_i}\sum_{j=1}^{m_i}\left[\hat{Y}_{ij,-k}
-\alpha_k\left(T_{ij}\right)\{a_x+b_x\left(X_{ijk}-x\right)\}\right]^2k_{h_{\mathrm{A}}}\left(X_{ijk}-x\right),\nonumber
\end{align}
where $x$ is any interior point of support  $\mathbb{S}_{k}$ of $X_{k}$,
and $\hat{Y}_{ij,-k}=Y_{ij}-\mathbf{Z}_{ij}^{\tau}\boldsymbol{\alpha}_0(T_{ij})-\sum_{l\neq k}\alpha_l(T_{ij})\tilde{\beta}_{l}(X_{ijl})$.
Then, the LLE of $\beta_k$  is given by $\hat{\beta}_k(x)=\hat{a}_x$.

However, both $\alpha_k$ and $\beta_k$ are unknown, implying the above-mentioned estimation methods
are infeasible.
To this end, we propose pilot estimations of additive component functions.
Similar to \cite{hu2019robust}, we view multiplicative term $\alpha_k\left(t\right)\beta_k\left(x\right)$
as a general bivariate function, say $g_k\left(t,x\right)$, and estimate it using tensor B-spline method.
Specifically, for any given $t$ and $x$,  the  tensor product is defined as
$\mathcal{T}\left(t,x\right)=\mathbf{B}_{k,\mathrm{A}}\left(x\right)
\otimes \boldsymbol{b}_{\mathrm{C}}\left(t\right)$,
where $\otimes $  means the Kronecker product of matrices or vectors,
and $\boldsymbol{b}_{\mathrm{C}}\left(t\right)$ and $\mathbf{B}_{k,\mathrm{A}}\left(x\right)$
denote the B-spline basis approximating $\alpha_k\left(t\right)$ and $\beta_k\left(x\right)$, respectively.

Then, we approximate $\alpha_{0l}\left(t\right)\approx\boldsymbol{\gamma}_{0l}^\tau \boldsymbol{b}_{\mathrm{C}}\left(t\right)$ for $l=0,...,q$,
and
$g_k\left(t,x_{k}\right)\approx\gamma_k^\tau\mathcal{T}_{k}\left(t,x_{k}\right)$
for $k=1,...,p$.
Solving the following optimization problem
\begin{equation}\label{object1}
\min_{\boldsymbol{\gamma}}\sum_{i=1}^{n}\frac{1}{m_{i}}\sum_{j=1}^{m_{i}}
\left[Y_{ij}-\boldsymbol{\gamma}_{0}^{\tau}\mathbf{Z}_{ij}\otimes\boldsymbol{b}_{\mathrm{C}}\left(T_{ij}\right)
-\sum_{k=1}^{p}\gamma_{k}^{\tau}\mathcal{T}_{k}\left(T_{ij},X_{ijk}\right)\right]^2,
\end{equation}
we got he estimator of $g_k$  as $\hat{g}_k\left(t,x_k\right)=\hat{\gamma}_k^{\tau}\mathcal{T}_{k}\left(t,x_{k}\right)$,
where $\hat{\gamma}_k$ is given by \eqref{object1}.
Furthermore, the identification condition $\mathrm{E}[\alpha_{k}(T_{ij})]=1$ implies
$\beta_k(x)=\int_a^bg_k(t,x)f(t)\mathrm{d}t$.
Hence,  a pilot estimator of additive component function $\beta_k$  can be given by
\begin{equation}\label{Ini}
\hat{\beta}_{k,\mathrm{P}}\left(x\right)
=\frac{1}{N}\sum_{i=1}^n\sum_{j=1}^{m_i}\hat{\gamma}_k^\tau\mathcal{T}_k\left(t_{ij},x\right), \quad k=1,...,p,
\end{equation}
where $N=\sum_{i=1}^{n}m_i$ is total observation, and subscript `P' means pilot estimator.

Now, we can define the PEBLLEs of varying-coefficient component functions and additive component functions.
\begin{itemize}
\item Substituting the pilot estimators $\hat{\beta}_{k,\mathrm{P}}$~($k=1,...,p$)
into the objective function~$Q_1$, we obtain the PEBLLE of $\alpha_{0l}$ for $l=0,...,q$ and $\alpha_k$ for $k=1,...,p$, and still denote them as $\hat{\alpha}_{0l}$ and $\hat{\alpha}_k$, respectively.
\item In the objective function $Q_2$, we take the PEBLLEs of varying-coefficient component functions
as their pilot estimations, and $\hat{\beta}_{l,\mathrm{P}}$~($l\neq k$) as the pilot estimators of additive component functions, and yield the PEBLLE of $\beta_k$, still write as $\hat{\beta}_k$.
\end{itemize}

\begin{remark}
{\rm Compared to the spline-based estimators of \cite{hu2019robust}, the PEBLLE
can provide asymptotic distribution with the specific expression of asymptotic bias,
and make inference on the confidence interval of component functions.
Meanwhile, our estimation methodologies adapt to both sparse and dense longitudinal/functional data,
and have wide application in the real word.}
\end{remark}

\section{Asymptotic Results}
\label{sec:theorems}

In this section, we will present the asymptotic distribution and convergence rate of PEBLLE
on a unified platform for different sampling plans.

\subsection{Asymptotic Properties of Varying-coefficient Component Functions}
Let $\bar{N}_{\mathrm{H}}=\big(\frac{1}{n}\sum_{i=1}^n\frac{1}{m_{i}}\big)^{-1}$ be
the harmonic mean of $\{m_{1},...,m_{n}\}$,
and denote the interior knots number of B-spline basis
$\boldsymbol{b}_{\mathrm{C}}\left(t\right)$ and $\mathbf{B}_{k,\mathrm{A}}\left(x\right)$~($k=1,...,p$)
as $K_{\mathrm{C}}$ and $K_{\mathrm{A}}$, respectively.
Then, based upon the  result of  Proposition \ref{pro}  presented in \ref{suppA},
Theorem \ref{Con-alp} shows the uniform convergence rates of
PEBLLEs of varying-coefficient component functions.

\begin{thm}\label{Con-alp}
Under Assumption (A1) -- (A6) and (A9),
if $K_{\mathrm{C}}K_{\mathrm{A}}=o\left(nNh_{\mathrm{C}}^{4}\right)$ and
$K_{\mathrm{A}}^{-r+1/2}+K_{\mathrm{C}}^{-r}=o\left(h_{\mathrm{C}}^{2}\right)$,
then we obtain that
{\small{\begin{align}
\sup_{t\in\left(a,b\right)}\left|\hat{\alpha}_{0l}\left(t\right)-\alpha_{0l}\left(t\right)\right|
={}&O_p\left(h_{\mathrm{C}}^{2}
+\sqrt{K_{\mathrm{A}}}\left(K_{\mathrm{C}}^{-r}+K_{\mathrm{A}}^{-r}\right)+\sqrt{\frac{\log{n}}{n}
\left(1+\frac{1}{\bar{N}_{\mathrm{H}}h_{\mathrm{C}}}\right)}\right),\nonumber\\
\sup_{t\in\left(a,b\right)}\left|\hat{\alpha}_{k}\left(t\right)-\alpha_{k}\left(t\right)\right|
={}&O_p\left(h_{\mathrm{C}}^{2}
+\sqrt{K_{\mathrm{A}}}\left(K_{\mathrm{C}}^{-r}+K_{\mathrm{A}}^{-r}\right)+\sqrt{\frac{\log{n}}{n}
\left(1+\frac{1}{\bar{N}_{\mathrm{H}}h_{\mathrm{C}}}\right)}\right),\nonumber
\end{align}}}
\end{thm}
where $l=0,...,q$ and $k=1,...,p$.

\begin{remark}
{\rm From Theorem \ref{Con-alp}, we notice that the variance term  obtains a nonparametric rate of convergence $\log{n}/(n\bar{N}_{\mathrm{H}}h_{\mathrm{C}})$ provided that $\bar{N}_{\mathrm{H}}/n^{\frac{1}{2r}}\to 0$ and $K_{\mathrm{C}}\asymp\left(n\bar{N}_{\mathrm{H}}\right)^{\frac{1}{2r+1}}$,
where ``$a \asymp b$'' means that $a$ and $b$ have the same order.
On the other hand, a parametric rate of convergence is implied if
$\bar{N}_{\mathrm{H}}/n^{\frac{1}{2r}}\to C$ ($0<C<\infty$)
and $K_{\mathrm{C}}\asymp n^{\frac{1}{2r}}$ or
$\bar{N}_{\mathrm{H}}/n^{\frac{1}{2r}}\to\infty$ and $K_{\mathrm{C}}=o(n^{\frac{1}{2r}})$.
}
\end{remark}

\begin{remark}\label{split}
{\rm Similar to \cite{zhang2016sparse}, we split data into sparse, dense or ultra dense
according to the ratio $\bar{N}_{\mathrm{H}}/n^{\frac{1}{2r}}$ tends to 0, a nonzero constant
or $\infty$ as $n\to\infty$.
In fact, we give a more general method of partitioning data in the sense that the same split
with \cite{zhang2016sparse} is used if $r=2$.
}
\end{remark}

Let $\boldsymbol{\alpha}\left(t\right)=\left\{\alpha_{00}\left(t\right),\alpha_{01}\left(t\right),...,
\alpha_{0q}\left(t\right),\alpha_1\left(t\right)...,\alpha_{p}\left(t\right)\right\}^{\tau}$,
and $\hat{\boldsymbol{\alpha}}\left(t\right)$ the PEBLLE of
$\boldsymbol{\alpha}\left(t\right)$.
Furthermore, we introduced the following symbols:
$\mathbf{F}_{ij}=\big(\mathbf{Z}_{ij}^{\tau},\boldsymbol{\beta}_{ij}^{\tau}\big)^{\tau}$ with $\boldsymbol{\beta}_{ij}=\left\{\beta_1\left(X_{ij1}\right),...,\beta_p\left(X_{ijp}\right)\right\}^{\tau}$,
$\mathbf{\Xi}\left(t\right)=\mathrm{E}\big[\mathbf{F}_{ij}\mathbf{F}_{ij}^{\tau}|T_{ij}=t\big]:=
\left[\mathbf{v}_{1}\left(t\right)\mathbf{v}_{2}(t)\right]$ and
 $G\left(t,t\right)=\lim_{t'\to t}G\left(t,t'\right)$ with
$G\left(t,t'\right)=\mathrm{E}\big[\mathbf{F}_{ij}\mathbf{F}_{ij'}^{\tau}|T_{ij}=t,T_{ij'}=t'\big]$.
In addition, we define
$\kappa=\int K^{2}\left(v\right)\mathrm{d}{v}$,
$\kappa_{2}=\int v^{2}K\left(v\right)\mathrm{d}v$,
$\kappa_{4}=\int v^{4}K\left(v\right)\mathrm{d}v$,
$\kappa_{22}=\int v^{2}K^{2}\left(v\right)\mathrm{d}v$,
and $g''$ denotes the second derivative of function $g$.

Theorem \ref{Dis-alp} presents a unified asymptotic normality of $\hat{\boldsymbol{\alpha}}\left(t\right)$,
which can be applied to sparse, dense and ultra dense cases of the data.

\begin{thm}\label{Dis-alp}
Under the assumption of (A1) -- (A9),  if
{\small{\begin{equation}\label{Lya-alp}
\frac{\max\Big\{\frac{1}{n^3h_{\mathrm{C}}^2}\sum_{i=1}^{n}\frac{1}{m_i^2},
\frac{1}{n^3h_{\mathrm{C}}}\sum_{i=1}^n\frac{1}{m_i^2}\left(m_i-1\right),
\frac{1}{n^3}\sum_{i=1}^{n}\left(1-\frac{1}{m_i}\right)\left(1-\frac{2}{m_i}\right)\Big\}}
{\left[\frac{1}{n\bar{N}_{\mathrm{H}}h_{\mathrm{C}}}+\frac{1}{n}\left(1
-\frac{1}{\bar{N}_{\mathrm{H}}}\right)\right]^{3/2}}\nonumber
\end{equation}}}
holds. Then, for any an interior $t$ in $\left(a,b\right)$, we obtain the asymptotic distribution of $\hat{\boldsymbol{\alpha}}(t)$ as below:
\begin{equation}\label{AsyDis-alp}
\Gamma_{\mathrm{C}}^{-1/2}(t)\left(\hat{\boldsymbol{\alpha}}\left(t\right)-\boldsymbol{\alpha}\left(t\right)
-\tfrac{1}{2}h_{\mathrm{C}}^{2}\kappa_{2}\Xi^{-1}\left(t\right)\rho_{1}\left(t\right)\right)
\xrightarrow{D}N\left(0,I_{p+q+1}\right),
\end{equation}
where $\rho_{1}\left(t\right)=\sum_{l=0}^{q}\alpha_{0l}''\left(t\right)v_{1l}\left(t\right)
+\sum_{k=0}^{p}\alpha_{k}''\left(t\right)v_{2k}\left(t\right)$
with $v_{1l}\left(t\right)$ being the $l$th column of $\mathbf{v}_1\left(t\right)$
and $v_{2k}\left(t\right)$ being the $k$th column of $\mathbf{v}_2\left(t\right)$, and
\[\Gamma_{\mathrm{C}}\left(t\right)=\frac{\kappa}{n\bar{N}_{\mathrm{H}}h_{\mathrm{C}}f_{\mathrm{T}}\left(t\right)}
\Sigma_{1,\mathrm{S}}\left(t\right)+\frac{1}{n}\left(1-\frac{1}{\bar{N}_{\mathrm{H}}}\right)
\Sigma_{1,\mathrm{D}}\left(t\right)\]
with $\Sigma_{1,\mathrm{S}}=\Xi^{-1}\left(t\right)\left(\gamma\left(t,t\right)+\sigma^2\left(t\right)\right)$
and~$\Sigma_{1,\mathrm{D}}=\Xi^{-1}\left(t\right)\gamma\left(t,t\right)G\left(t,t\right)\Xi^{-1}\left(t\right)$.
\end{thm}

According to the method of partitioning data defined in Remark \ref{split} and \eqref{AsyDis-alp},
Corollary \ref{Cor-alp} lists the asymptotic distributions for sparse, dense and ultra dense cases of the data as follows.

\begin{corollary}\label{Cor-alp}
Suppose that the conditions of Theorem \ref{Dis-alp} hold and $t$ is a fixed interior point on the interval
 $\left(a,b\right)$.
\begin{itemize}
\item [$\mathrm{\left(i\right)}$] Sparsity Case~($\bar{N}_{\mathrm{H}}/n^{\frac{1}{2r}}\to 0$). If $h_{\mathrm{C}}\asymp \left(n\bar{N}_{\mathrm{H}}\right)^{-1/\left(2r+1\right)}$, then
{\small{\begin{equation}\label{S-alp}
\sqrt{n\bar{N}_{\mathrm{H}}h_{\mathrm{C}}f_{\mathrm{T}}\left(t\right)}
\left(\hat{\boldsymbol{\alpha}}\left(t\right)-\boldsymbol{\alpha}\left(t\right)
-\tfrac{1}{2}h_{\mathrm{C}}^2\kappa_2\Xi^{-1}\left(t\right)\rho_{1}\left(t\right)\right)
\xrightarrow{D} N\left(0,\kappa\Sigma_{1,\mathrm{S}}\right).
\end{equation}}}
\item [$\mathrm{\left(ii\right)}$] Dense Case~($\bar{N}_{\mathrm{H}}/n^{\frac{1}{2r}}\to C_1<\infty$).
If $h_{\mathrm{C}}=O\left(n^{-1/(2r)}\right)$, then
{\small{\begin{equation}\label{D-alp}
\sqrt{n}\left(\hat{\boldsymbol{\alpha}}\left(t\right)-\boldsymbol{\alpha}\left(t\right)
-\tfrac{1}{2}h_{\mathrm{C}}^2\kappa_2\Xi^{-1}\left(t\right)\rho_{1}\left(t\right)\right)
\xrightarrow{D} N\left(0,\frac{\kappa}{f_{\mathrm{T}}(t)C_1}\Sigma_{1,\mathrm{S}}
+\Sigma_{1,\mathrm{D}}\right).
\end{equation}}}
\item [$\mathrm{\left(iii\right)}$] Ultra Dense Case~($\bar{N}_{\mathrm{H}}/n^{\frac{1}{2r}}\to \infty$).
If $h_{\mathrm{C}}=o\left(n^{-1/\left(2r\right)}\right)$, then
{\small{\begin{equation}\label{UD-alp}
\sqrt{n}\left(\hat{\boldsymbol{\alpha}}\left(t\right)-\boldsymbol{\alpha}\left(t\right)
-\tfrac{1}{2}h_{\mathrm{C}}^2\kappa_2\Xi^{-1}\left(t\right)\rho_{1}\left(t\right)\right)
\xrightarrow{D} N\left(0,\Sigma_{1,\mathrm{D}}\right).
\end{equation}}}
\end{itemize}
\end{corollary}

Let  $\hat{\Xi}^{-1}\left(t\right)$,  $\hat{\rho}_{1}\left(t\right)$,
$\hat{\gamma}\left(t,t\right)$, $\hat{\sigma}^{2}\left(t\right)$,
$\hat{f}_{T}\left(t\right)$, $\hat{G}\left(t,t\right)$, $\hat{v}_{1l}\left(t\right)$ and $\hat{v}_{2l}\left(t\right)$  are
kernel smoothing of $\Xi^{-1}\left(t\right)$,  $\rho_{1}\left(t\right)$ , $\gamma\left(t,t\right)$,
$\sigma^{2}\left(t\right)$, $f_{T}\left(t\right)$, $G\left(t,t\right)$, $v_{1l}\left(t\right)$ and $v_{2l}\left(t\right)$.
Then, the naive consistent estimators of asymptotic bias $\rho_{1}\left(t\right)$ and asymptotic variance $\Gamma_{\mathrm{C}}\left(t\right)$ are given by
$\hat{\rho}_1\left(t\right)=\sum_{l=0}^{q}\hat{\alpha}_{0l}''\left(t\right)\hat{v}_{1l}(t)
+\sum_{k=1}^{p}\hat{\alpha}_{k}''\left(t\right)\hat{v}_{2k}\left(t\right)$
and
\[\hat{\Gamma}_{\mathrm{C}}\left(t\right)=
\frac{\kappa}{n\bar{N}_{\mathrm{H}}\hat{h}_{\mathrm{C}}\hat{f}_{\mathrm{T}}\left(t\right)}
\hat{\Sigma}_{1,\mathrm{S}}\left(t\right)+\frac{1}{n}\left(1-\frac{1}{\bar{N}_{\mathrm{H}}}\right)
\hat{\Sigma}_{1,\mathrm{D}}\left(t\right),\]
where $\hat{\Sigma}_{1,\mathrm{S}}=\hat{\Xi}^{-1}\left(t\right)\left(\hat{\gamma}\left(t,t\right)
+\hat{\sigma}^{2}\left(t\right)\right)$
and $\hat{\Sigma}_{1,\mathrm{D}}=\hat{\Xi}^{-1}\left(t\right)
\hat{\gamma}\left(t,t\right)\hat{G}\left(t,t\right)\hat{\Xi}^{-1}\left(t\right)$.

Based upon \eqref{AsyDis-alp}, we can construct a
$\left(1-\alpha\right)\%$ confidence interval of varying-coefficient component functions as below
\begin{equation}\label{con-alp1}
\begin{split}
{}&\hat{\alpha}_{0l}\left(t\right)-\tfrac{1}{2}h_{\mathrm{C}}^2\kappa_{2}
\left(\hat{\Xi}^{-1}\left(t\right)\hat{\rho}_{1}\left(t\right)\right)_{l+1}
\pm z_{1-\alpha/2}\left(\hat{\Gamma}_{\mathrm{C}}^{1/2}\left(t\right)\right)_{l+1,l+1},\\
{}&\hat{\alpha}_{k}\left(t\right)-\tfrac{1}{2}h_{\mathrm{C}}^2\kappa_{2}\left(\hat{\Xi}^{-1}
\left(t\right)\hat{\rho}_{1}(t)\right)_{q+1+k}
\pm z_{1-\alpha/2}\left(\hat{\Gamma}_{\mathrm{C}}^{1/2}\left(t\right)\right)_{q+k+1,q+k+1},
\end{split}
\end{equation}
where $z_{1-\alpha/2}$ is the $1-\alpha/2$ standard normal quantile,
 the subscript $k$ denotes the $k$-th element of involved vector,
and the subscript $\left(k,k\right)$ means the $k$-th diagonal element of a given matrix.
Note that \eqref{con-alp1} is a unified confidence interval  suitable for sparse, dense and
ultra dense cases of the data.

\subsection{Asymptotic Properties of Additive Component Functions}

In this subsection, we focus on the asymptotic results of PEBLLE of additive component functions. Theorem \ref{Con-bet} gives the uniform rates of convergence of $\hat{\beta}_k$.

\begin{thm}\label{Con-bet}
Suppose that (A1) -- (A6) and (A9) hold. If $K_{\mathrm{C}}K_{\mathrm{A}}=o\left(nNh_{\mathrm{C}}^{4}\right)$ and
$K_{\mathrm{A}}^{-r}+K_{\mathrm{C}}^{-r}=o\left(h_{\mathrm{C}}^{2}\right)$,
and $x$ is any interior in $\mathbb{S}_k$, then
$\sup_{x\in\mathbb{S}_k}|\hat{\beta}_{k}\left(x\right)-\beta_{k}\left(x\right)|$ is bounded by
\begin{align}
O_p\left(h_{\mathrm{A}}^{2}+h_{\mathrm{C}}^{2}
+\sqrt{K_{\mathrm{A}}}\left(K_{\mathrm{C}}^{-r}+K_{\mathrm{A}}^{-r}\right)+
\sqrt{\frac{\log{n}}{n}\left(1+\frac{1}{\bar{N}_{\mathrm{H}}h_{\mathrm{A}}}\right)}\right).\nonumber\label{converge=beta}
\end{align}
\end{thm}

Denote $\mu_{k}=\mathrm{E}\left[\alpha_{k}^{2}\left(T_{ij}\right)\right]$,
$\psi_{k,1}=\mathrm{E}\left[\alpha_{k}^{2}\left(T_{ij}\right)\left\{\gamma\left(T_{ij},T_{ij}\right)
+\sigma^{2}(T_{ij})\right\}\right]$
and
$\psi_{k,2}=\mathrm{E}\left[\alpha_{k}\left(T_{ij}\right)\alpha_{k}\left(T_{ij'}\right)\gamma\left(T_{ij},T_{ij'}\right)\right]$.
Theorem \ref{Dis-bet} presents the asymptotic normality of $\hat{\beta}_{k}$  on a unified formwork
for different types of data.

\begin{thm}\label{Dis-bet}
Under the condition (A1) -- (A9), if $h_{\mathrm{C}}=o\left(h_{\mathrm{A}}\right)$ and
{\small{\begin{equation}\label{Lya-alp}
\frac{\max\left\{\frac{1}{n^3h_{\mathrm{A}}^2}\sum_{i=1}^{n}\frac{1}{m_i^2},\frac{1}{n^3h_{\mathrm{A}}}
\sum_{i=1}^n\frac{1}{m_i^2}\left(m_i-1\right),\frac{1}{n^3}\sum_{i=1}^{n}\left(1-\frac{1}{m_i}\right)
\left(1-\frac{2}{m_i}\right)\right\}}
{\left[\frac{1}{n\bar{N}_{\mathrm{H}}h_{\mathrm{A}}}+\frac{1}{n}
\left(1-\frac{1}{\bar{N}_{\mathrm{H}}}\right)\right]^{3/2}}\nonumber
\end{equation}}}
hold. Then, for any an interior $x$ in $\mathbb{S}_k$, we have
\begin{equation}\label{AsyDis-bet}
\Gamma_{\mathrm{A}}^{-1/2}\left(x\right)\left(\hat{\beta}_k\left(x\right)-\beta_k\left(x\right)
-\tfrac{1}{2}\beta_k''\left(x\right)h_{\mathrm{A}}^{2}\kappa_{2}/\mu_k\right)\xrightarrow{D}N\left(0,1\right),
\end{equation}
where $\Gamma_{\mathrm{A},k}\left(x\right)=\frac{\kappa\psi_{k,1}}{n\bar{N}_{\mathrm{H}}
h_{\mathrm{A}}f_{_{\mathrm{X}_k}}\left(x\right)}+\frac{1}{n}\left(1-\frac{1}{\bar{N}_{\mathrm{H}}}\right)
\frac{\psi_{k,2}}{\mu_k^2}$.
\end{thm}

As a corollary, we get different asymptotic results for  sparse, dense and ultra dense data.

\begin{corollary}\label{Cor-beta}
Suppose that the conditions of Theorem \ref{Dis-bet} hold and $x$ is a fixed interior point in $\mathbb{S}_k$.
\begin{itemize}
\item [$\mathrm{\left(i\right)}$] Sparsity Case. If 
$h_{\mathrm{A}}\asymp \left(n\bar{N}_{\mathrm{H}}\right)^{-\frac{1}{2r+1}}$, then it follows that
{\small{\begin{equation}\label{S-bet}
\sqrt{n\bar{N}_{\mathrm{H}}h_{\mathrm{A}}f_{\mathrm{X_k}}\left(x\right)}\left(\hat{\beta}_k\left(x\right)
-\beta_k\left(x\right)-\tfrac{1}{2}\beta_k''\left(x\right)h_{\mathrm{A}}^{2}\kappa_{2}/\mu_k\right)
\xrightarrow{D} N\left(0,\kappa\psi_{k,1}\right).
\end{equation}}}
\item [$\mathrm{\left(ii\right)}$] Dense Case. If 
$h_{\mathrm{A}}=O\left(n^{-\frac{1}{2r}}\right)$, then
{\small{\begin{equation}\label{D-bet}
\sqrt{n}\left(\hat{\beta}_k\left(x\right)-\beta_k\left(x\right)-\tfrac{1}{2}\beta_k''\left(x\right)
h_{\mathrm{A}}^{2}\kappa_{2}/\mu_k\right)
\xrightarrow{D} N\left(0,\frac{\kappa\psi_{k,1}}{f_{X_{k}}\left(x\right)C_1}+\frac{\psi_{k,2}}{\mu_k^2}\right).
\end{equation}}}
\item [$\mathrm{\left(iii\right)}$] Ultra Dense Case. If
$h_{\mathrm{A}}=o\left(n^{-\frac{1}{2r}}\right)$, then
{\small{\begin{equation}\label{UD-bet}
\sqrt{n}\left(\hat{\beta}_k\left(x\right)-\beta_k\left(x\right)
-\tfrac{1}{2}\beta_k''\left(x\right)h_{\mathrm{A}}^{2}\kappa_{2}/\mu_k\right)
\xrightarrow{D} N\left(0,\psi_{k,2}/\mu_k^2\right).
\end{equation}}}
\end{itemize}
\end{corollary}

Let $\hat{\mu}_k$, $\hat{f}_{_{X_{k}}}(x)$  and $\hat{\psi}_{k,j},j=1,2$ be consistent estimators of
 $\mu_k$, $f_{_{X_{k}}}(x)$ and $\psi_{k,j}$. Then, the asymptotic variance $\Gamma_{\mathrm{A}}(x)$ can be consistently estimated by
\[\hat{\Gamma}_{\mathrm{A}}(x)=\frac{\kappa\hat{\psi}_{k,1}}{n\bar{N}_{\mathrm{H}}\hat{h}_{\mathrm{A}}\hat{f}_{_{\mathrm{X}_k}}(x)}+\frac{1}{n}\Big(1-\frac{1}{\bar{N}_{\mathrm{H}}}\Big)\frac{\hat{\psi}_{k,2}}{\hat{\mu}_k^2},\]
which gives a $(1-\alpha)\%$ pointwise confidence interval of $\beta_k$ in a unified forms for
sparse, dense and ultra dense data. That is,
\begin{equation}\label{NS-beta}
\hat{\beta}_{k}(x)-\tfrac{1}{2}\beta_{k}''(x)h_{\mathrm{A}}^2\kappa_{2}/\hat{\mu}_k
\pm z_{1-\alpha/2}\hat{\Gamma}_{\mathrm{A}}^{-1/2}(x).
\end{equation}

\section{Testing of Model Specification}
\label{sec:testing}
For the sake of parsimony,  it is essential to test time-varying property of varying-coefficient component functions
and  to test linearity of additive component functions.

\subsection{Time-varying Testing of Varying-coefficient Component Functions}
In this subsection, we propose a consistent testing to  judge whether the varying-coefficient component functions are really time-varying or not.
It is a problem of model selection between a general Semi-VCAM and a submodel  PLAM or an AM
in the practical applications.

We denote $\delta_{ij}=\nu_{i}\left(T_{ij}\right)+\varepsilon_{ij}$ in Semi-VCAM \eqref{M_VCAM}, and
consider a mixed-effect nonparametric  model
$Y_{ij}=m\left(T_{ij},\mathbf{Z}_{ij},\mathbf{X}_{ij}\right)+\delta_{ij}$, where
$m\left(t,\mathbf{z},\mathbf{x}\right)=\mathrm{E}\left[Y_{ij}|T_{ij}=t,\mathbf{Z}_{ij}
=\mathbf{z},\mathbf{X}_{ij}=\mathbf{x}\right]$.
The time-varying testing postulates $m$ as
\[
m\left(t,\mathbf{z},\mathbf{x}\right)
=\mathbf{z}^{\tau}\boldsymbol{a}_0+\sum_{k=1}^{p}a_k\beta_k\left(x_k\right)
:=g_0\left(\mathbf{z},\mathbf{x};\mathbf{a},\boldsymbol{\beta}\left(\mathbf{x}\right)\right)\]
under null hypothesis $H_{0,\mathrm{C}}$,  where
$\mathbf{a}=\left(\boldsymbol{a}_0^{\tau},a_1,...,a_k\right)^{\tau}$ is a unknown constant vector,
and $\boldsymbol{\beta}\left(\mathbf{x}\right)=\left(\beta_{1}(x_1),...,\beta_{p}(x_p)\right)^{\tau}$.
Whilst under  alternative hypothesis $H_{1,\mathrm{C}}$, $m$ is  the regression function
of  Semi-VCAM  \eqref{M_VCAM},
denoted as $g\left(t,\mathbf{z},\mathbf{x};\boldsymbol{\alpha}\left(t\right),
\boldsymbol{\beta}\left(\mathbf{x}\right)\right)$.
Then, the interested hypothesis is given as below
\begin{equation}\label{alp-test-all}
\begin{split}
{}&H_{0,\mathrm{C}}:m\left(t,\mathbf{z},\mathbf{x}\right)=g_0\left(t,\mathbf{z},\mathbf{x};\boldsymbol{\alpha},
\boldsymbol{\beta}\left(\mathbf{x}\right)\right)\ \text{a.s.}\\
\leftrightarrow{}&H_{1,\mathrm{C}}:m\left(t,\mathbf{z},\mathbf{x}\right)=g\left(t,\mathbf{z},\mathbf{x};
\boldsymbol{\alpha}\left(t\right),\boldsymbol{\beta}\left(\mathbf{x}\right)\right)\ \text{a.s}.
\end{split}
\end{equation}

Under $H_{0,\mathrm{C}}$,  we replace  $\beta_{k}\left(x\right)$ with PEBLLE
$\hat{\beta}_{k}\left(x\right)$, and obtain the parametric estimator of  vector $\mathbf{a}$ as follows
\begin{equation}\label{alpha0}
\tilde{\mathbf{a}}=\left(\sum_{i=1}^{n}\hat{S}_i\hat{S}_i^{\tau}\right)^{-1
}\sum_{i=1}^{n}\hat{S}_i\mathbf{Y}_i,
\end{equation}
where
$\hat{S}_i=(\hat{S}_{i1},...,\hat{S}_{im_i})^{\tau}$ with
$\hat{S}_{ij}=\big(\mathbf{Z}_{ij}^{\tau},\hat{\beta}_{1}(X_{ij1}),...,\hat{\beta}_{p}(X_{ijp})\big)^{\tau}$,
and $\mathbf{Y}_i=\left(Y_{i1},...,Y_{im_i}\right)^{\tau}$.

For the $i$-th subject and the $j$-th subject,
we introduce the weight matrix
$W_{ij}=\big(w_{ij}^{\left(l,v\right) }\big)_{m_i\times m_j}$, where $w_{ij}^{\left(l,v\right)}
=k_{h_{\mathrm{C}}}\left(T_{il},T_{jv}\right)K_{h_{\mathrm{A}}}\left(X_{il},X_{jv}\right)$,
with $k_{h}\left(t,z\right)=k\left(\frac{t-z}{h}\right)$ and $K_{h}\left(\mathbf{x}_1,\mathbf{x}_2\right)=\Pi_{i=1}^{p}k_{h}\left(x_{1i},x_{2i}\right)$
for $\mathbf{x}_{k}=\left(x_{k1},...,x_{kp}\right)$, $k=1,2$.
Let
$\hat{e}_{ij}=Y_{ij}-g_0(T_{ij},\mathbf{Z}_{ij},\mathbf{X}_{ij};\tilde{\mathbf{a}},
\hat{\boldsymbol{\beta}}\left(\mathbf{X}_{ij}\right))$, and we
propose a testing statistic based upon the quadratic form of residuals as follows
\begin{equation}\label{test-alp}
\hat{J}_n=\frac{1}{n^2\bar{N}_{\mathrm{H}}^2|H|}\sum_{i=1}^{n}\sum_{j\neq i}^{n}\boldsymbol{\hat{e}}_i^{\tau}W_{ij}\boldsymbol{\hat{e}}_{j},
\end{equation}
where $\mathbf{\hat{e}}_i=\left(\hat{e}_{i1},...,\hat{e}_{im_i}\right)^{\tau}$
and $|H|=h_{\mathrm{C}}h_{\mathrm{A}}^p$.

Furthermore, we assumes additional conditions as follows.
\begin{itemize}
\item [(T1)] $\lim_{n\to\infty}\frac{1}{n}\sum_{i=1}^{n}\mathrm{E}\left[S_iS_i^{\tau}\right]=\Omega$, and $\left\Vert\Omega\right\Vert_{F}$
is bounded away from zero and infinity, where $S_i$ is the analogue of $\hat{S}_i$ after $\hat{\beta}_k$
being replaced by $\beta_k$, and $\left\Vert\cdot\right\Vert_{F}$ is the Frobenius norm of the involved matrix.
\item [(T2)] $\bar{N}_{\mathrm{H}}^2|H|\to 0$, $n\bar{N}_{\mathrm{H}}\sqrt{|H|}\to\infty$
and $n\bar{N}_{\mathrm{H}}\sqrt{|H|}h_{\mathrm{A}}^4\to 0$.
\end{itemize}

Let $\bar{N}_2=\frac{1}{n}\sum_{i=1}^{n}m_i^2$ , Theorem \ref{null-dis-alp} and \ref{alter-dis-alp} present  the asymptotic distribution of the proposed test statistic $\hat{J}_n$ under $H_{0,\mathrm{C}}$ and $H_{1,\mathrm{C}}$, respectively.

\begin{thm}\label{null-dis-alp}
Under Assumption (A1) -- (A8) and (T1) -- (T2), it holds that
\[\frac{n^2\bar{N}_{\mathrm{H}}^2}{\sqrt{N^2-n\bar{N}_{2}}}\sqrt{|H|}\hat{J}_n/\hat{\sigma}_1\xrightarrow{D}N(0,1)\]
under $H_{0,\mathrm{C}}$,
where
\[\hat{\sigma}_1^2=\frac{1}{n^2\bar{N}_{\mathrm{H}}|H|}\sum_{i=1}^{n}\sum_{j\neq i}^{n}
\sum_{l=1}^{m_i}\sum_{v=1}^{m_j}\hat{e}_{il}^2\hat{e}_{jv}^2\left(w_{ij}^{\left(l,v\right)}\right)^2\]
is a consistent estimator of the asymptotic variance of
$\frac{n^2\bar{N}_{\mathrm{H}}^2}{\sqrt{N^2-n\bar{N}_{2}}}\sqrt{|H|}\hat{J}_n$, i.e.,
\[\sigma_1^2=\kappa^{p+1}\mathrm{E}\left[\{\gamma\left(T,T\right)
+\sigma^{2}\left(T\right)\}^2f_{\mathrm{T}}\left(T\right)\right]
\Pi_{k=1}^{p}\mathrm{E}\left[f_{k}\left(X_{k}\right)\right].\]
\end{thm}

\begin{thm}\label{alter-dis-alp}
Under the conditions of Theorem \ref{null-dis-alp}, If  $H_{1,\mathrm{C}}$ holds, then $Pr\big(n\bar{N}_{\mathrm{H}}\sqrt{|H|}\hat{J}_n/\hat{\sigma}_1\ge M_n\big)\to 1$ as $n\to\infty$,
where $M_n$ is any non-stochastic positive sequence such that $M_n=o\big(n\bar{N}_{\mathrm{H}}\sqrt{|H|}\big)$.
\end{thm}

\subsection{Linearity Testing of Additive Component Functions}
In this subsection, we check whether each additive component function in Semi-VCAM \eqref{M_VCAM}
reduces to a linear form, which yields a more parsimonious VCM.

Let $h_0(t,\mathbf{z},\mathbf{x};\boldsymbol{\alpha}(t))=\mathbf{z}^{\tau}\boldsymbol{\alpha}_0(t)+\sum_{k=1}^{p}\alpha_k(t)x_k$.
It is expected to test
\begin{equation}\label{test-beta}
\begin{split}
 {}&H_{0,\mathrm{A}}:m\left(t,\mathbf{z},\mathbf{x}\right)
 =h_0\left(t,\mathbf{z},\mathbf{x};\boldsymbol{\alpha}\left(t\right)\right)\ \text{a.s.}\\
\leftrightarrow {}&H_{1,\mathrm{A}}:m\left(t,\mathbf{z},\mathbf{x}\right)
 =g\left(t,\mathbf{z},\mathbf{x};\boldsymbol{\alpha}\left(t\right),
 \boldsymbol{\beta}\left(\mathbf{x}\right)\right)\ \text{a.s.}
 \end{split}
\end{equation}

Denote $\boldsymbol{\tilde{\alpha}}\left(t\right)$ as the LLE of $\boldsymbol{\alpha}\left(t\right)$
under null hypothesis $H_{0,\mathrm{A}}$.
Then, the testing statistics is given by
\begin{equation}\label{test-bet}
\hat{I}_n=\frac{1}{n^2\bar{N}_{\mathrm{H}}^2|H|}\sum_{i=1}^{n}\sum_{j\neq i}^{n}\boldsymbol{\hat{\varsigma}}_i^{\tau}W_{ij}\boldsymbol{\hat{\varsigma}}_{j},
\end{equation}
where $\boldsymbol{\hat{\varsigma}}_{i}=(\hat{\varsigma}_{i1},...,\hat{\varsigma}_{im_i})^{\tau}$
with $\hat{\varsigma}_{ij}=Y_{ij}-h_0\left(T_{ij},\mathbf{Z}_{ij},\mathbf{X}_{ij};
\boldsymbol{\tilde{\alpha}}\left(T_{ij}\right)\right)$.
The asymptotic distributions of $\hat{I}_n$ under $H_{0,\mathrm{A}}$ and $H_{1,\mathrm{A}}$ are
presented in the following two theorems.
\begin{thm}\label{null-dis-beta}
Under the conditions of Theorem \ref{null-dis-alp},  it follows that
\[\frac{n^2\bar{N}_{\mathrm{H}}^2}{\sqrt{N^2-n\bar{N}_{2}}}\sqrt{|H|}\hat{I}_n/\hat{\sigma}_1\xrightarrow{D}N(0,1)\]
under $H_{0,\mathrm{A}}$.
\end{thm}

\begin{thm}\label{alter-dis-beta}
Suppose that the conditions of Theorem \ref{null-dis-alp} holds. Then  under $H_{1,\mathrm{A}}$, we have $Pr(n\bar{N}_{\mathrm{H}}\sqrt{|H|}\hat{I}_n/\hat{\sigma}_1\ge E_n)\to 1$ as $n\to\infty$,
where $E_n$ is any non-stochastic positive sequence such that $E_n=o(n\bar{N}_{\mathrm{H}}\sqrt{|H|})$.
\end{thm}

\section{Implementation}
\label{sec:Implement}

In this section, we address the practical issues that arise in the newly-proposed
methodologies.

\begin{itemize}
\item B-spline method of pilot estimation
\end{itemize}

As a common practice in spline smoothing,
we predetermine the order of the B-spline functions and then
 select optimal interior knots number  through  BIC criterion
\begin{equation}\label{mybic}
\mathrm{BIC}\left(K_{\mathrm{C}},K_{\mathrm{A}}\right)
=\log{\left(\mathrm{RSS}\right)}+\mathcal{N}\log{n}/n,\nonumber
\end{equation}
where {\small{$\mathrm{RSS}=\frac{1}{n}\sum_{i=1}^n\frac{1}{m_i}\sum_{j=1}^{m_{i}}\left[Y_{ij}
-\boldsymbol{\gamma}_{0}^{\tau}\mathbf{Z}_{ij}\otimes\boldsymbol{b}_{\mathrm{C}}\left(T_{ij}\right)
-\sum_{k=1}^{p}\gamma_{k}^{\tau}\mathcal{T}_{k}\left(T_{ij},X_{ijk}\right)\right]^2$}} and
$\mathcal{N}=\left(q+1\right)J_{\mathrm{C}}+pJ_{\mathrm{C}}J_{\mathrm{A}}$,
with $J_{\mathrm{C}}$ and $J_{\mathrm{A}}$ being the dimension of B-spline basis space
$\boldsymbol{b}_{\mathrm{C}}\left(t\right)$ and $\mathbf{B}_{k,\mathrm{A}}\left(x_k\right)$.
Then the optimal interior knots number is given by
$(\hat{K}_{\mathrm{C}},\hat{K}_{\mathrm{A}})
=\arg\min\mathrm{BIC}\left(K_{\mathrm{C}},K_{\mathrm{A}}\right)$.

\begin{itemize}
\item LLE based on pilot estimation
\end{itemize}

In local linear smoothing, we use Epanechnikov  kernel function $k\left(u\right)=0.75\left(1-u^2\right)I_{|u|\leq 1}$, and select the optimal bandwidths using  ``leave-one-out" cross-validation procedure suggested by \cite{rice1991}.
Define the subject-based cross-validation (CV) criterion as below
\begin{equation}\label{CV}
\mathrm{CV}\left(h_{\mathrm{C}},h_{\mathrm{A}}\right)=\sum_{i=1}^{n}\frac{1}{m_{i}}\sum_{j=1}^{m_{i}}
\left[Y_{ij}-\mathbf{Z}_{ij}^{\tau}\hat{\boldsymbol{\alpha}}_{0,-i}\left(T_{ij}\right)
-\sum_{k=1}^{p}\hat{\alpha}_{k,-i}\left(T_{ij}\right)\hat{\beta}_{k,-i}\left(X_{ijk}\right)\right]^{2},\nonumber
\end{equation}
where the subscript ``$-i$'' represents the estimator using the data with all repeated measurements except the $i$th subject.
The optimal bandwidth is the unique minimizer of  $\mathrm{CV}(h_{\mathrm{C}},h_{\mathrm{A}})$.

In simulation studies, we also can use the average squared error (ASE) as follows
\begin{align}\label{ASE}
\mathrm{ASE}\left(h_{\mathrm{C}},h_{\mathrm{A}}\right)
={}&\sum_{i=1}^{n}\frac{1}{m_{i}}\sum_{j=1}^{m_{i}}
\Big[\mathbf{Z}_{ij}^{\tau}\left(\boldsymbol{\alpha}_{0}\left(T_{ij}\right)-\hat{\boldsymbol{\alpha}}_{0}\left(T_{ij}\right)\right)
+\sum_{k=1}^{p}\alpha_{k}\left(T_{ij}\right)\beta_{k}\left(X_{ijk}\right)\nonumber\\
{}&\qquad\qquad\quad-\sum_{k=1}^{p}\hat{\alpha}_{k}\left(T_{ij}\right)\hat{\beta}_{k}\left(X_{ijk}\right)\Big]^2.
\end{align}
Similar to Remark 2.3 of   \cite{wu2000}, it is not difficult to show that the CV bandwidths approximately minimize
ASE.

\section{Numerical Studies}
\label{sec:numerical}

\subsection{Simulation Studies}
\label{sec:simulation}

In this subsection, we consider simulation examples to investigate the finite-sample performance
of the proposed estimation method in Section \ref{sec:est}  and the testing procedure in Section \ref{sec:testing}.

\begin{example}
Here we consider a mixed-effects Semi-VCAM.
 Let $T_{ij}$ are uniformly distributed on $[0,1]$,
 $Z_{ij}$ are i.i.d. Bernoulli random variable with the probability of success $p=0.5$,
 and $X_{ij}=U_i \left(1+T_{ij}\right) + \vartheta_{ij}$,  where $U_i\sim U\left(-0.4,0.4\right)$
and $\vartheta_{ij}\sim N\left(0,0.2^2\right)$.
The response $Y_{ij}$ is generated by a mixed-effects Semi-VCAM  as below
\[Y_{ij}=\alpha_{00}\left(T_{ij}\right)+\alpha_{01}\left(T_{ij}\right)Z_{ij}+\alpha_1\left(T_{ij}\right)\beta_1\left(X_{ij}\right)
+\nu_{i}\left(T_{ij}\right)+\varepsilon_{ij},\]
for $i=1,...,n$ and $j=1,...,m$,
where the measurement error $\varepsilon_{ij}$ are i.i.d from $N\left(0,1\right)$,
and the subject-specific random trajectory
$\nu_{i}\left(T_{ij}\right)=\eta_{i1}+\sqrt{2}\eta_{i2}\sin{\left(2\pi T_{ij}\right)}+\sqrt{2}\eta_{i3}\cos{\left(2\pi T_{ij}\right)}$
with $\eta_{ij}\sim N\left(0,w_j\right)$ for $j=1,2,3$  and $\left(w_1,w_2,w_3\right)=\left(0.6,0.2,0.2\right)$.
The univariate smooth component functions are given by
$\alpha_{00}\left(t\right)=6t$, $\alpha_{01}\left(t\right)=2.5\cos{\left(2\pi t\right)}$, $\alpha_{1}\left(t\right)=\frac{t\left(1-t\right)}{\int_{0}^{1}t\left(1-t\right)\mathrm{d}t}$
and $\beta\left(x\right)=4.5\sin{\left(0.4\pi x\right)}-\mathrm{E}\left[4.5\sin{\left(0.4\pi X\right)}\right]$.

We select  20 equally-spaced points  on the range of $T_{ij}$ and $X_{ij}$, and
define the mean prediction integrated squared error (MPISE) based on $Q$ replications,
\[
\mathrm{MPISE}\left(f\right)=\frac{1}{Q}\sum_{q=1}^{Q}\int\left[\hat{f}_{q}\left(u\right)
 -f\left(u\right)\right]^{2}\mathrm{d}u,\ \
 \]
 where $\hat{f}_{q}$ is the PEBLLE of the estimated function $f$ in the $q$-th replication.
  Under different combinations of $n$ and $m$, based upon $Q=300$ Monte Carlo replications,
 Table \ref{tag:est} gives the  MPISEs of
 PEBLLE of component functions,  and the standard deviation is shown in parentheses.
 We also list the optimal bandwidths  according to  \eqref{ASE}.
 The result  exhibits a good finite-sample performance whatever the data is sparse or dense.
It is also found that MPISEs  decrease markedly as the total observations increase.

\begin{table}[h]
\centering
\caption{The MPISEs(standard deviation in parentheses) of component functions in Example 1.
}
\label{tag:est}
\vspace{3mm}
\begin{tabular}{cccccccc}
\toprule
$n$&$m$&$\hat{h}_{\mathrm{C}}$&$\hat{h}_{\mathrm{A}}$&
$\hat{\alpha}_{00}$ &$\hat{\alpha}_{01}$&$\hat{\alpha}_1$ &$\hat{\beta}$\\
\hline
\multirow{10}{*}{50}&\multirow{2}{*}{5}&\multirow{2}{*}{0.1763}&\multirow{2}{*}{0.4132}
&0.0880&0.1794&0.0048&0.1529\\
&&&&\scriptsize{(0.0605)}&\scriptsize{(0.1129)}&\scriptsize{(0.0028)}&\scriptsize{(0.1122)}\\
&\multirow{2}{*}{10}&\multirow{2}{*}{0.1600}&\multirow{2}{*}{0.3950}
&0.0603&0.0986&0.0028&0.0957\\
&&&&\scriptsize{(0.0400)}&\scriptsize{(0.0585)}&\scriptsize{(0.0016)}&\scriptsize{(0.0646)}\\
&\multirow{2}{*}{30}&\multirow{2}{*}{0.1447}&\multirow{2}{*}{0.3500}
&0.0278&0.0439&0.0019&0.0821\\
&&&&\scriptsize{(0.0194)}&\scriptsize{(0.0245)}&\scriptsize{(0.0012)}&\scriptsize{(0.0574)}\\
&\multirow{2}{*}{50}&\multirow{2}{*}{0.1237}&\multirow{2}{*}{0.3395}
&0.0276&0.0278&0.0016&0.0662\\
&&&&\scriptsize{(0.0193)}&\scriptsize{(0.0137)}&\scriptsize{(0.0010)}&\scriptsize{(0.0516)}\\
&\multirow{2}{*}{100}&\multirow{2}{*}{0.1132}&\multirow{2}{*}{0.2921}
&0.0264&0.0182&0.0014&0.0641\\
&&&&\scriptsize{(0.0263)}&\scriptsize{(0.0092)}&\scriptsize{(0.0010)}&\scriptsize{(0.0441)}\\
\midrule
\multirow{10}{*}{100}&\multirow{2}{*}{10}&\multirow{2}{*}{0.1553}&\multirow{2}{*}{0.3553}
&0.0311&0.0655&0.0023&0.0657\\
&&&&\scriptsize{(0.0200)}&\scriptsize{(0.0369)}&\scriptsize{(0.0013)}&\scriptsize{(0.0395)}\\
&\multirow{2}{*}{30}&\multirow{2}{*}{0.1500}&\multirow{2}{*}{0.2831}
&0.0193&0.0318&0.0016&0.0597\\
&&&&\scriptsize{(0.0153)}&\scriptsize{(0.0129)}&\scriptsize{(0.0009)}&\scriptsize{(0.0405)}\\
&\multirow{2}{*}{60}&\multirow{2}{*}{0.1111}&\multirow{2}{*}{0.2278}
&0.0174&0.0170&0.0009&0.0553\\
&&&&\scriptsize{(0.0147)}&\scriptsize{(0.0075)}&\scriptsize{(0.0008)}&\scriptsize{(0.0443)}\\
&\multirow{2}{*}{100}&\multirow{2}{*}{0.0550}&\multirow{2}{*}{0.2200}
&0.0157&0.0088&0.0006&0.0365\\
&&&&\scriptsize{(0.0110)}&\scriptsize{(0.0033)}&\scriptsize{(0.0004)}&\scriptsize{(0.0284)}\\
&\multirow{2}{*}{150}&\multirow{2}{*}{0.0556}&\multirow{2}{*}{0.1778}
&0.0144&0.0063&0.0006&0.0275\\
&&&&\scriptsize{(0.0013)}&\scriptsize{(0.0026)}&\scriptsize{(0.0006)}&\scriptsize{(0.0180)}\\
\bottomrule
\end{tabular}
\end{table}

Figure \ref{fig:1}  visualizes the PEBLLE for $(n,m)=(100,10)$.
The solid curve plots true component function,
the dashed line figures the PEBLLE,
and the dash-dotted lines give 95\% confidence bands based on the asymptotic distribution.
The figure shows that our estimator is close to the true function even under the medium total observations $N=1000$.

\begin{figure}[h!]
\centering
\begin{tabular}{@{}c@{}c}
\includegraphics[width=0.45\linewidth]{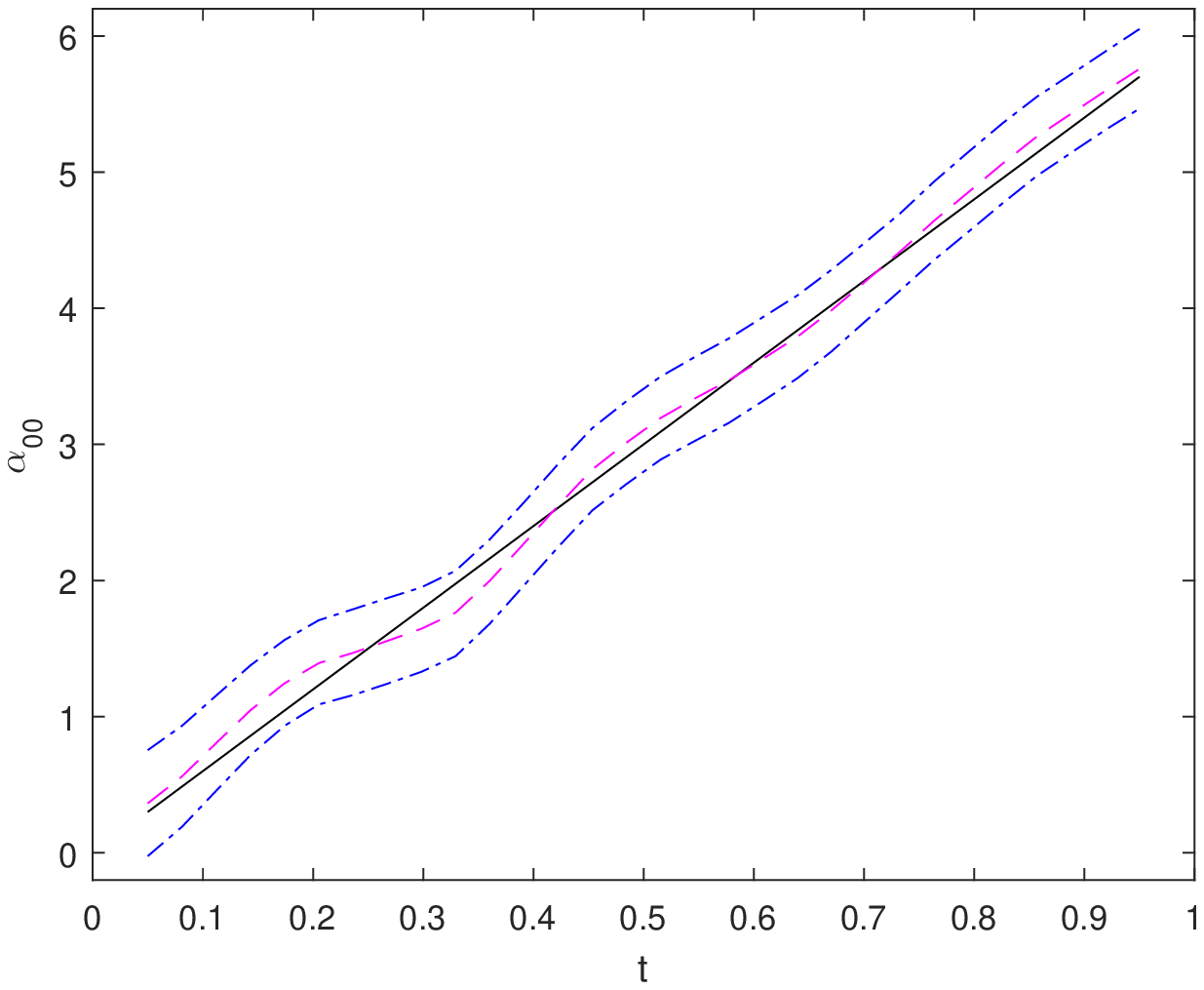} \ &
\includegraphics[width=0.45\linewidth]{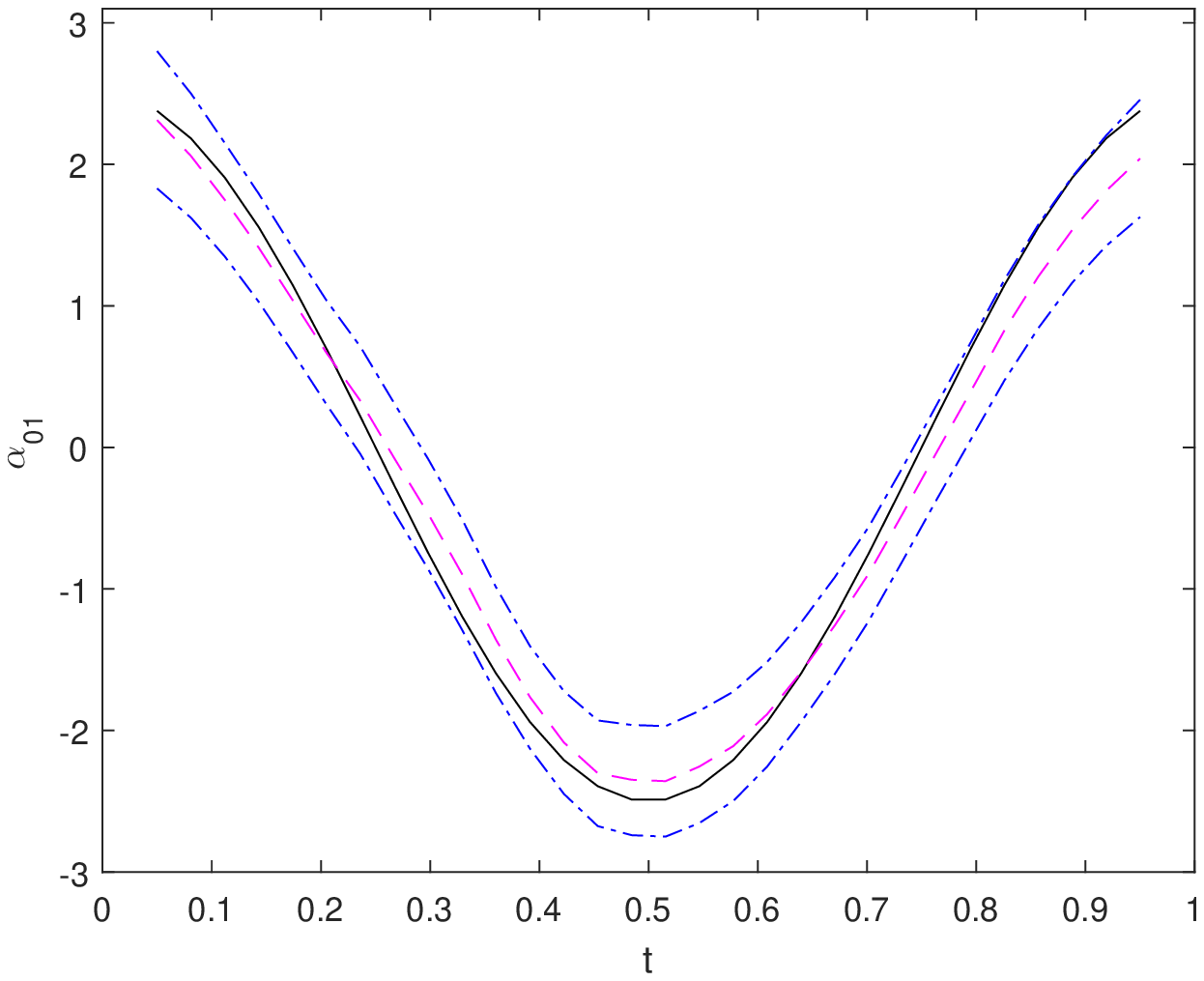} \ \\
{\small(a) Estimation of $\alpha_{00}(t)$} & {\small(b) Estimation of $\alpha_{01}(t)$} \\
\includegraphics[width=0.45\linewidth]{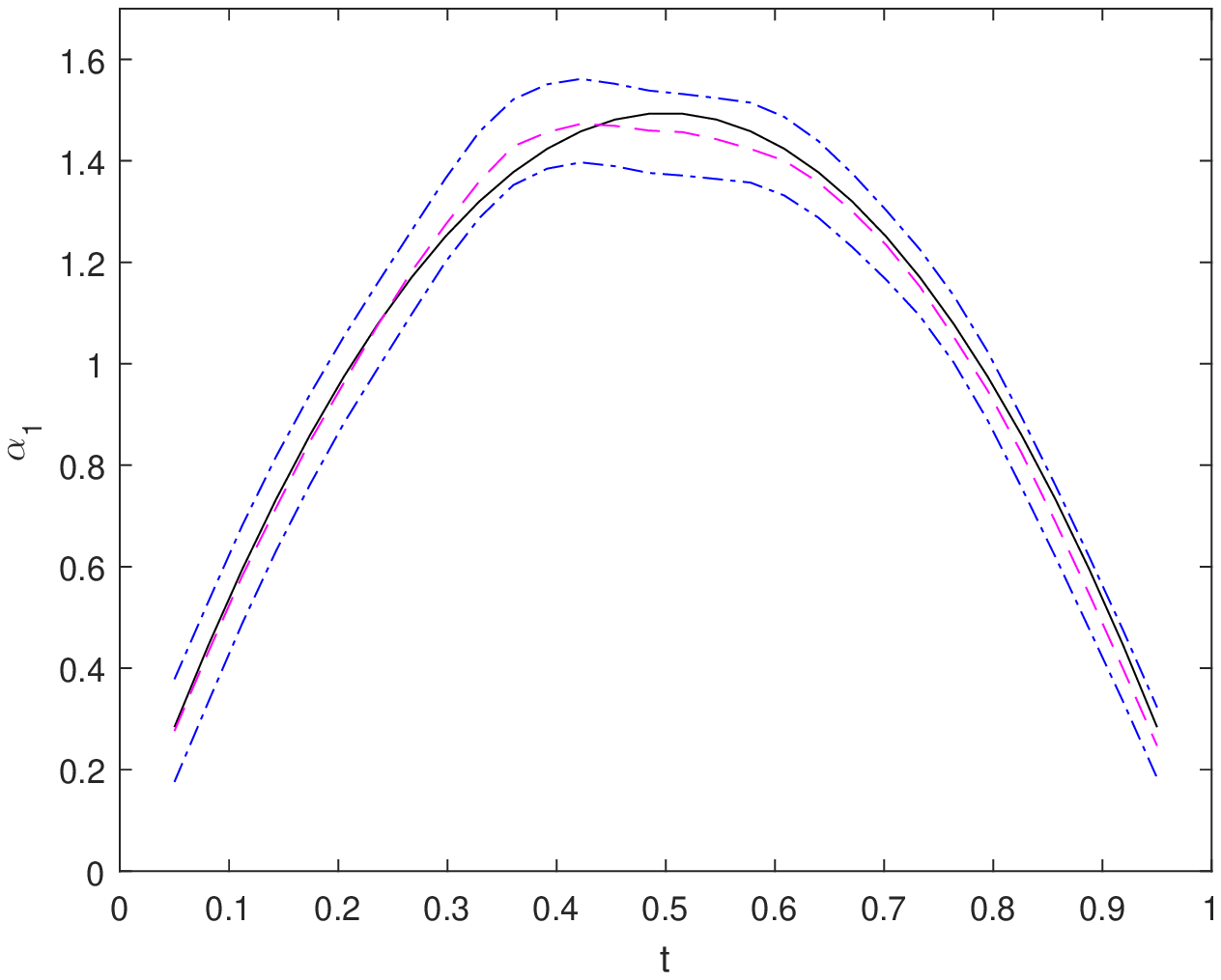} \ &
\includegraphics[width=0.45\linewidth]{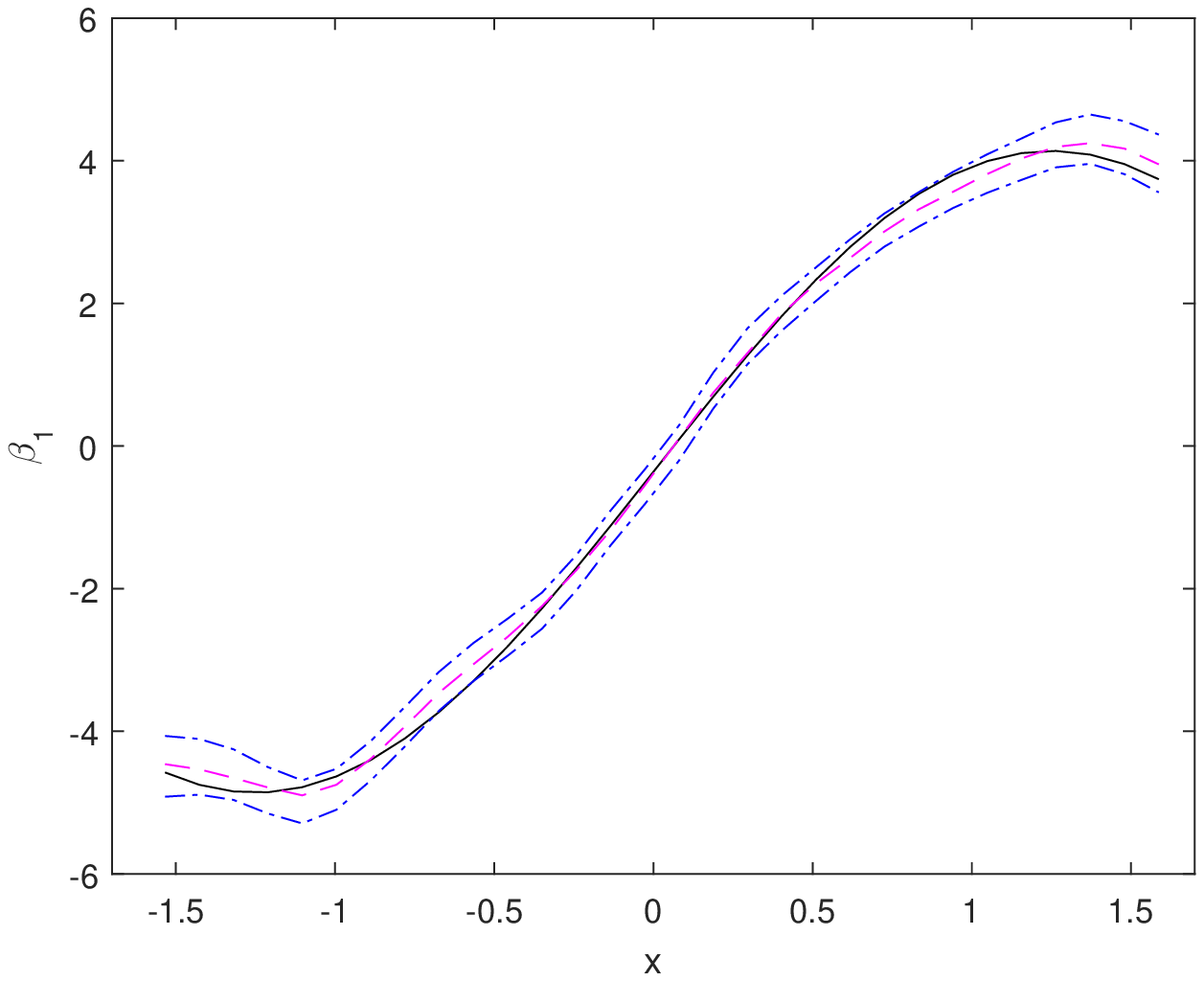} \ \\
{\small(c) Estimation of $\alpha_{1}(t)$} & {\small(d) Estimation of $\beta_{1}(x)$} \\
\end{tabular}
\caption{Estimation of  component functions in Example 1.
The solid curve represents the true function,
  and the dashed line plots the PEBLLE, and the dash-dotted lines gives the 95\%  pointwise
  confidence bands based on \eqref{con-alp1} and \eqref{NS-beta}.
\label{fig:1}}
\end{figure}

We also investigate the performance of asymptotic distribution given in Theorems \ref{Dis-alp} and \ref{Dis-bet}.
After doing 300 Monte Carlo replications, we compare the average empirical coverage percentages (AECPs) based on four methods,
that is, the unified method (U) given in \eqref{AsyDis-alp} and \eqref{AsyDis-bet}, sparse method (S) in \eqref{S-alp} and \eqref{S-bet},
dense method (D) in \eqref{D-alp} and \eqref{D-bet}, and ultra dense method (UD) in \eqref{UD-alp} and \eqref{UD-bet}.
We take $n=50,100$  and $m=5,10,30,80,200$.
Table \ref{tag:cover90}  and \ref{tag:cover95} list the AECPs and the average empirical length (AEL) of  confidence interval under the significance level 90\% and 95\%, respectively.
From the resultant tables, we make a conclusion  that:
 \begin{itemize}
 \item [(1)] the AECPs of unified method (bold tags in tables) are superior to the other three methods, whatever the data is sparse, dense or ultra dense;
 \item [(2)] the AECPs of sparse method decrease as $m$ grows, and they are inferior to  the unified method even for sparse data;
 \item [(3)] the AECPs of dense and ultra dense method increase as $m$ grows, and they are comparable to  that of the  unified method.
 \end{itemize}

\begin{table}[h!]
\caption{The AECPs and AELs~(in parentheses) of four methods with level 90\% in Example 1.}
\label{tag:cover90}
\centering
\vspace{0.5em}
\begin{tabular}{cccccccccc}
\hline\hline
\vspace{-0.2em}
\multirow{2}{*}{$m$}&\multirow{2}{*}{Fun}
&\multicolumn{4}{c}{$n=50$ } & \multicolumn{4}{c}{$n=100$} \\
\cmidrule(r){3-6}\cmidrule(l){7-10}
&&U(\%)&S(\%)&D(\%)&UD(\%)&U(\%)&S(\%)&D(\%)&UD(\%)\\
\midrule
\multirow{8}{*}{5}&\multirow{2}{*}{$\alpha_{00}$}
&\textbf{87.63}&80.98&80.93&64.08&\textbf{88.63}&83.97&81.05&60.80\\
&
&\scriptsize{(0.9016)}&\scriptsize{(0.7539)}&\scriptsize{(0.7671)}&\scriptsize{(0.5309)}
&\scriptsize{(0.7384)}&\scriptsize{(0.6467)}&\scriptsize{(0.6142)}&\scriptsize{(0.3958)}\\
&\multirow{2}{*}{$\alpha_{01}$}
&\textbf{86.57}&79.32&79.22&62.23&\textbf{86.83}&81.52&78.63&57.37\\
&
&\scriptsize{(1.2580)}&\scriptsize{(1.0501)}&\scriptsize{(1.0721)}&\scriptsize{(0.7443)}
&\scriptsize{(1.0919)}&\scriptsize{(0.9576)}&\scriptsize{(0.9068)}&\scriptsize{(0.5814)}\\
&\multirow{2}{*}{$\alpha_{1}$}
&\textbf{86.08}&79.00&79.52&64.42&\textbf{87.82}&85.80&81.17&58.14\\
&
&\scriptsize{(0.3500)}&\scriptsize{(0.2909)}&\scriptsize{(0.2994)}&\scriptsize{(0.2092)}
&\scriptsize{(0.3007)}&\scriptsize{(0.2626)}&\scriptsize{(0.2508)}&\scriptsize{(0.6510)}\\
&\multirow{2}{*}{$\beta_{1}$}
&\textbf{87.53}&85.67&81.68&44.95&\textbf{88.02}&86.48&84.88&38.95\\
&
&\scriptsize{(1.0994)}&\scriptsize{(1.0530)}&\scriptsize{(0.9188)}&\scriptsize{(0.3071)}
&\scriptsize{(1.0756)}&\scriptsize{(1.0301)}&\scriptsize{(0.9057)}&\scriptsize{(0.2154)}\\
\midrule
\multirow{8}{*}{10}&\multirow{2}{*}{$\alpha_{00}$}
&\textbf{88.40}&72.55&85.88&77.98&\textbf{88.48}&74.90&84.88&75.50\\
&
&\scriptsize{(0.7374)}&\scriptsize{(0.4958)}&\scriptsize{(0.6884)}&\scriptsize{(0.5648)}
&\scriptsize{(0.5533)}&\scriptsize{(0.3873)}&\scriptsize{(0.5049)}&\scriptsize{(0.4048)}\\
&\multirow{2}{*}{$\alpha_{01}$}
&\textbf{86.63}&68.90&83.28&73.93&\textbf{87.20}&70.90&82.95&72.35\\
&
&\scriptsize{(1.0573)}&\scriptsize{(0.7118)}&\scriptsize{(0.9868)}&\scriptsize{(0.8088)}
&\scriptsize{(0.7926)}&\scriptsize{(0.5556)}&\scriptsize{(0.7229)}&\scriptsize{(0.5788)}\\
&\multirow{2}{*}{$\alpha_{1}$}
&\textbf{88.00}&71.70&85.35&76.75&\textbf{88.15}&74.65&84.85&75.08\\
&
&\scriptsize{(0.2770)}&\scriptsize{(0.1871)}&\scriptsize{(0.2584)}&\scriptsize{(0.2114)}
&\scriptsize{(0.2057)}&\scriptsize{(0.1435)}&\scriptsize{(0.1880)}&\scriptsize{(0.1512)}\\
&\multirow{2}{*}{$\beta_{1}$}
&\textbf{88.25}&84.43&81.15&52.58&\textbf{88.68}&83.58&85.25&60.63\\
&
&\scriptsize{(0.8836)}&\scriptsize{(0.8200)}&\scriptsize{(0.7146)}&\scriptsize{(0.3007)}
&\scriptsize{(0.4573)}&\scriptsize{(0.3999)}&\scriptsize{(0.4129)}&\scriptsize{(0.2228)}\\
\midrule
\multirow{8}{*}{30}&\multirow{2}{*}{$\alpha_{00}$}
&\textbf{88.90}&50.65&88.20&84.35&\textbf{89.58}&53.10&89.25&85.53\\
&
&\scriptsize{(0.6279)}&\scriptsize{(0.2601)}&\scriptsize{(0.6218)}&\scriptsize{(0.5762)}
&\scriptsize{(0.4638)}&\scriptsize{(0.1965)}&\scriptsize{(0.4611)}&\scriptsize{(0.4235)}
\\
&\multirow{2}{*}{$\alpha_{01}$}
&\textbf{88.15}&47.75&87.75&84.20&\textbf{88.63}&46.37&88.37&84.10\\
&
&\scriptsize{(0.8892)}&\scriptsize{(0.3691)}&\scriptsize{(0.8805)}&\scriptsize{(0.8156)}
&\scriptsize{(0.6545)}&\scriptsize{(0.2773)}&\scriptsize{(0.6506)}&\scriptsize{(0.5975)}\\
&\multirow{2}{*}{$\alpha_{1}$}
&\textbf{88.05}&52.70&87.40&85.20&\textbf{88.82}&54.35&88.47&85.75\\
&
&\scriptsize{(0.2374)}&\scriptsize{(0.0984)}&\scriptsize{(0.2352)}&\scriptsize{(0.2180)}
&\scriptsize{(0.1688)}&\scriptsize{(0.0715)}&\scriptsize{(0.1678)}&\scriptsize{(0.1542)}\\
&\multirow{2}{*}{$\beta_{1}$}
&\textbf{88.50}&71.65&85.60&77.85&\textbf{88.63}&72.62&84.85&75.30\\
&
&\scriptsize{(0.4139)}&\scriptsize{(0.2755)}&\scriptsize{(0.3775)}&\scriptsize{(0.3110)}
&\scriptsize{(0.3014)}&\scriptsize{(0.2089)}&\scriptsize{(0.2761)}&\scriptsize{(0.2190)}\\
\midrule
\multirow{8}{*}{80}&\multirow{2}{*}{$\alpha_{00}$}
&\textbf{89.01}&39.67&88.42&87.16&\textbf{89.61}&38.70&88.88&87.85\\
&
&\scriptsize{(0.5937)}&\scriptsize{(0.1793)}&\scriptsize{(0.5851)}&\scriptsize{(0.5667)}
&\scriptsize{(0.4408)}&\scriptsize{(0.1309)}&\scriptsize{(0.4365)}&\scriptsize{(0.4203)}\\
&\multirow{2}{*}{$\alpha_{01}$}
&\textbf{89.49}&38.71&89.07&87.93&\textbf{89.87}&35.09&89.56&89.11\\
&
&\scriptsize{(0.8367)}&\scriptsize{(0.2523)}&\scriptsize{(0.8248)}&\scriptsize{(0.7988)}
&\scriptsize{(0.6178)}&\scriptsize{(0.1835)}&\scriptsize{(0.6118)}&\scriptsize{(0.5891)}\\
&\multirow{2}{*}{$\alpha_{1}$}
&\textbf{88.71}&41.89&88.13&87.07&\textbf{88.90}&40.00&88.41&87.63\\
&
&\scriptsize{(0.2198)}&\scriptsize{(0.0662)}&\scriptsize{(0.2167)}&\scriptsize{(0.2098)}
&\scriptsize{(0.1593)}&\scriptsize{(0.0475)}&\scriptsize{(0.1578)}&\scriptsize{(0.1519)}\\
&\multirow{2}{*}{$\beta_{1}$}
&\textbf{88.91}&55.58&88.20&85.93&\textbf{89.40}&53.54&88.70&86.72\\
&
&\scriptsize{(0.3471)}&\scriptsize{(0.1583)}&\scriptsize{(0.3315)}&\scriptsize{(0.3105)}
&\scriptsize{(0.2444)}&\scriptsize{(0.1080)}&\scriptsize{(0.2390)}&\scriptsize{(0.2204)}\\

\midrule
\multirow{8}{*}{200}&\multirow{2}{*}{$\alpha_{00}$}
&\textbf{89.50}&28.40&89.20&89.05&\textbf{89.75}&32.45&89.55&89.30\\
&
&\scriptsize{(0.5926)}&\scriptsize{(0.1166)}&\scriptsize{(0.5849)}&\scriptsize{(0.5618)}
&\scriptsize{(0.4192)}&\scriptsize{(0.0979)}&\scriptsize{(0.4072)}&\scriptsize{(0.4097)}\\
&\multirow{2}{*}{$\alpha_{01}$}
&\textbf{89.47}&34.73&89.25&89.13&\textbf{89.60}&34.80&89.55&89.25\\
&
&\scriptsize{(0.8311)}&\scriptsize{(0.1636)}&\scriptsize{(0.8238)}&\scriptsize{(0.7959)}
&\scriptsize{(0.6050)}&\scriptsize{(0.1376)}&\scriptsize{(0.5968)}&\scriptsize{(0.5539)}\\
&\multirow{2}{*}{$\alpha_{1}$}
&\textbf{89.20}&29.93&88.90&88.70&\textbf{89.65}&36.35&89.30&89.05\\
&
&\scriptsize{(0.2069)}&\scriptsize{(0.0406)}&\scriptsize{(0.2056)}&\scriptsize{(0.2032)}
&\scriptsize{(0.1419)}&\scriptsize{(0.0368)}&\scriptsize{(0.1412)}&\scriptsize{(0.1384)}\\
&\multirow{2}{*}{$\beta_{1}$}
&\textbf{89.60}&38.60&89.13&89.04&\textbf{89.80}&43.15&89.35&89.20\\
&
&\scriptsize{(0.3219)}&\scriptsize{(0.0986)}&\scriptsize{(0.3163)}&\scriptsize{(0.3072)}
&\scriptsize{(0.2358)}&\scriptsize{(0.0917)}&\scriptsize{(0.2111)}&\scriptsize{(0.2003)}\\
\hline\hline
\end{tabular}
\end{table}

\begin{table}
\caption{The AECPs and AELs (in parentheses) of of four methods  with level 95\% in Example 1.}
\label{tag:cover95}
\centering
\vspace{0.5em}
\begin{tabular}{cccccccccc}
\hline\hline
\vspace{-0.2em}
\multirow{2}{*}{$m$}&\multirow{2}{*}{Fun}
&\multicolumn{4}{c}{$n=50$ } & \multicolumn{4}{c}{$n=100$} \\
\cmidrule(r){3-6}\cmidrule(l){7-10}
&&U(\%)&S(\%)&D(\%)&UD(\%)&U(\%)&S(\%)&D(\%)&UD(\%)\\
\midrule
\multirow{8}{*}{5}&\multirow{2}{*}{$\alpha_{00}$}
&\textbf{93.15}&88.93&88.23&72.70&\textbf{93.68}&90.45&88.50&69.23\\
&&\scriptsize{(1.0818)}&\scriptsize{(0.9006)}&\scriptsize{(0.9160)}&\scriptsize{(0.6467)}
&\scriptsize{(0.8599)}&\scriptsize{(0.7479)}&\scriptsize{(0.7303)}&\scriptsize{(0.4270)}\\
&\multirow{2}{*}{$\alpha_{01}$}
&\textbf{92.68}&87.73&87.03&70.88&\textbf{93.08}&88.22&86.12&66.52\\
&
&\scriptsize{(1.5106)}&\scriptsize{(1.2561)}&\scriptsize{(1.2804)}&\scriptsize{(0.9062)}
&\scriptsize{(1.2863)}&\scriptsize{(1.1242)}&\scriptsize{(1.0779)}&\scriptsize{(0.6945)}\\
&\multirow{2}{*}{$\alpha_{1}$}
&\textbf{91.60}&87.12&85.92&71.92&\textbf{93.12}&90.10&87.65&70.23\\
&
&\scriptsize{(0.4249)}&\scriptsize{(0.3520)}&\scriptsize{(0.3613)}&\scriptsize{(0.2573)}
&\scriptsize{(0.3540)}&\scriptsize{(0.3082)}&\scriptsize{(0.2979)}&\scriptsize{(0.1945)}\\
&\multirow{2}{*}{$\beta_{1}$}
&\textbf{92.30}&91.00&87.95&52.08&\textbf{93.30}&92.25&90.18&43.37\\
&
&\scriptsize{(1.2933)}&\scriptsize{(1.2373)}&\scriptsize{(1.0916)}&\scriptsize{(0.3655)}
&\scriptsize{(0.9537)}&\scriptsize{(0.9195)}&\scriptsize{(0.8595)}&\scriptsize{(0.2563)}\\
\midrule
\multirow{8}{*}{10}&\multirow{2}{*}{$\alpha_{00}$}
&\textbf{93.50}&81.28&89.93&82.13&\textbf{94.13}&85.53&89.35&81.13\\
&
&\scriptsize{(0.9012)}&\scriptsize{(0.6247)}&\scriptsize{(0.8171)}&\scriptsize{(0.6706)}
&\scriptsize{(0.6933)}&\scriptsize{(0.5056)}&\scriptsize{(0.6030)}&\scriptsize{(0.4853)}\\
&\multirow{2}{*}{$\alpha_{01}$}
&\textbf{93.50}&82.28&89.98&82.40&\textbf{93.90}&86.60&90.65&82.18\\
&
&\scriptsize{(1.2926)}&\scriptsize{(0.8966)}&\scriptsize{(1.1718)}&\scriptsize{(0.9611)}
&\scriptsize{(0.9955)}&\scriptsize{(0.7274)}&\scriptsize{(0.8649)}&\scriptsize{(0.6946)}\\
&\multirow{2}{*}{$\alpha_{1}$}
&\textbf{93.15}&82.48&90.40&82.73&\textbf{93.23}&84.05&88.58&80.50\\
&
&\scriptsize{(0.3390)}&\scriptsize{(0.2357)}&\scriptsize{(0.3072)}&\scriptsize{(0.2516)}
&\scriptsize{(0.2561)}&\scriptsize{(0.1862)}&\scriptsize{(0.2232)}&\scriptsize{(0.1801)}\\
&\multirow{2}{*}{$\beta_{1}$}
&\textbf{93.63}&90.15&87.68&60.20&\textbf{94.33}&90.53&92.43&70.48\\
&
&\scriptsize{(0.9368)}&\scriptsize{(0.8567)}&\scriptsize{(0.7700)}&\scriptsize{(0.3599)}
&\scriptsize{(0.5288)}&\scriptsize{(0.4582)}&\scriptsize{(0.4908)}&\scriptsize{(0.2658)}\\
\midrule
\multirow{8}{*}{30}&\multirow{2}{*}{$\alpha_{00}$}
&\textbf{93.88}&68.25&92.12&89.30&\textbf{94.15}&64.30&94.00&90.75\\
&
&\scriptsize{(0.7657)}&\scriptsize{(0.3638)}&\scriptsize{(0.7305)}&\scriptsize{(0.6761)}
&\scriptsize{(0.5495)}&\scriptsize{(0.2506)}&\scriptsize{(0.5376)}&\scriptsize{(0.4906)}\\
&\multirow{2}{*}{$\alpha_{01}$}
&\textbf{93.90}&67.45&92.90&90.20&\textbf{94.05}&59.25&93.05&90.05\\
&
&\scriptsize{(1.0900)}&\scriptsize{(0.5187)}&\scriptsize{(1.0402)}&\scriptsize{(0.9621)}
&\scriptsize{(0.7753)}&\scriptsize{(0.3535)}&\scriptsize{(0.7585)}&\scriptsize{(0.6922)}\\
&\multirow{2}{*}{$\alpha_{1}$}
&\textbf{93.80}&72.90&92.05&89.75&\textbf{94.20}&68.75&93.15&89.95\\
&
&\scriptsize{(0.2952)}&\scriptsize{(0.1404)}&\scriptsize{(0.2818)}&\scriptsize{(0.2605)}
&\scriptsize{(0.2012)}&\scriptsize{(0.0916)}&\scriptsize{(0.1969)}&\scriptsize{(0.1798)}\\
&\multirow{2}{*}{$\beta_{1}$}
&\textbf{94.10}&81.00&92.20&86.00&\textbf{94.55}&82.05&92.45&83.85\\
&
&\scriptsize{(0.4793)}&\scriptsize{(0.3082)}&\scriptsize{(0.4438)}&\scriptsize{(0.3708)}
&\scriptsize{(0.3590)}&\scriptsize{(0.2498)}&\scriptsize{(0.3256)}&\scriptsize{(0.2602)}\\
\midrule
\multirow{8}{*}{80}&\multirow{2}{*}{$\alpha_{00}$}
&\textbf{94.45}&45.35&93.80&92.80&\textbf{94.60}&49.25&93.30&92.05\\
&
&\scriptsize{(0.7298)}&\scriptsize{(0.2136)}&\scriptsize{(0.7207)}&\scriptsize{(0.6933)}
&\scriptsize{(0.5183)}&\scriptsize{(0.1602)}&\scriptsize{(0.5107)}&\scriptsize{(0.4924)}\\
&\multirow{2}{*}{$\alpha_{01}$}
&\textbf{94.25}&52.80&94.20&93.80&\textbf{94.75}&51.40&94.35&93.70\\
&
&\scriptsize{(1.0268)}&\scriptsize{(0.3003)}&\scriptsize{(1.0141)}&\scriptsize{(0.9440)}
&\scriptsize{(0.7273)}&\scriptsize{(0.2250)}&\scriptsize{(0.7165)}&\scriptsize{(0.6908)}\\
&\multirow{2}{*}{$\alpha_{1}$}
&\textbf{94.10}&51.70&92.75&91.65&\textbf{94.65}&52.90&93.05&91.85\\
&
&\scriptsize{(0.2635)}&\scriptsize{(0.0769)}&\scriptsize{(0.2602)}&\scriptsize{(0.2525)}
&\scriptsize{(0.1966)}&\scriptsize{(0.0609)}&\scriptsize{(0.1937)}&\scriptsize{(0.1867)}\\
&\multirow{2}{*}{$\beta_{1}$}
&\textbf{94.50}&58.15&94.00&92.45&\textbf{94.80}&65.40&94.00&93.15\\
&
&\scriptsize{(0.4093)}&\scriptsize{(0.1845)}&\scriptsize{(0.3938)}&\scriptsize{(0.3674)}
&\scriptsize{(0.2902)}&\scriptsize{(0.1313)}&\scriptsize{(0.2805)}&\scriptsize{(0.2601)}\\
\midrule
\multirow{8}{*}{200}&\multirow{2}{*}{$\alpha_{00}$}
&\textbf{94.50}&30.95&94.30&94.10&\textbf{94.85}&34.25&94.65&94.45\\
&
&\scriptsize{(0.7169)}&\scriptsize{(0.1342)}&\scriptsize{(0.7132)}&\scriptsize{(0.6843)}
&\scriptsize{(0.5122)}&\scriptsize{(0.1005)}&\scriptsize{(0.4998)}&\scriptsize{(0.4827)}\\
&\multirow{2}{*}{$\alpha_{01}$}
&\textbf{94.80}&39.65&94.40&94.25&\textbf{94.95}&40.70&94.85&94.40\\
&
&\scriptsize{(1.0003)}&\scriptsize{(0.1876)}&\scriptsize{(0.9950)}&\scriptsize{(0.9325)}
&\scriptsize{(0.7217)}&\scriptsize{(0.1407)}&\scriptsize{(0.7083)}&\scriptsize{(0.6885)}\\
&\multirow{2}{*}{$\alpha_{1}$}
&\textbf{94.35}&35.20&94.15&94.00&\textbf{94.75}&41.45&94.55&94.30\\
&
&\scriptsize{(0.2517)}&\scriptsize{(0.0507)}&\scriptsize{(0.2503)}&\scriptsize{(0.2469)}
&\scriptsize{(0.1842)}&\scriptsize{(0.0353)}&\scriptsize{(0.1834)}&\scriptsize{(0.1809)}\\
&\multirow{2}{*}{$\beta_{1}$}
&\textbf{94.75}&47.40&94.30&94.15&\textbf{94.85}&47.35&94.60&94.35\\
&
&\scriptsize{(0.3838)}&\scriptsize{(0.1116)}&\scriptsize{(0.3778)}&\scriptsize{(0.3580)}
&\scriptsize{(0.2712)}&\scriptsize{(0.0809)}&\scriptsize{(0.2683)}&\scriptsize{(0.2595)}\\
\hline\hline
\end{tabular}
\end{table}
\end{example}

\begin{example}
Now we  investigate the performance of hypothesis testing constructed in Section \ref{sec:testing}.
To this end, we consider the following two DGP:
\begin{itemize}
\item DGP I: In this case, we test the time-varying property of varying-coefficient component functions,
that is to decide whether a PLAM is sufficient.
 We take the same settings with Example 1 for $T_{ij}$, $X_{ij}$, $Z_{ij}$, $\varepsilon_{ij}$, $\nu_i$ and $\beta_1\left(x\right)$.
 The time-varying testing of conditional mean function
$m\left(t,z,x\right)$ is as follows
{\small{\begin{equation}\label{time-test}
H_0:m\left(t,z,x\right)=g_0\left(z,x\right) \ a.s. \ \leftrightarrow H_1:m\left(t,z,x\right)=g_1\left(t,z,x\right) \ a.s.,
\end{equation}}}
where $g_0\left(z,x\right)=6+2.5z+\beta_1\left(x\right)$
and \[g_1\left(t,z,x\right)=g_0\left(z,x\right)+\theta\left( t+ z\cos{\left(2\pi t\right)}+ t\left(1-t\right)\beta_1\left(x\right)\right),\]
with $\theta =$ 0.2, 0.4, 0.6, 0.8, 1.0.
\item DGP II: Here we consider  the linearity testing of additive component functions
to judge whether a VCM is sufficient.
Let $T_{ij}$, $Z_{ij}$, $\alpha_{00}\left(t\right)$ and $\alpha_{01}\left(t\right)$ be given in Example 1,
$\alpha_1\left(t\right)=\sin{\left(\pi t\right)}/\int_{0}^{1}\sin{\left(\pi t\right)}\mathrm{d}t$,
and $X_{ij}=U_i \left(1+T_{ij}\right) + \vartheta_{ij}$,  where $U_i\sim U\left(-0.5,0.5\right)$
and $\vartheta_{ij}\sim N\left(0,1\right)$.
The interested hypothesis is given by
{\small{\begin{equation}\label{linear-test}
H_0:m\left(t,z,x\right)=h_0\left(t,z,x\right)\ \text{a.s.}\leftrightarrow
H_1:m\left(t,z,x\right)=h_1\left(t,z,x\right)\ \text{a.s.,}
\end{equation}}}
where $h_0\left(t,z,x\right)=\alpha_{00}\left(t\right)+\alpha_{01}\left(t\right)z+\alpha_{1}\left(t\right)x$
and $h_1\left(t,z,x\right)=h_0\left(t,z,x\right)+ 1.5 \theta\alpha_{1}\left(t\right)\sin{\left(\pi x\right)}$
with $\theta =$ 0.2, 0.4, 0.6, 0.8, 1.0.
\end{itemize}

We consider different combinations of $n=30,50,100$  and $m=$ 5, 10, 30, 60, 100, and
generate $Q=300$ Monte Carlo replications
and $B=300 $ bootstrap samples for each simulated data set.
 Under 5\% and 10\% significance levels,
 based upon  bootstrap critical value, Table \ref{tag:cover-alp} and \ref{tag:cover-bet} present power of hypothesis \eqref{time-test} and \eqref{linear-test} for different deviation parameters $\theta$ ranging from 0 to 1
 with the span of 0.2, respectively.
The results show that the proposed testing procedure all performs well for sparse, dense and dense data.
In fact, the power for $\theta=0$ is size of hypothesis, which is close to the theoretical significance level 0.05 or 0.1.
As expected, the power increases to one as $\theta$ ascends whatever significance levels
and sampling plans.
Moreover,
Figures \ref{fig:pow-alp} and \ref{fig:pow-beta} plot the rejection rates of  testing \eqref{time-test}
and \eqref{linear-test} at the 5\% and 10\% significance levels for some combinations of $n$ and $m$,
respectively.

\begin{table}
\caption{Power of testing \eqref{time-test}  under confidence level $\alpha=5\%$ and 10\%.}
\label{tag:cover-alp}
\centering
\vspace{0.5em}
\begin{tabular}{ccccccccc}
\hline\hline
\vspace{-0.2em}
\multirow{2}{*}{$\alpha$}&\multirow{2}{*}{$\theta$}&\multicolumn{7}{c}{$(n,m)$ }\\
\cmidrule(r){3-9}
&&(50,5)&(50,10)&(100,5)&(100,10)&(50,30)&(30,60)&(30,100)\\
\midrule
\multirow{6}{*}{5\%}&0&0.067&0.055&0.050&0.040&0.055&0.060&0.050\\
                              &0.2&0.233&0.300&0.265&0.370&0.270&0.225&0.230\\
                              &0.4&0.500&0.690&0.725&0.885&0.875&0.710&0.770\\
                              &0.6&0.830&0.955&0.985&1.000&1.000&0.995&1.000\\
                              &0.8&0.970&1.000&1.000&1.000&1.000&1.000&1.000\\
                              &1.0&0.997&1.000&1.000&1.000&1.000&1.000&1.000\\

\cline{1-9}
\multirow{6}{*}{10\%}&0&0.113&0.095&0.095&0.085&0.100&0.085&0.090\\
                                &0.2&0.333&0.410&0.360&0.505&0.345&0.290&0.240\\
                                &0.4&0.610&0.755&0.815&0.930&0.920&0.765&0.810\\
                                &0.6&0.857&0.965&0.990&1.000&1.000&0.995&1.000\\
                                &0.8&0.980&1.000&1.000&1.000&1.000&1.000&1.000\\
                               &1.0&1.000&1.000&1.000&1.000&1.000&1.000&1.000\\
\hline\hline
\end{tabular}
\end{table}

\begin{table}
\caption{Size and power of test  \eqref{linear-test} under confidence level $\alpha=5\%$ and 10\%.}
\label{tag:cover-bet}
\centering
\vspace{0.5em}
\begin{tabular}{ccccccccc}
\hline\hline
\vspace{-0.2em}
\multirow{2}{*}{$\alpha$}&\multirow{2}{*}{$\theta$}&\multicolumn{7}{c}{$(n,m)$ }\\
\cmidrule(r){3-9}
&&(50,5)&(50,10)&(100,5)&(100,10)&(50,30)&(30,60)&(30,100)\\
\midrule
\multirow{6}{*}{5\%}&0&0.047&0.057&0.050&0.045&0.040&0.045&0.060\\
                              &0.2&0.107&0.120&0.173&0.205&0.200&0.200&0.180\\
                              &0.4&0.203&0.557&0.477&0.825&0.907&0.850&0.807\\
                              &0.6&0.653&0.983&0.937&1.000&1.000&0.990&0.993\\
                              &0.8&0.920&1.000&1.000&1.000&1.000&1.000&1.000\\
                              &1.0&0.997&1.000&1.000&1.000&1.000&1.000&1.000\\
\cline{1-9}
\multirow{6}{*}{10\%}&0   &0.080&0.083&0.083&0.090&0.0933&0.090&0.100\\
                                &0.2&0.157&0.197&0.250&0.350&0.327&0.360&0.260\\
                                &0.4&0.303&0.670&0.610&0.925&0.933&0.930&0.873\\
                                &0.6&0.740&0.990&0.967&1.000&1.000&1.000&1.000\\
                                &0.8&0.947&1.000&1.000&1.000&1.000&1.000&1.000\\
                                &1.0&0.997&1.000&1.000&1.000&1.000&1.000&1.000\\
\hline\hline
\end{tabular}
\end{table}

\begin{figure}[h!]
\centering
\begin{tabular}{@{}c@{}c@{}c}
\includegraphics[width=0.34\linewidth]{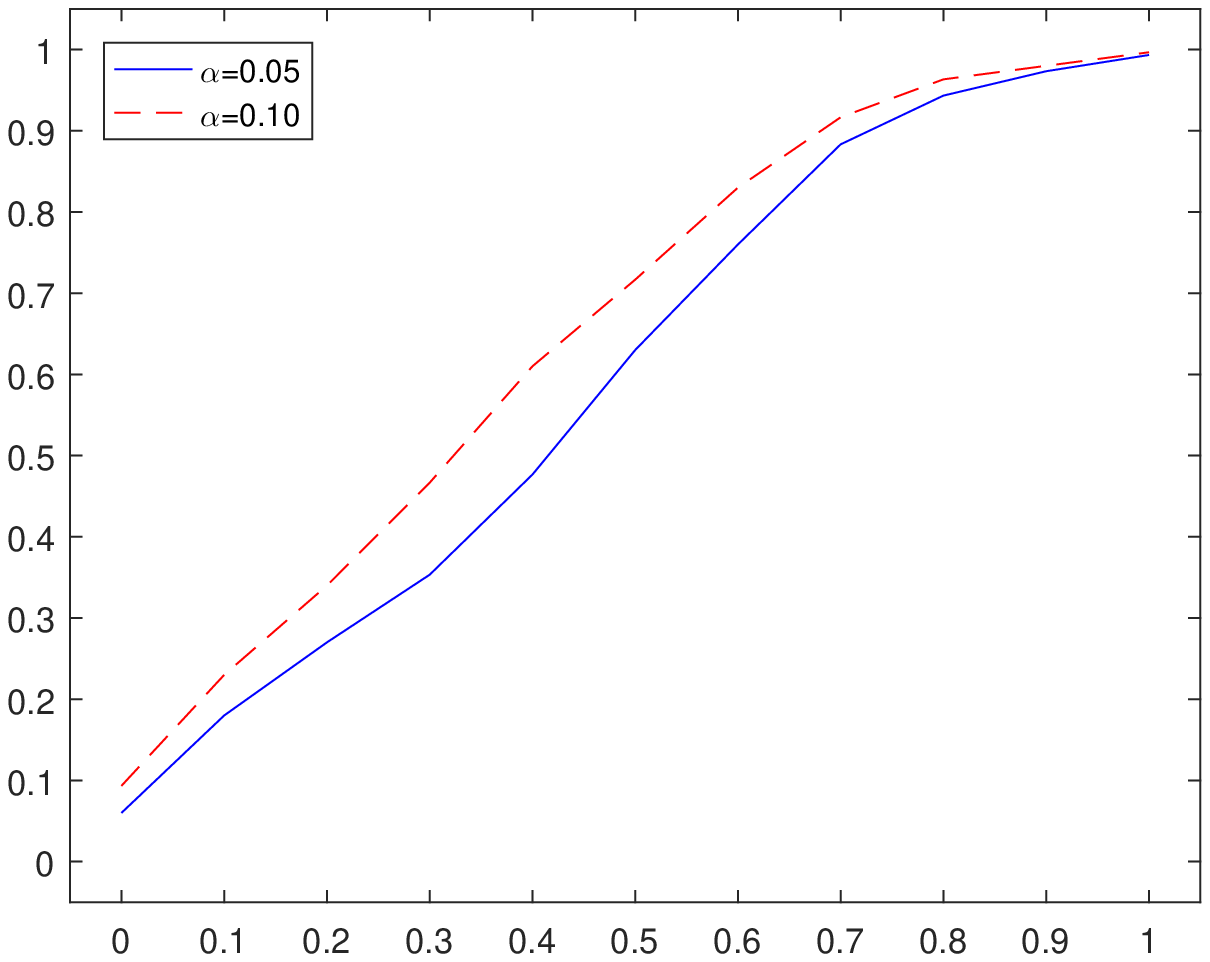} \ &
\includegraphics[width=0.34\linewidth]{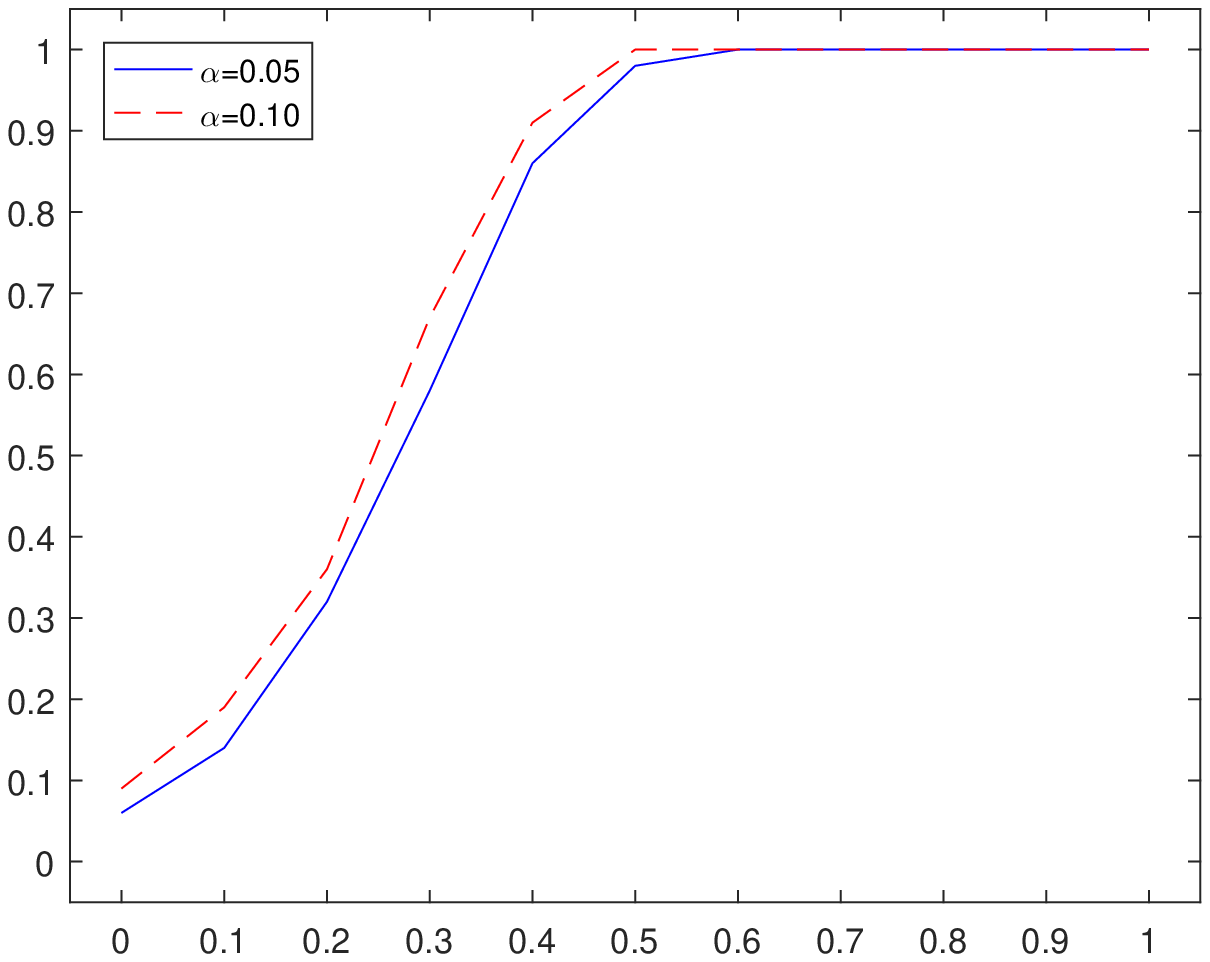} \ &
\includegraphics[width=0.34\linewidth]{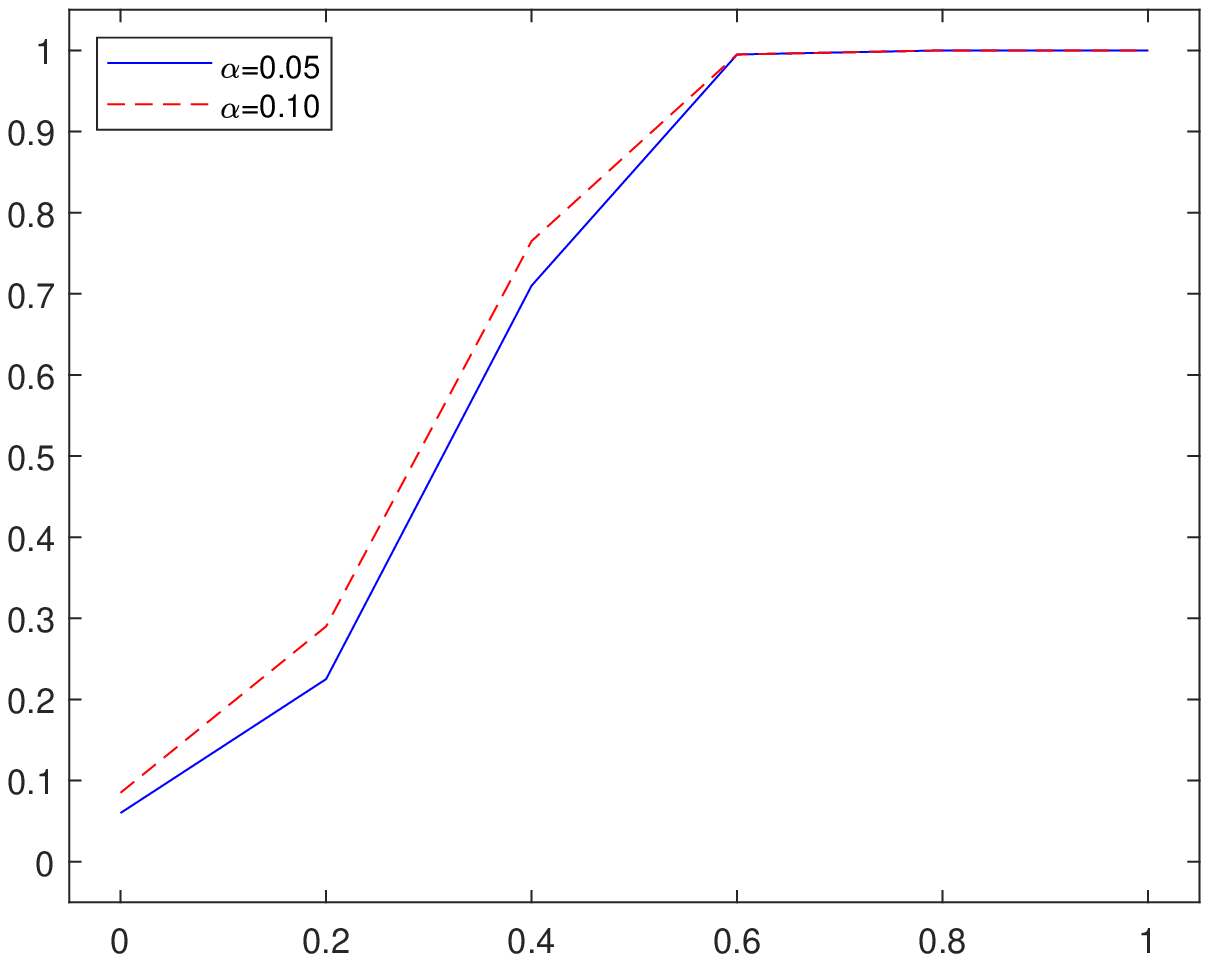} \ \\
{\tiny($a$) Power for $(n,m)=(50,5)$} & {\tiny($b$) Power for $(n,m)=(50,30)$}
&{\tiny($c$) Power for $(n,m)=(30,60)$}\\
\includegraphics[width=0.34\linewidth]{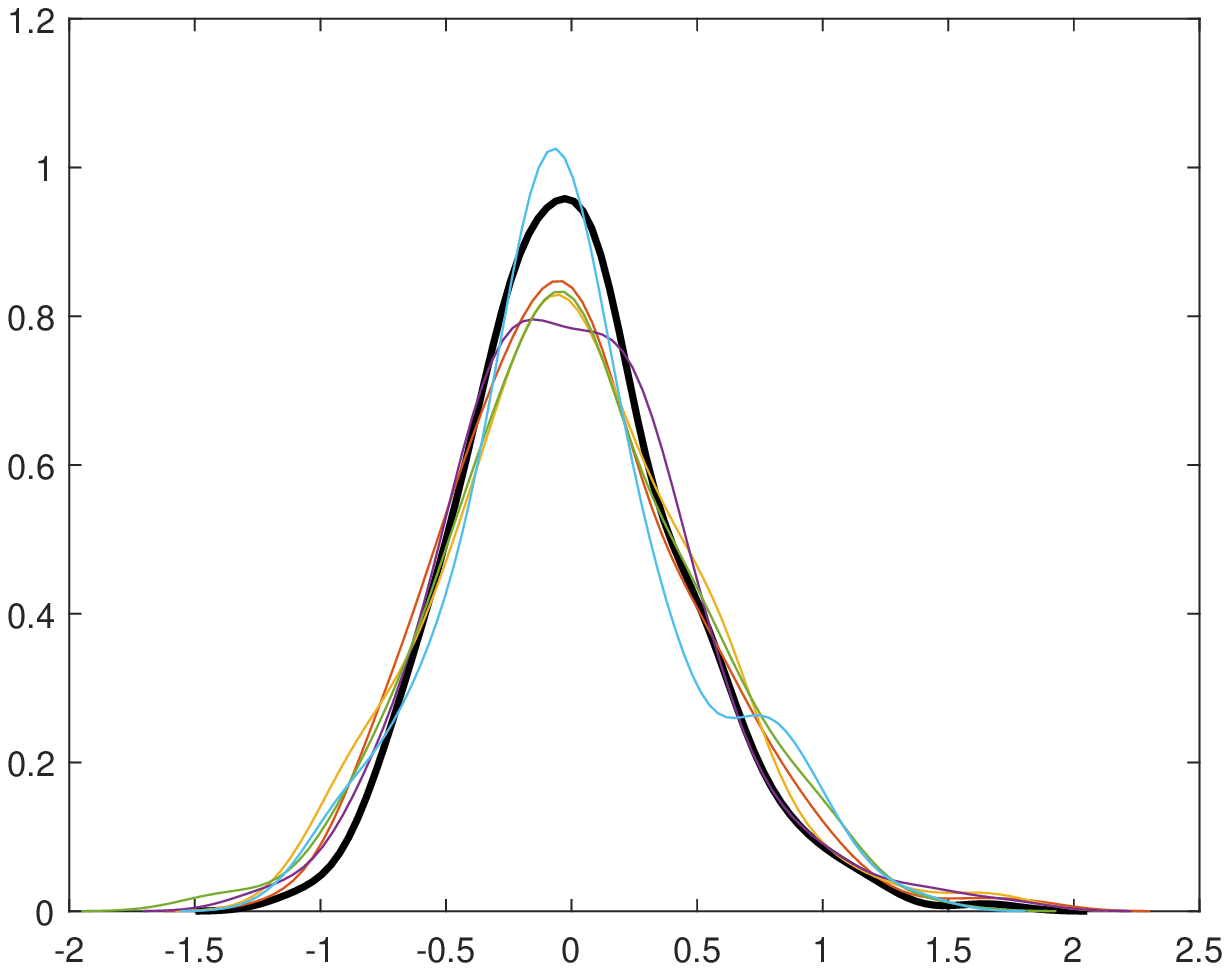} \ &
\includegraphics[width=0.34\linewidth]{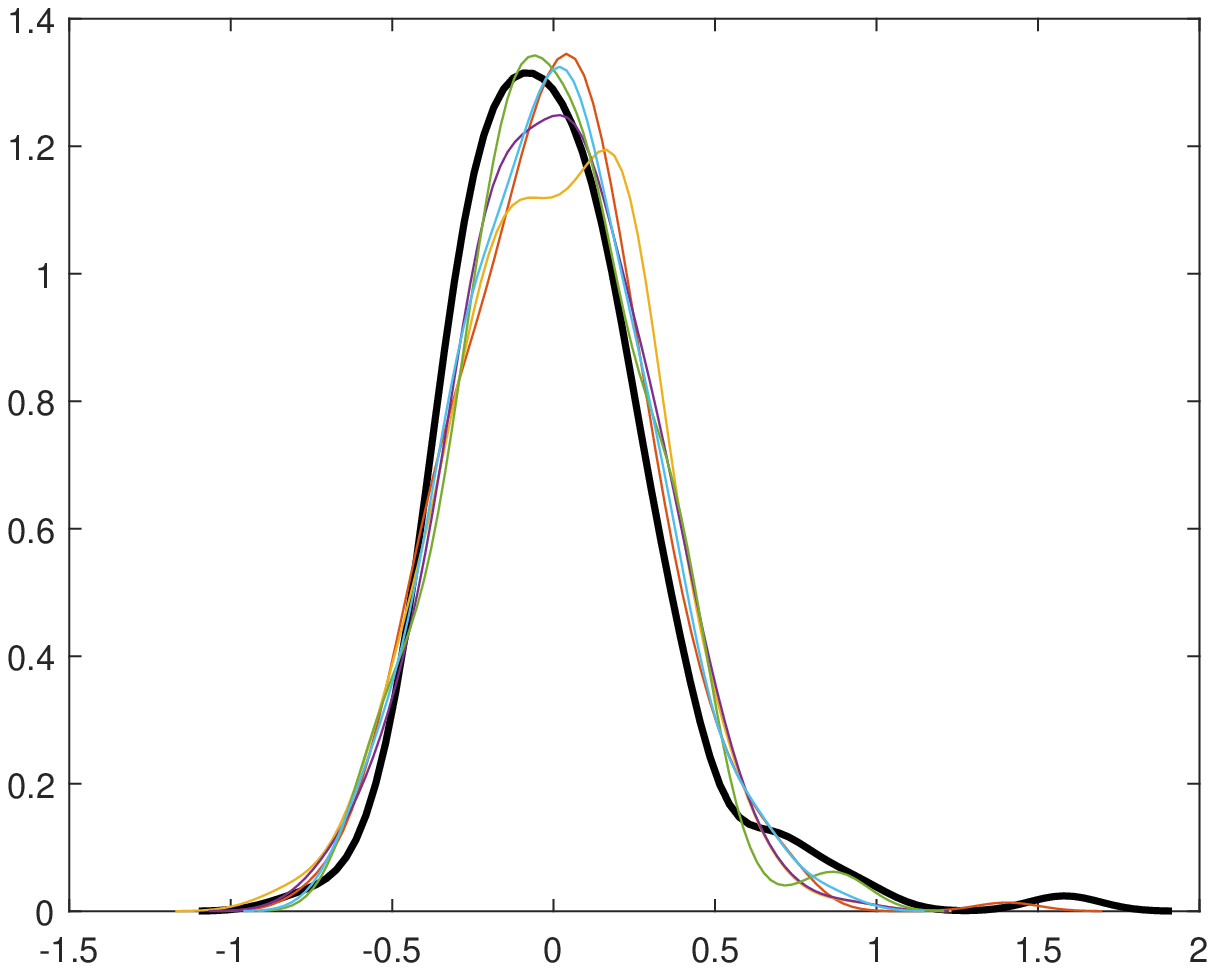} \ &
\includegraphics[width=0.34\linewidth]{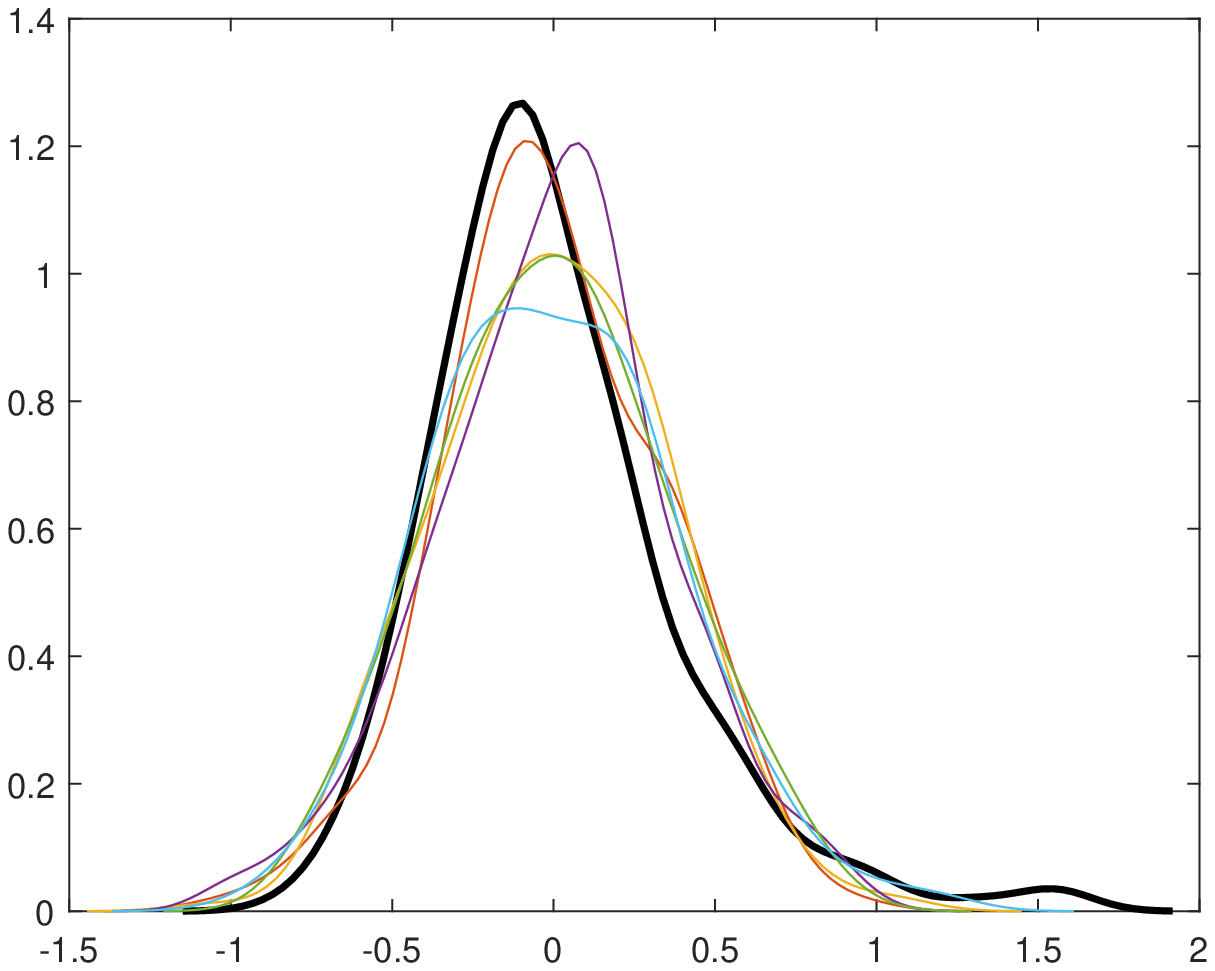} \ \\
{\tiny($a'$) Density for $(n,m)=(50,5)$ } & {\tiny($b'$) Density for $(n,m)=(50,30)$}
& {\tiny($c'$) Density for $(n,m)=(30,60)$}\\
\end{tabular}
\caption{Power of time-varying testing \eqref{time-test} in Example 2.
For three combinations of $n$ and $m$,
($a$) -($c$) figure power at the level $\alpha=0.05$ and 0.1; whist ($a'$) -($c'$) give the simulated density of
standardized test statistics (thick black) and five bootstrap approximations (thin).}
\label{fig:pow-alp}
\end{figure}

\begin{figure}[h!]
\centering
\begin{tabular}{@{}c@{}c@{}c}
\includegraphics[width=0.34\linewidth]{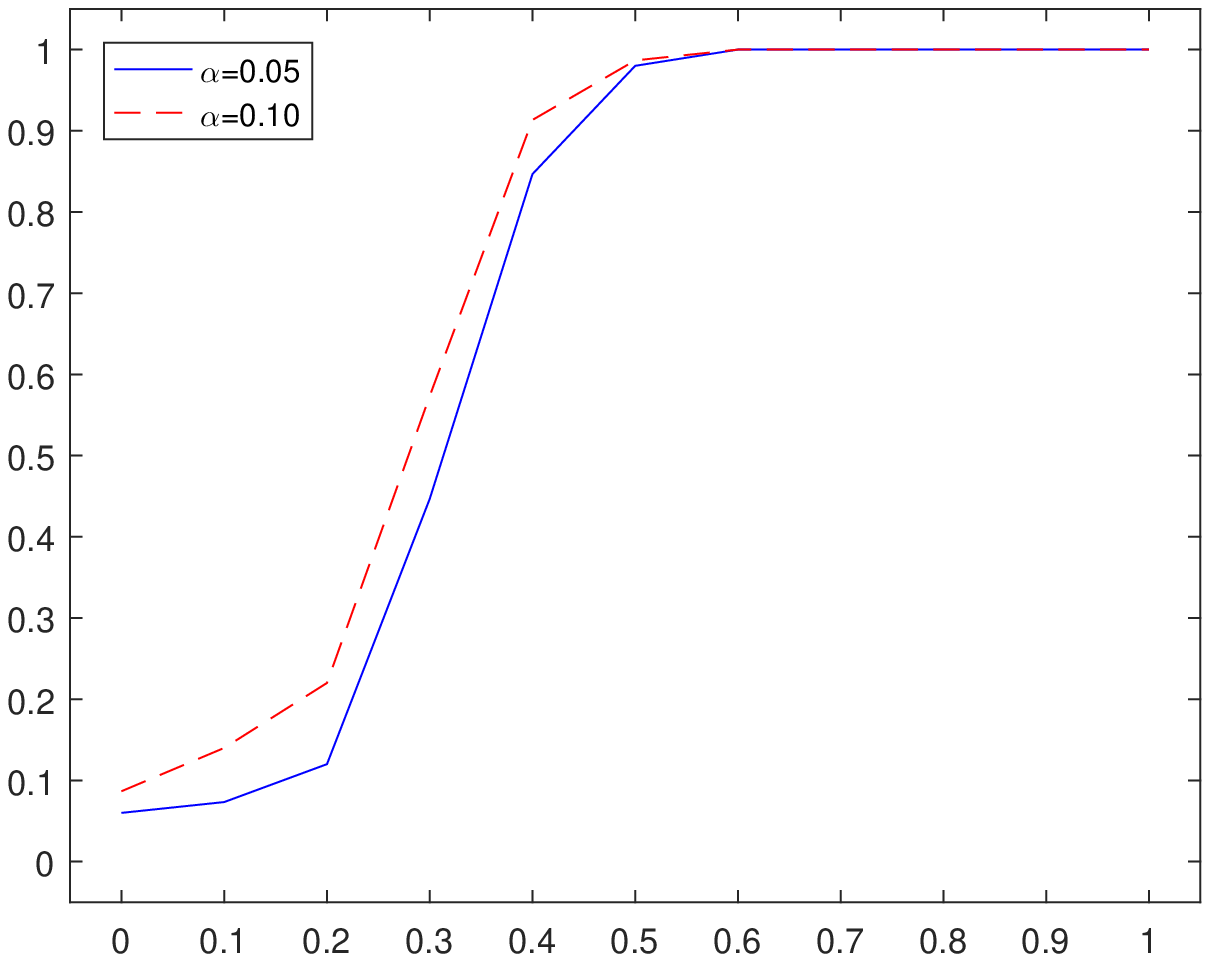} \ &
\includegraphics[width=0.34\linewidth]{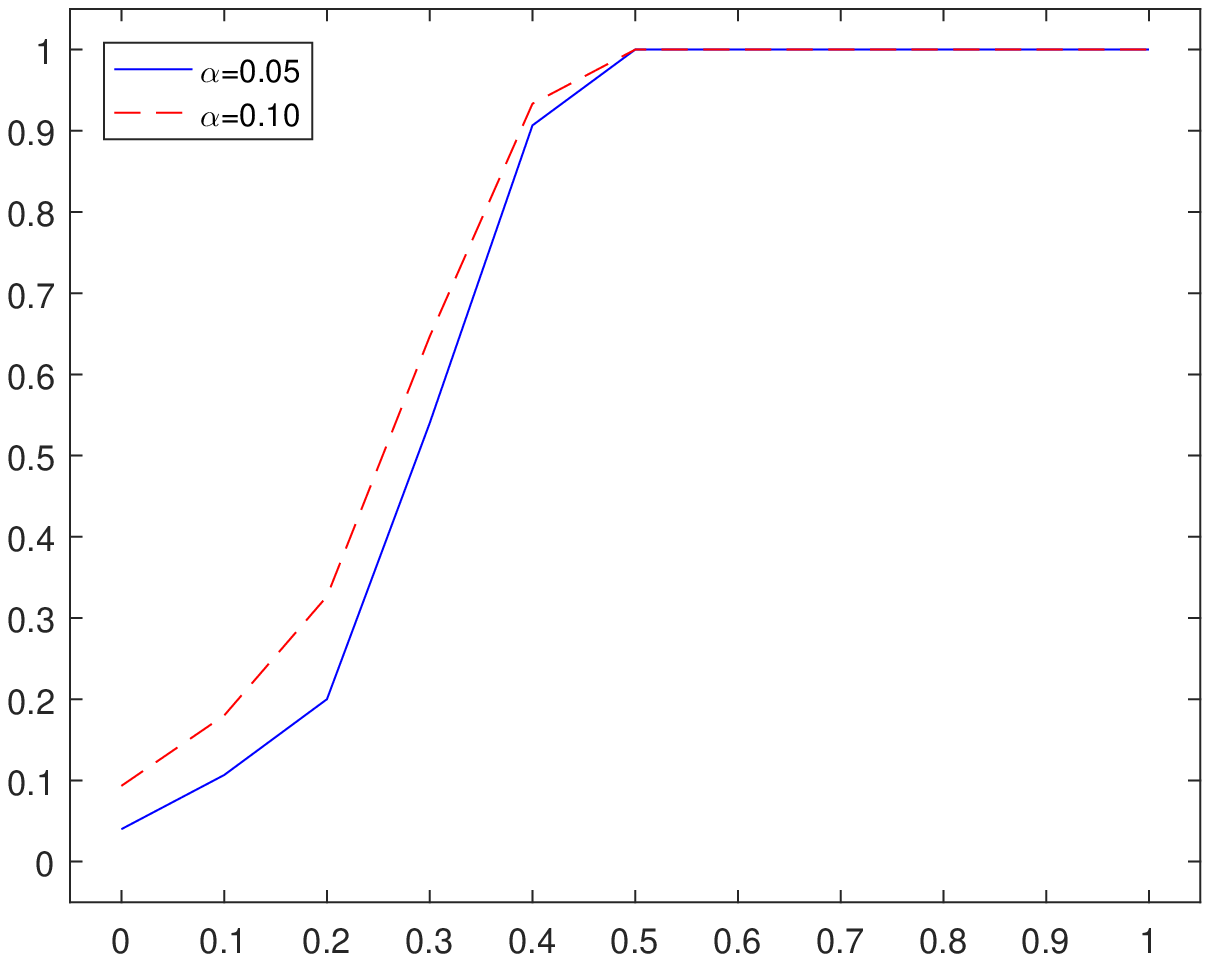} \ &
\includegraphics[width=0.34\linewidth]{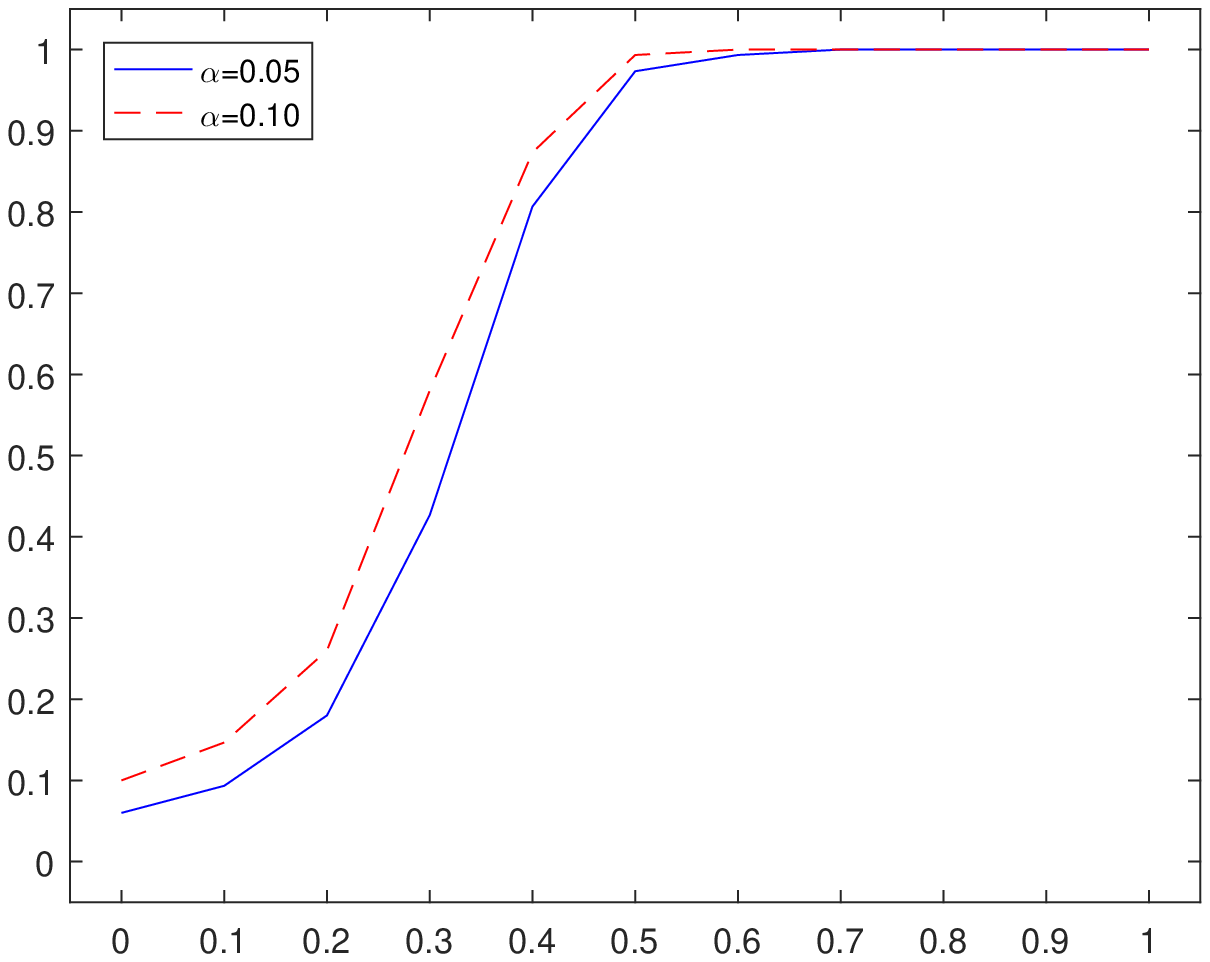} \ \\
{\tiny($a$) Power for $(n,m)=(100,10)$} & {\tiny($b$) Power for $(n,m)=(50,30)$}
&{\tiny($c$) Power for $(n,m)=(30,100)$}\\
\includegraphics[width=0.34\linewidth]{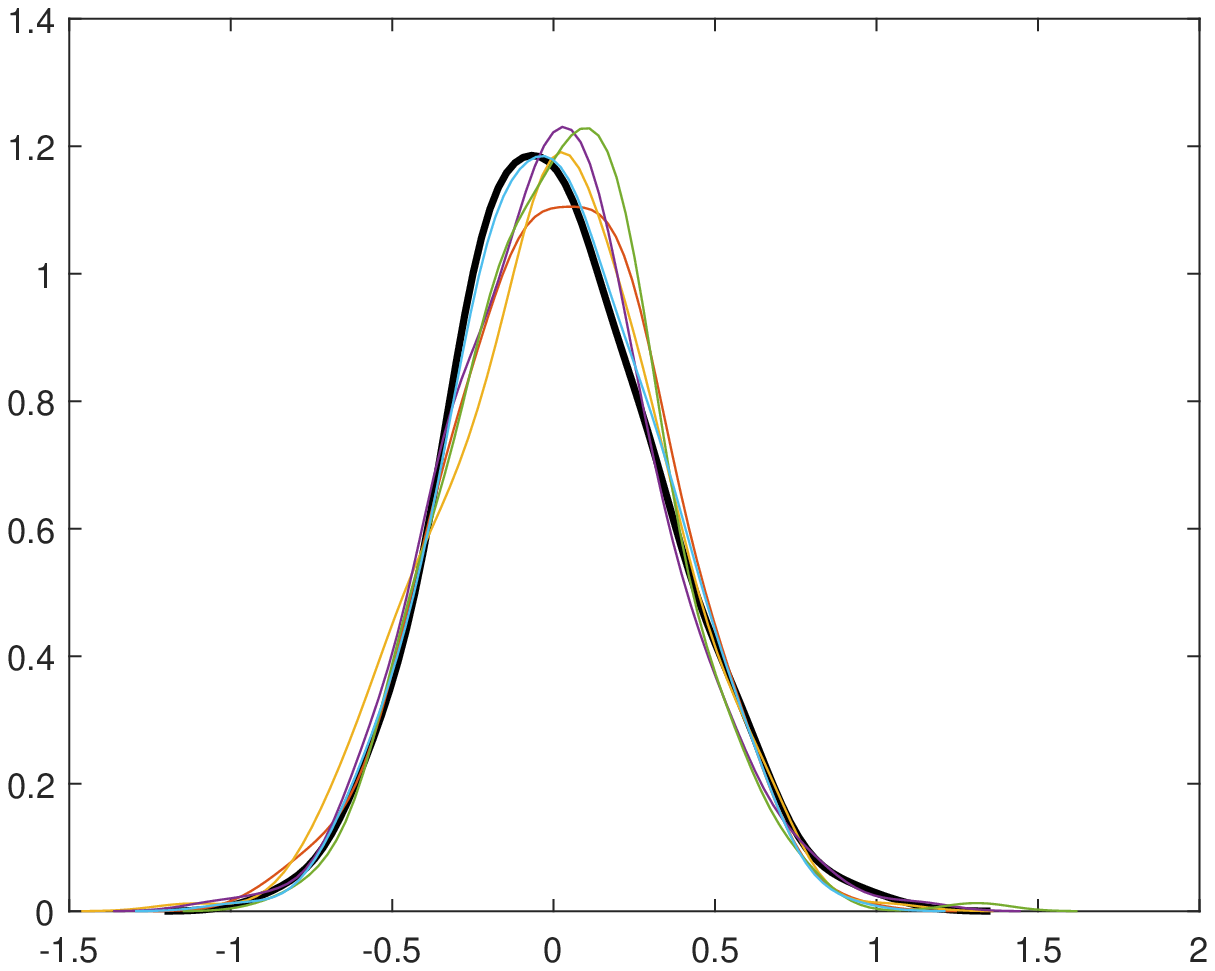}\ &
\includegraphics[width=0.34\linewidth]{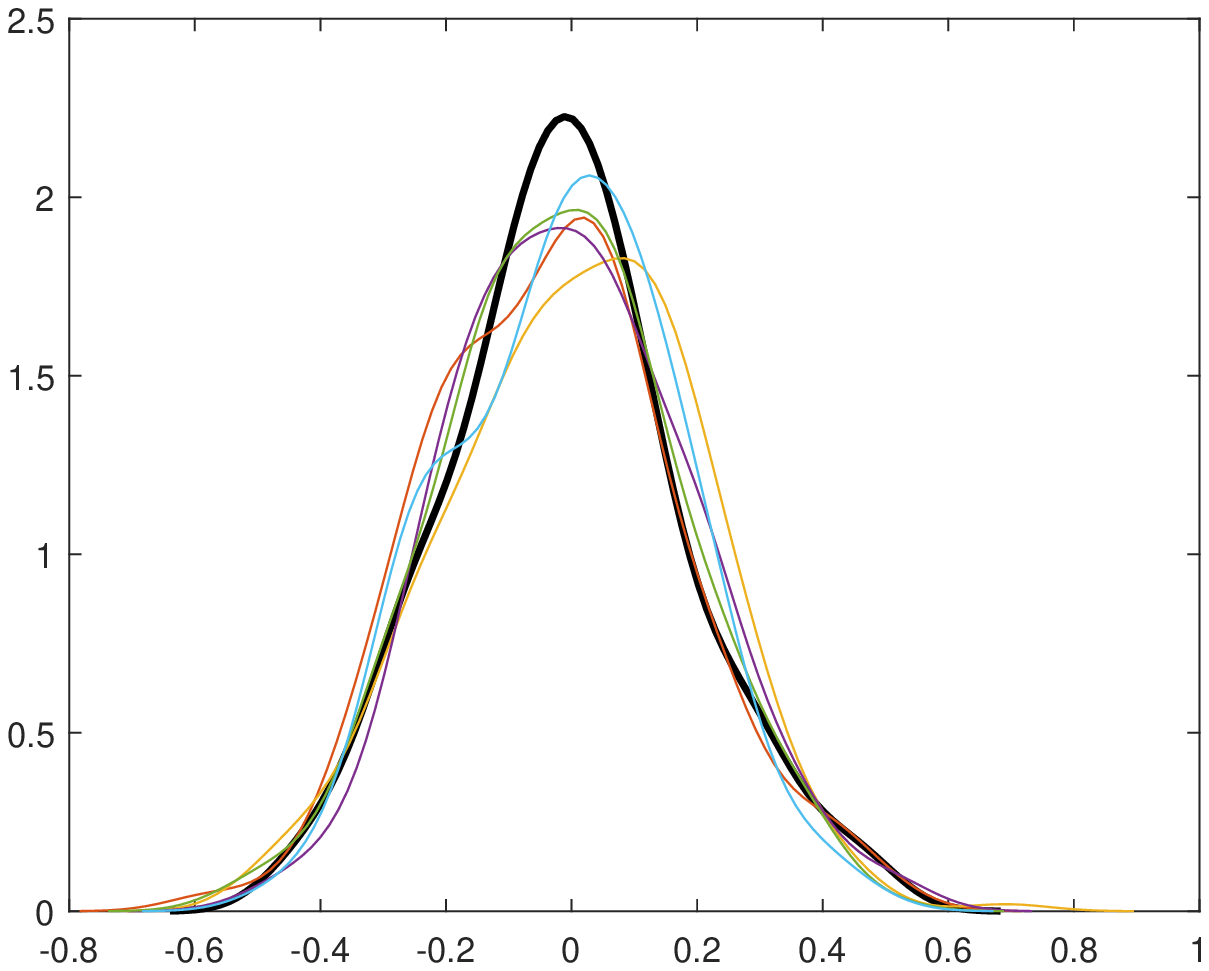}\ &
\includegraphics[width=0.34\linewidth]{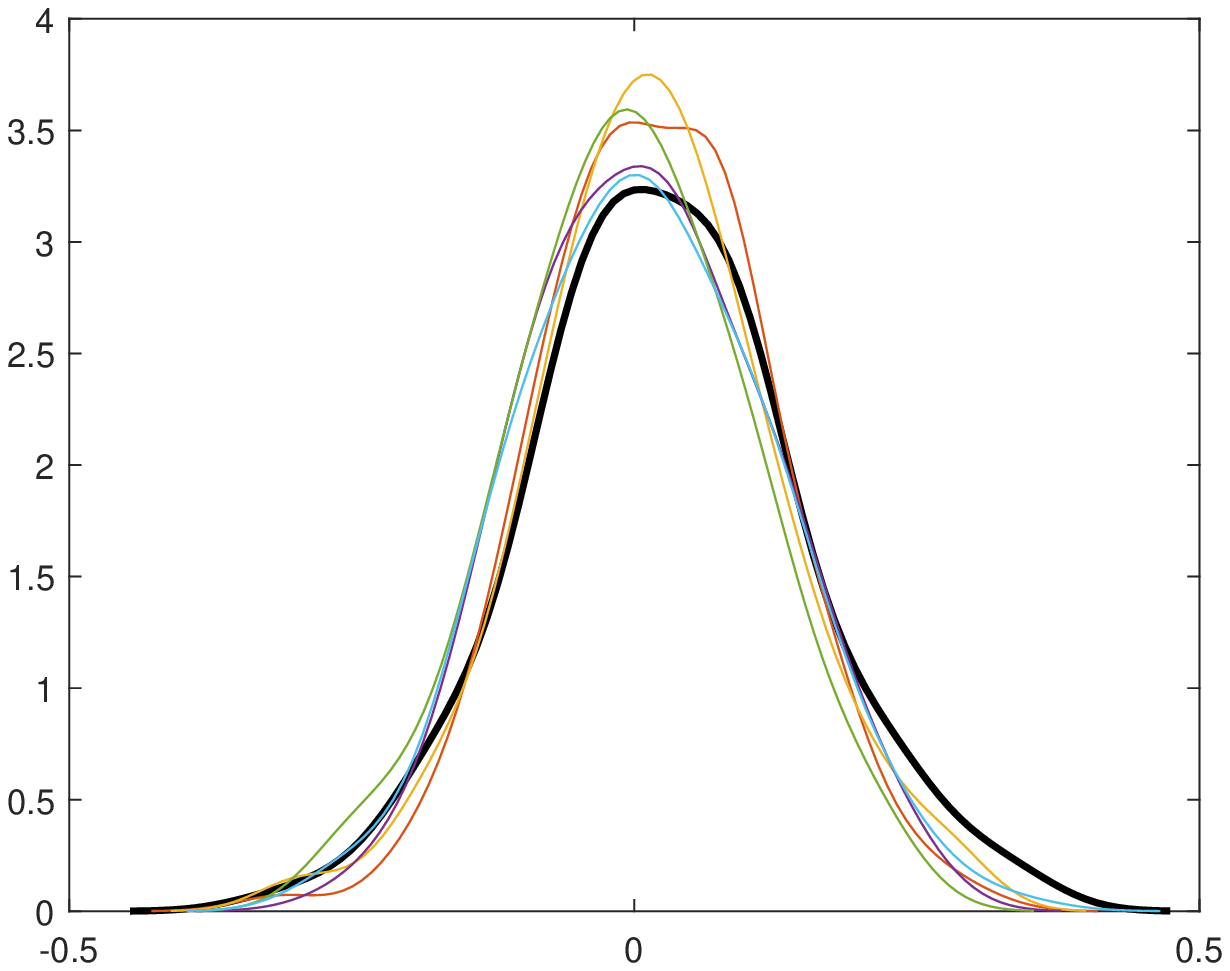}\ \\
{\tiny($a'$) Density for $(n,m)=(100,10)$ } & {\tiny($b'$) Density for $(n,m)=(50,30)$} & {\tiny($c'$)
Density for $(n,m)=(30,100)$} \\
\end{tabular}
\caption{Power of linearity testing \eqref{linear-test} in Example 2.
For three combinations of $n$ and $m$,
($a$) -($c$) figure power at the level $\alpha=0.05$ and 0.1; whist ($a'$) -($c'$) give the simulated density of
standardized test statistics (thick black) and five bootstrap approximations (thin).}
\label{fig:pow-beta}
\end{figure}
\end{example}

\subsection{Real Data Analysis}
\label{sec:simulation}

\begin{example}
Now we apply our method  to the new coronavirus disease (COVID-19)  mentioned
in Section \ref{sec:Intro}.  We collected the daily cumulative confirmed cases ($Z_{i,t}$) and
the daily cumulative cured cases from {\color{blue}{https://github.com/CSSEGISandData/COVID-19}},
the daily movement population from Wuhan to other provinces
({\color{blue}{https://qianxi.baidu.com/}}),
the maximum daily temperature ({\color{blue}{http://www.weather.com.cn}})
and the population data ({\color{blue}{https://zh.wikipedia.org/wiki/}}).

The response variable GRCC, denoted by $Y_{i,t}$,  is
measured by $\log{\left(Z_{i,t}\right)}-\log{\left(Z_{i,t-1}\right)}$, which is presented
in Figure \ref{fig:covid} (a) for 29 provinces in China from January 23th to April 8th.
We notice that the big values of GRCC (above 0.5)  mainly concentrate on the period
from January 23th to February 3th.
It is a strong evidence that the intervention policy of China's government plays a positive
 role in controlling the spread of Coronavirus disease.

To explore the influence factor of GRCC, we used five covariates:
$X_{1,it}$  being the movement population from Wuhan (MPFW), which is measured by
the proportion of the population moving from Wuhan to the $i$th province out of moving out population
at day $t-14$;
$X_{2,it}$ the daily cumulative cured cases (CUCC) at day $t-1$;
$X_{3,it}$ the daily cumulative confirmed cases (CFCC) at day $t-1$;
$X_{4,it}$  the maximum daily temperature at day $t$, and
$X_{5,i}$ the population of $i$th province.

We normalize the covariate $X_{1,it}$, and make the
logarithm transformation for $X_{k,it},\ k=2,3$ and $X_{5,i}$.
Based on 500 bootstrap sampling, we do the time-varying testing \eqref{alp-test-all}
 and linearity testing \eqref{test-beta},
obtaining the $p$ values 0.028 and 0.038, respectively.
Therefore, we reject the AM and VCM at significant level 0.05,
and adopt the general model as below:
\begin{equation}\label{model-covid}
Y_{it}=\alpha_{0}\left(t/T\right)+\sum_{k=1}^{4}\alpha_k\left(t/T\right)\beta_{k}\left(X_{k,it}\right)
+\alpha_5\left(t/T\right)\beta_{5}\left(X_{5,i}\right),
\end{equation}
where $i=1,...,29$, $t=1,...,T$ with $T=77$.
Figure \ref{fig:covid} gives the PEBLLE  of component functions,
and 95\% pointwise confidence bands according to \eqref{con-alp1} and \eqref{NS-beta}.

\begin{figure}[h!]
\centering
\begin{tabular}{@{}c@{}c@{}c}
\includegraphics[width=0.35\linewidth]{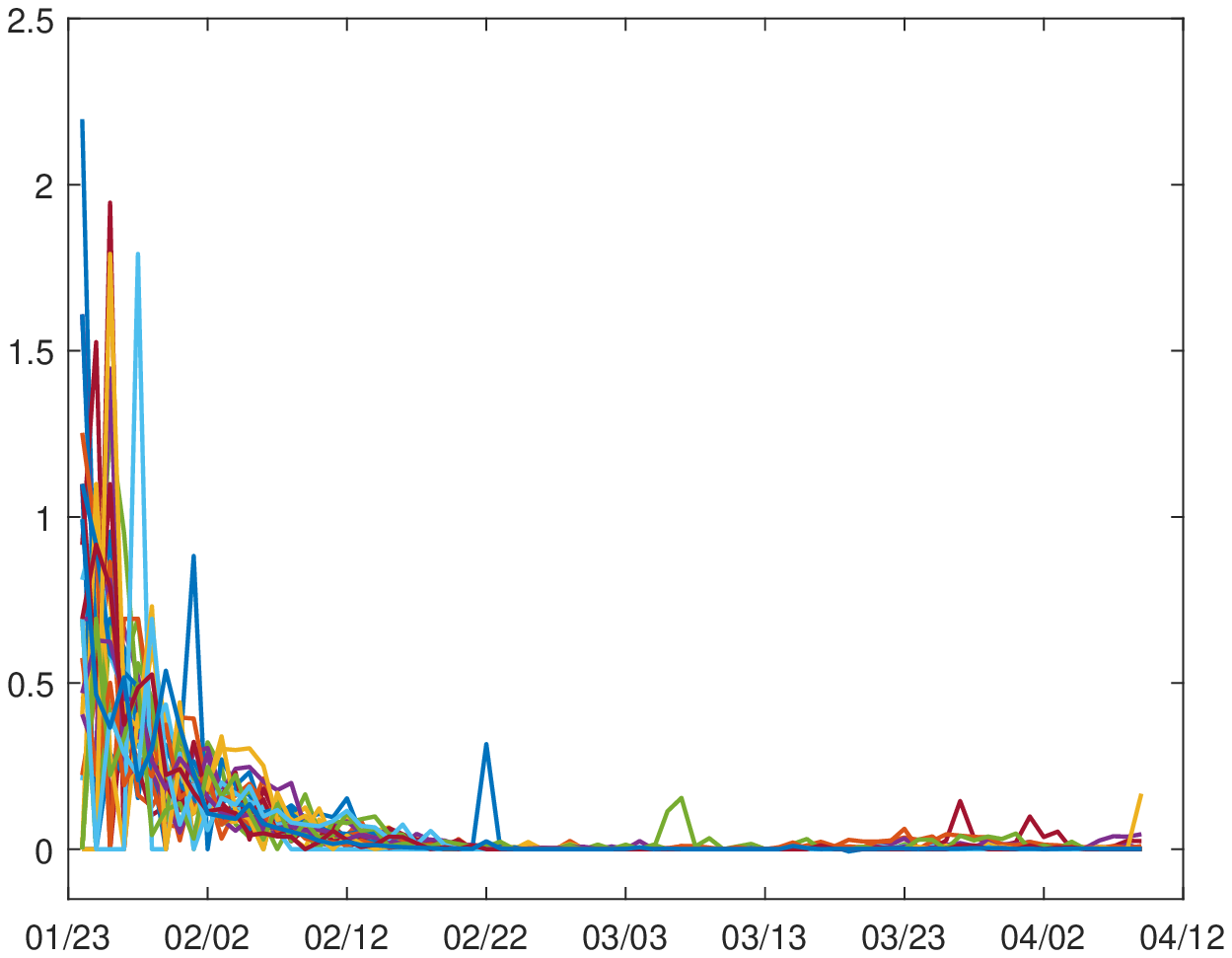} \ &
\includegraphics[width=0.35\linewidth]{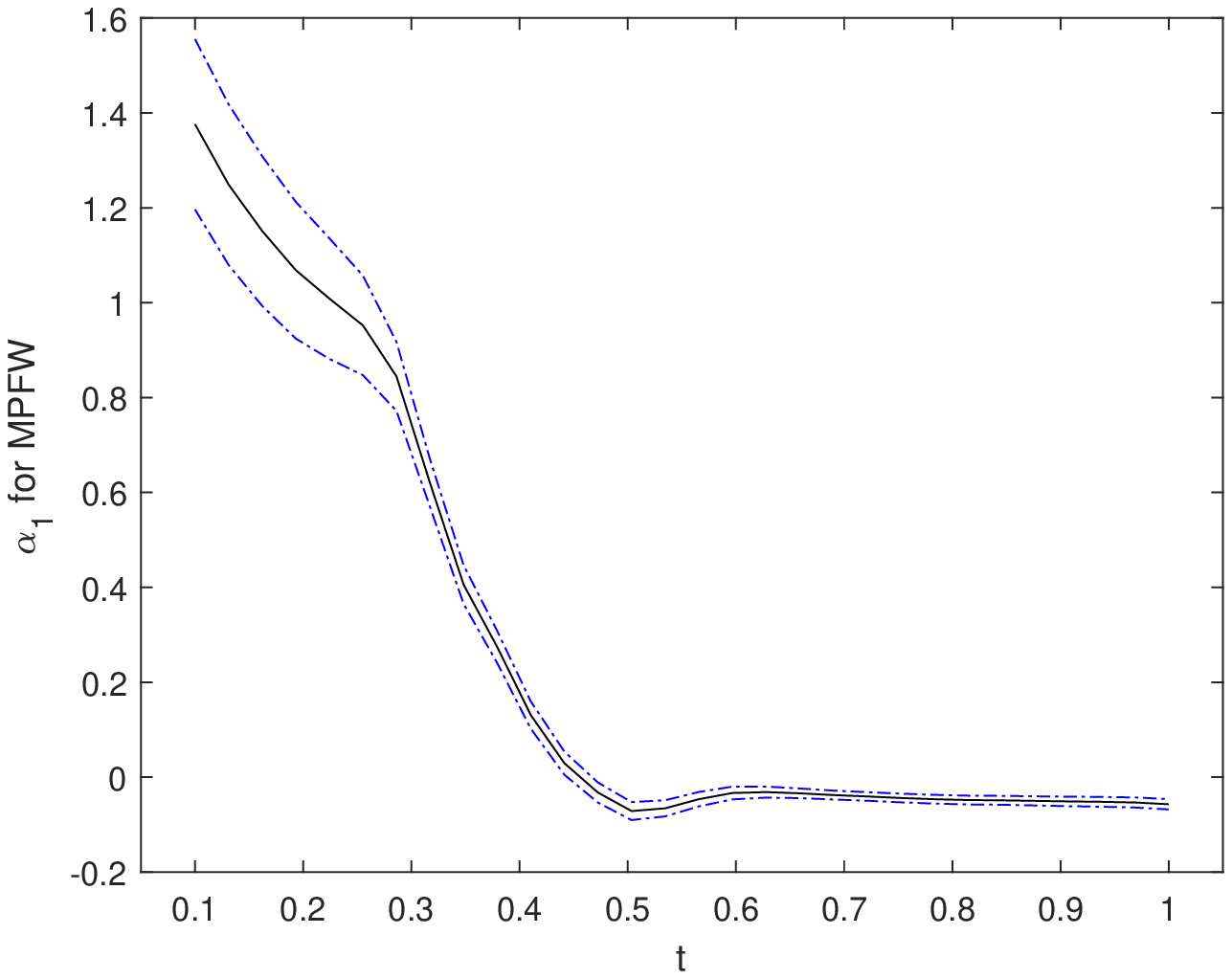} \ &
\includegraphics[width=0.35\linewidth]{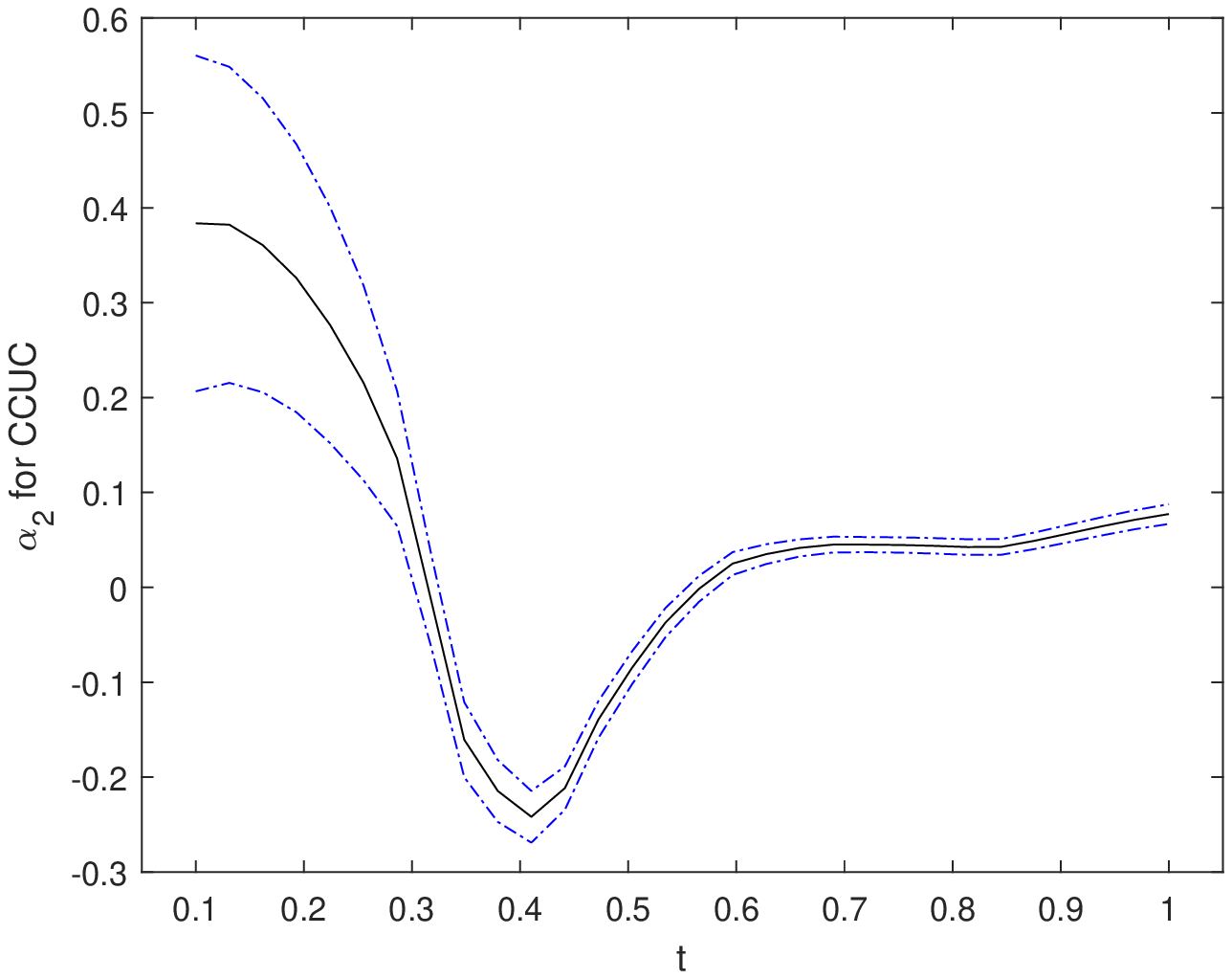} \ \\
{\tiny(a) GRCC}& {\tiny($c$) $\alpha_1$} & {\tiny($d$) $\alpha_2$} \\
\includegraphics[width=0.35\linewidth]{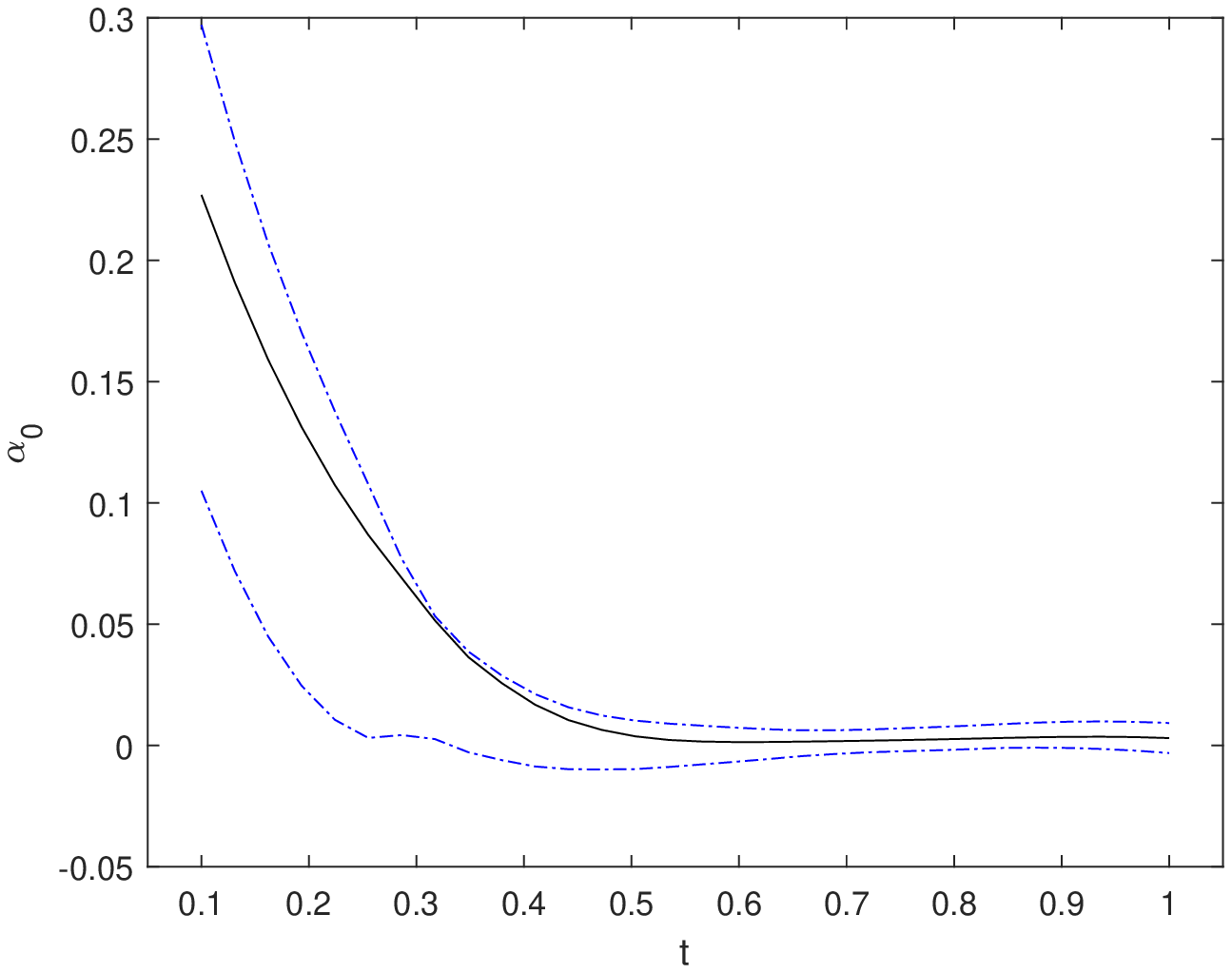} \ &
\includegraphics[width=0.35\linewidth]{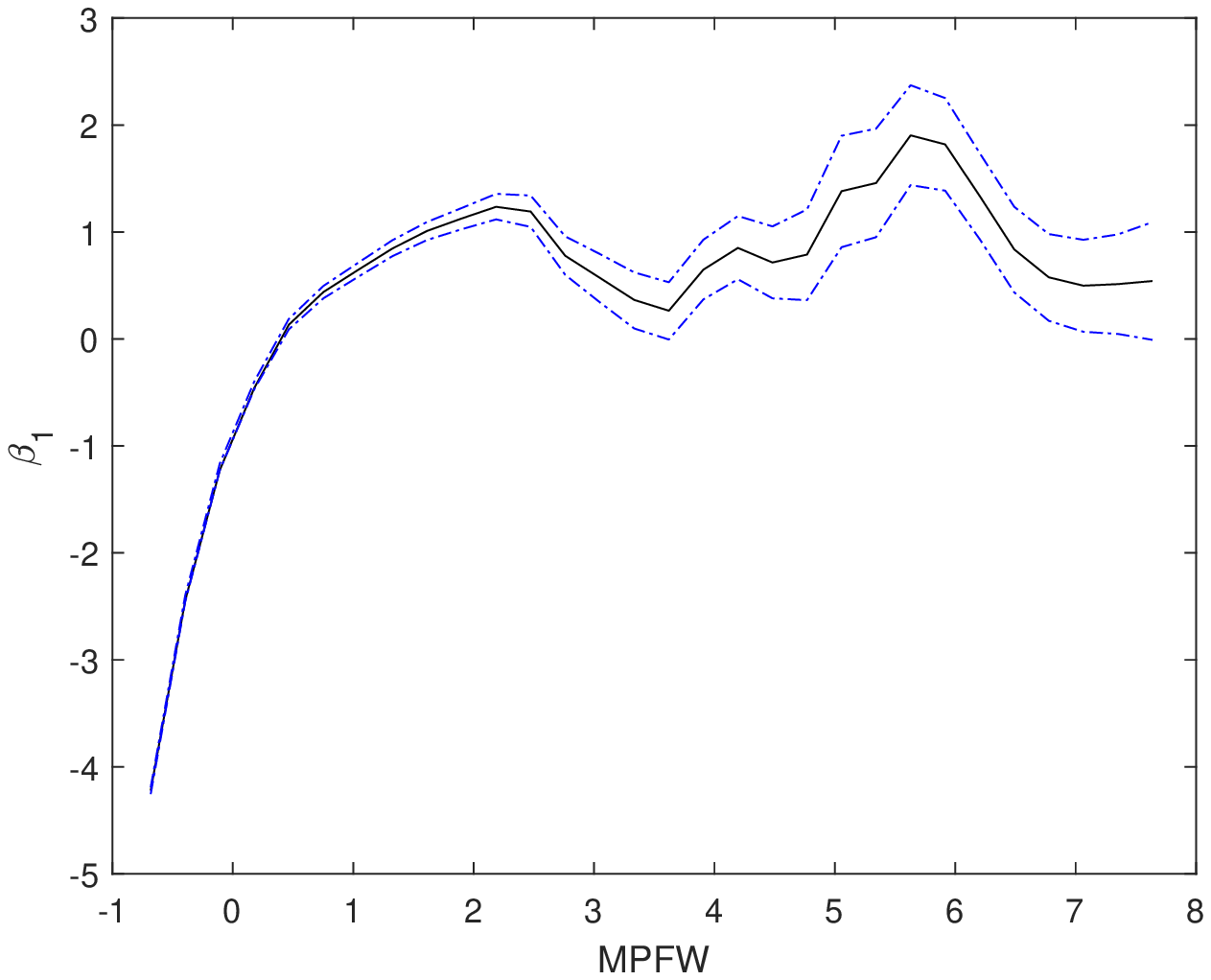} \ &
\includegraphics[width=0.35\linewidth]{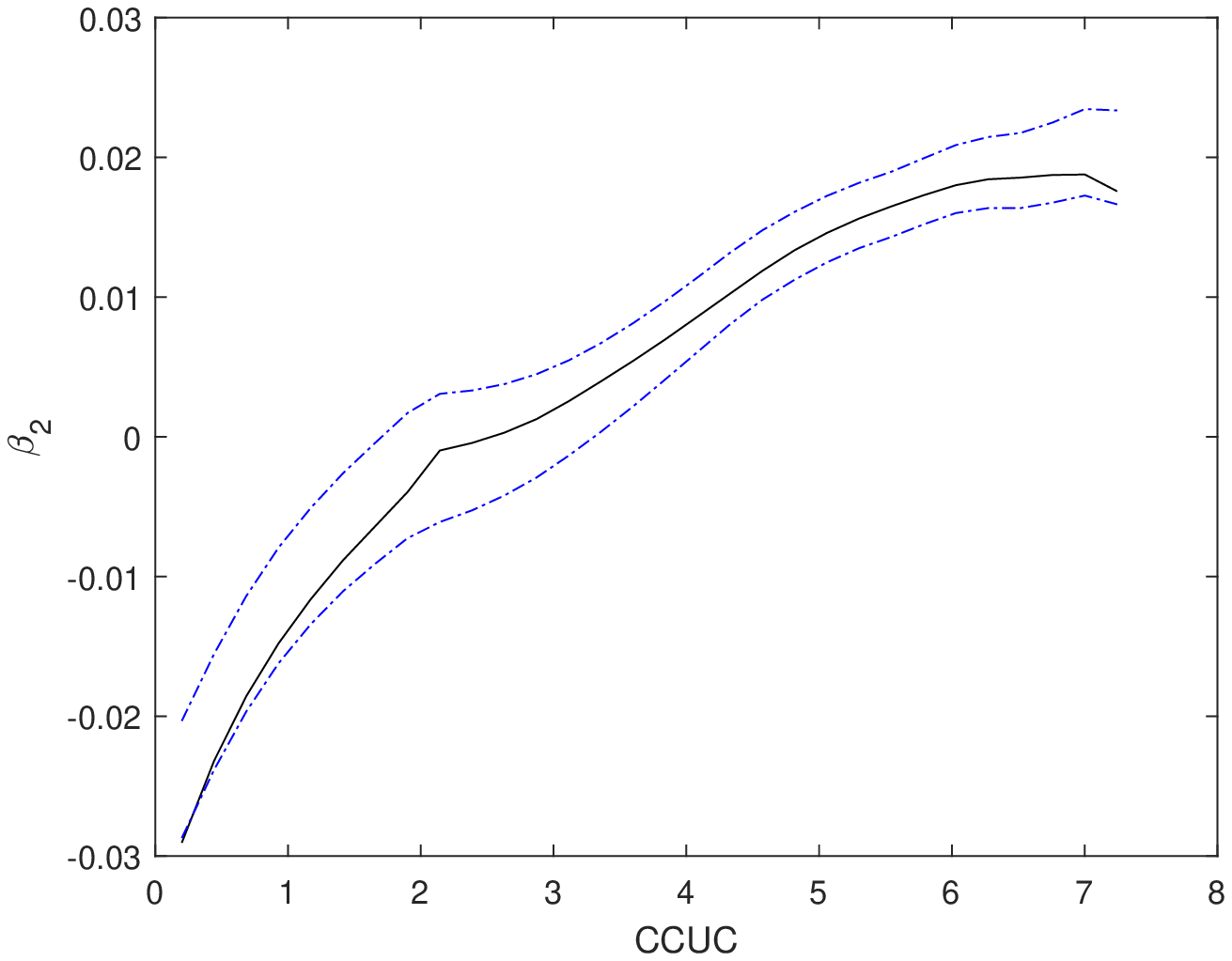} \ \\
{\tiny(b) $\alpha_0$}& {\tiny($c'$) $\beta_1$} & {\tiny($d'$) $\beta_2$} \\
\includegraphics[width=0.35\linewidth]{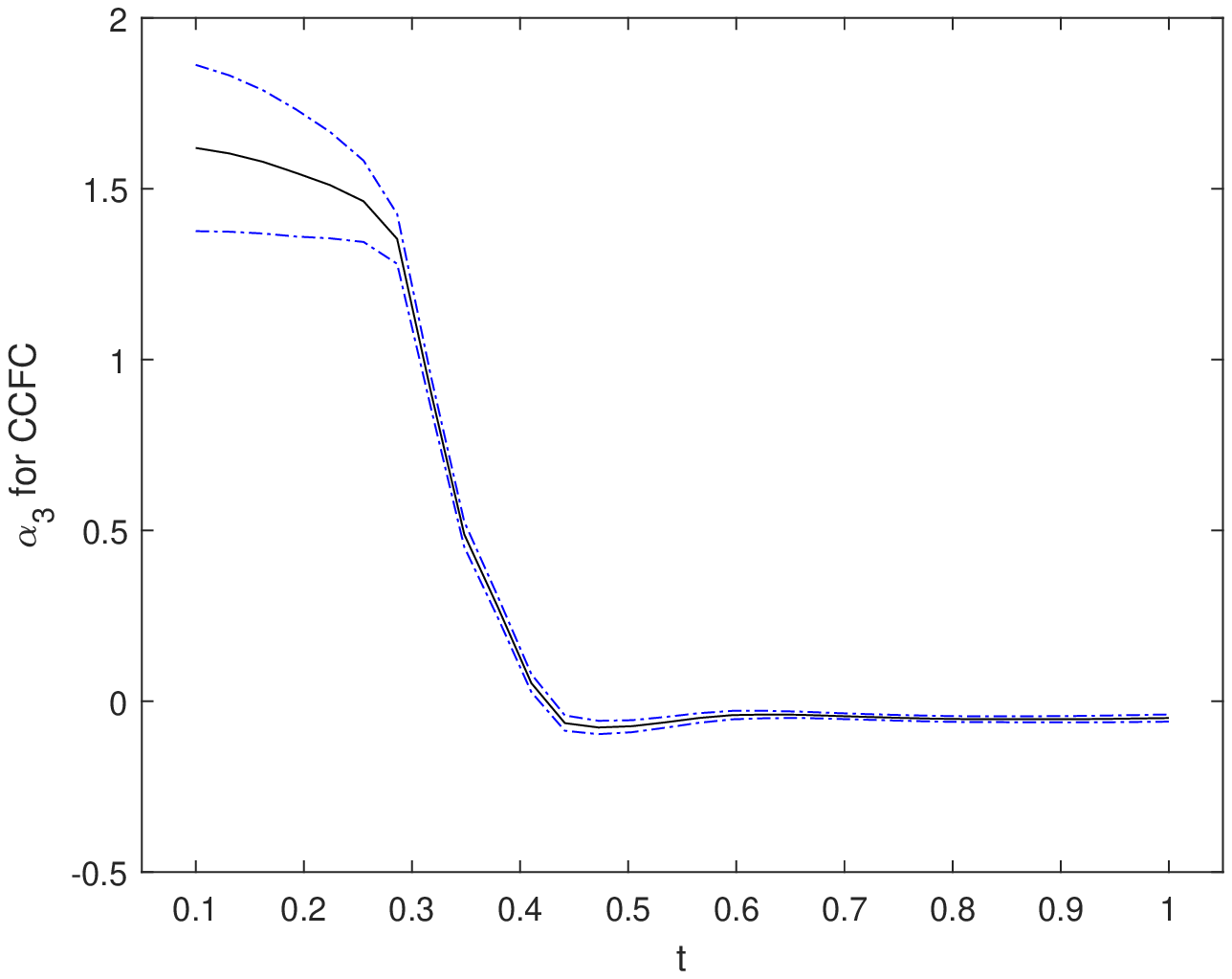} \ &
\includegraphics[width=0.35\linewidth]{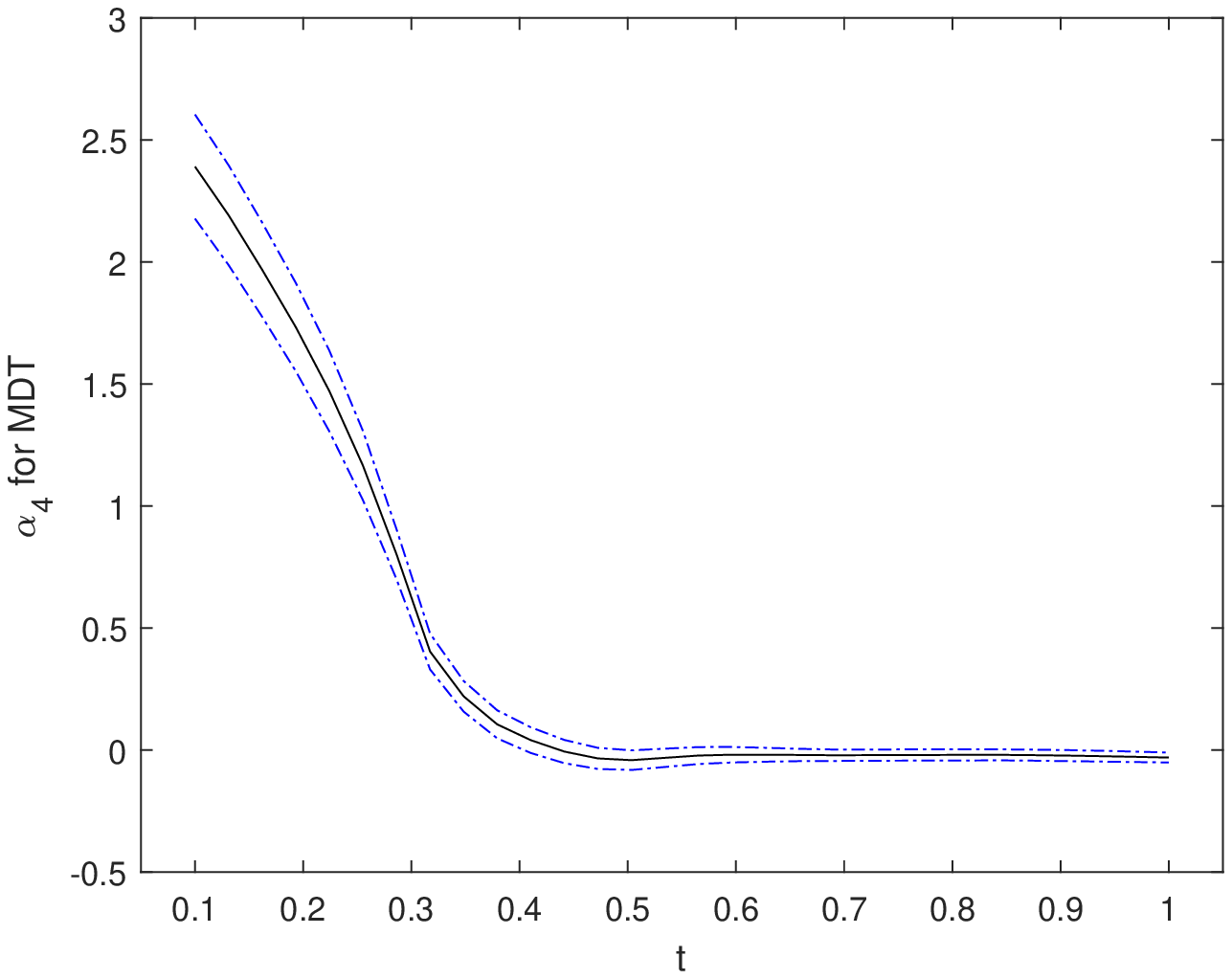} \ &
\includegraphics[width=0.35\linewidth]{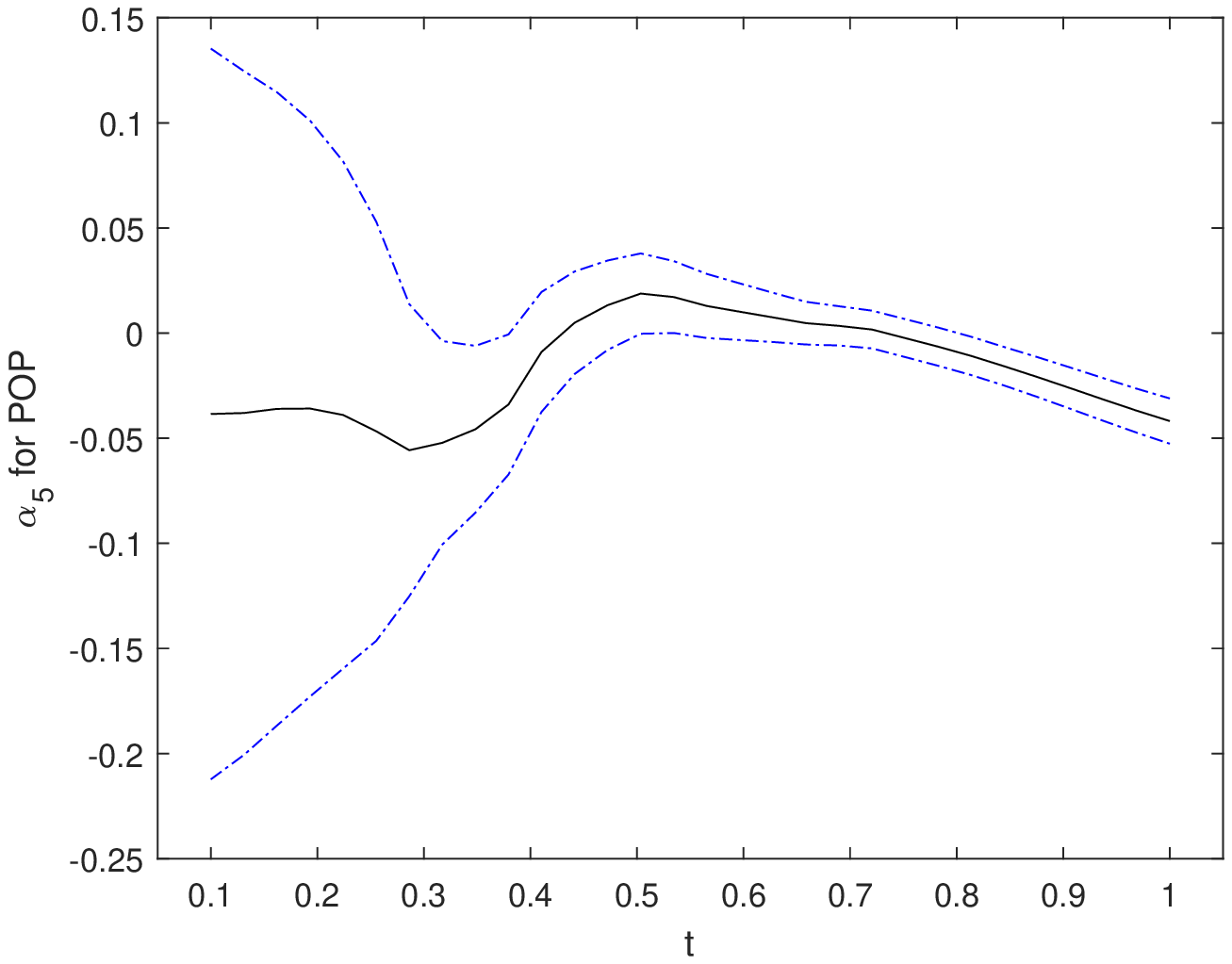} \ \\
{\tiny(e) $\alpha_3$} & {\tiny(f) $\alpha_4$}  & {\tiny(g) $\alpha_5$} \\
\includegraphics[width=0.35\linewidth]{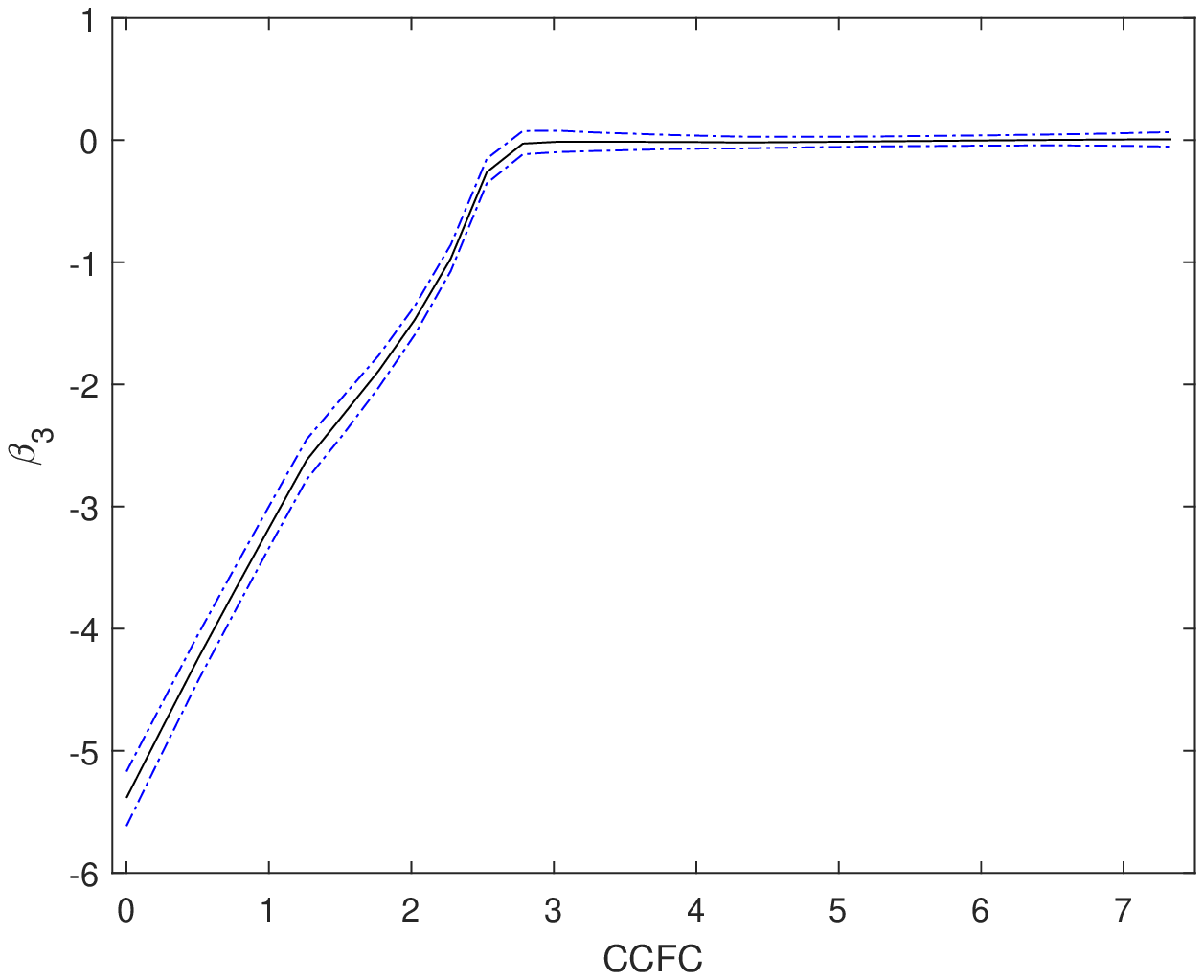} \ &
\includegraphics[width=0.35\linewidth]{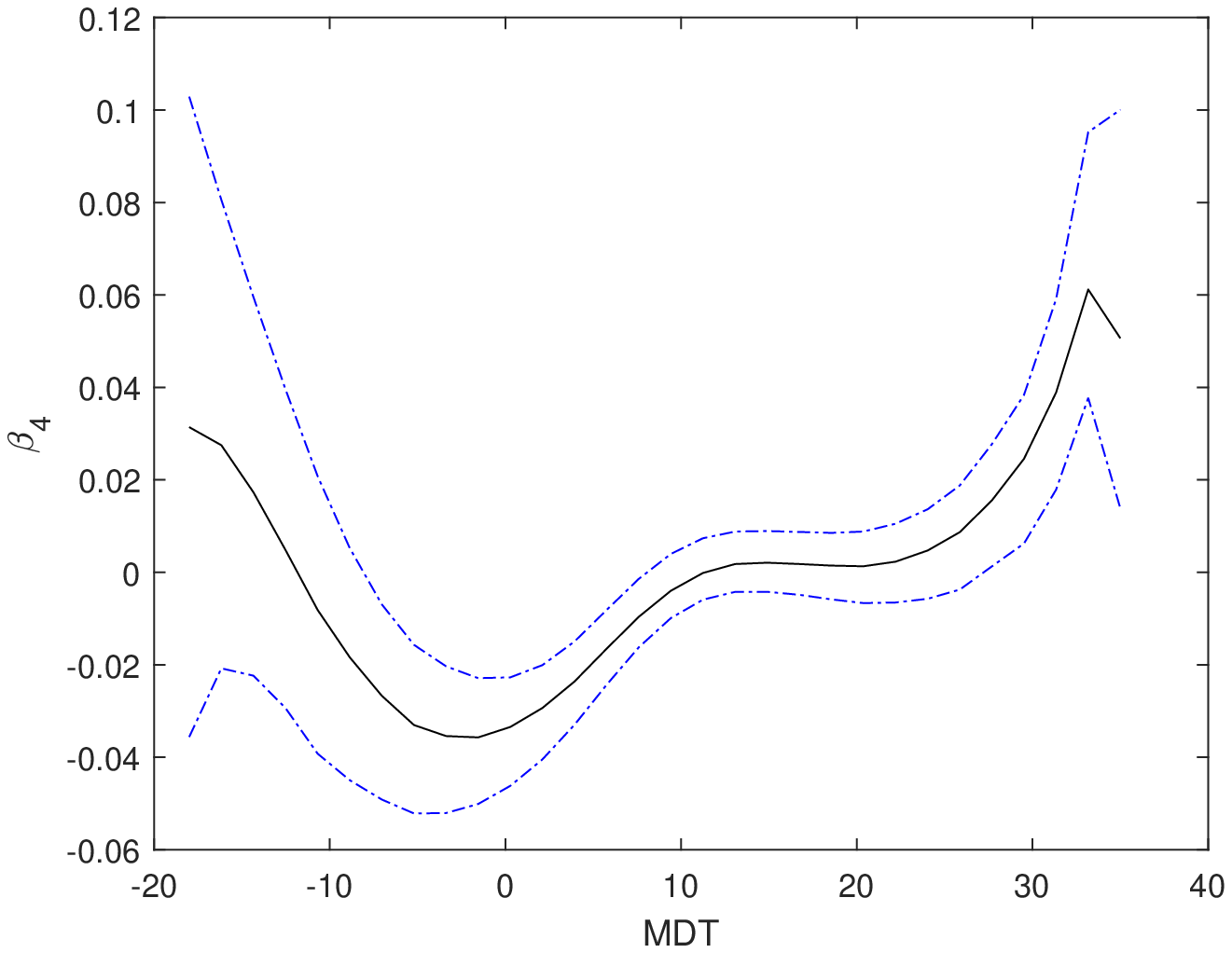} \ &
\includegraphics[width=0.35\linewidth]{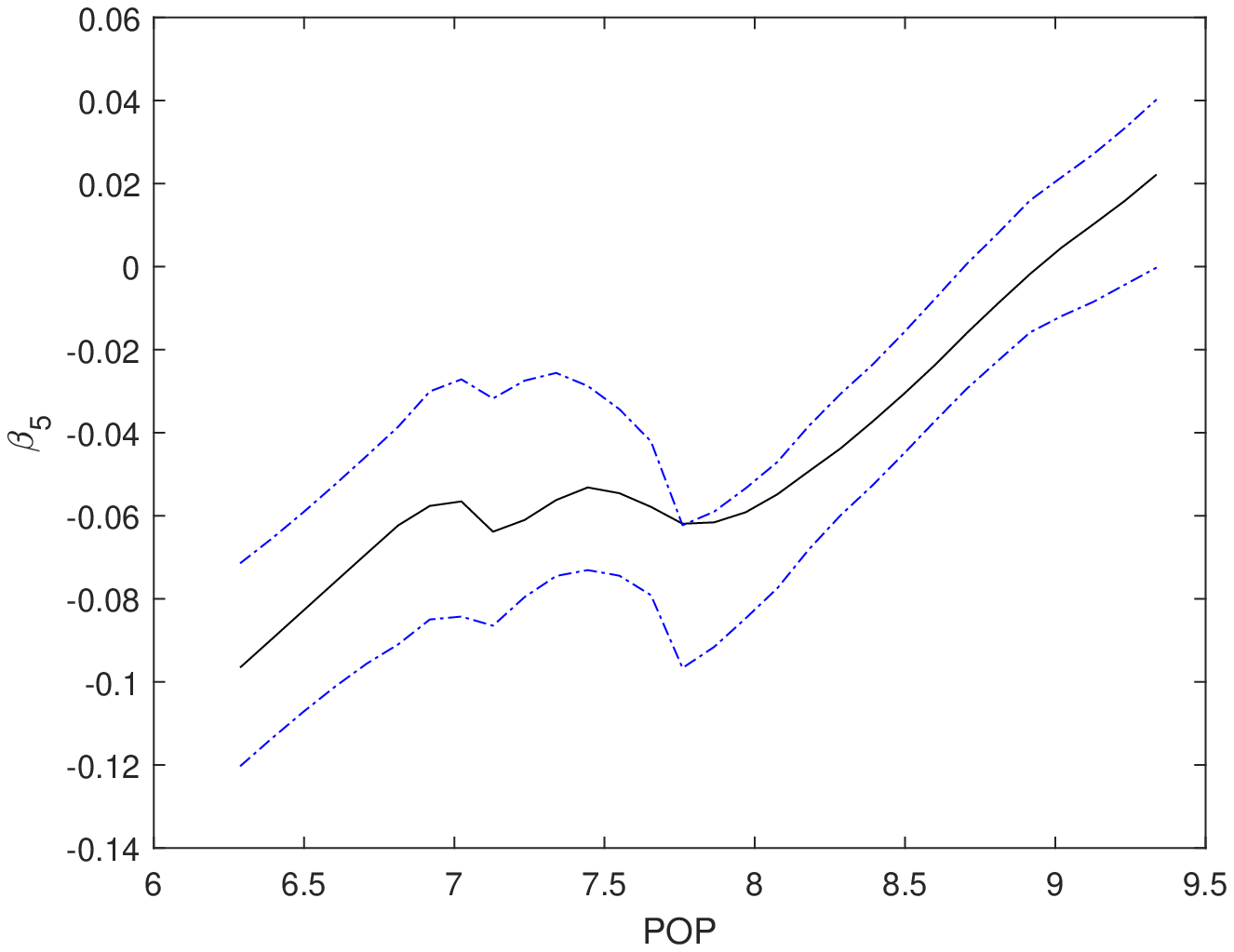} \ \\
{\tiny($e'$) $\beta_3$} & {\tiny($f'$) $\beta_4$}  & {\tiny($g'$)$\beta_5$} \\
\end{tabular}
\caption{Analysis Results for COVID-19 Data}
\label{fig:covid}
\end{figure}

From Figure \ref{fig:covid}, we conclude that the trend term $\alpha_0$,
varying-coefficient function $\alpha_1$ for MPFW,
$\alpha_3$ for CCFC and $\alpha_4$ for MDT have similar properties, i.e., they drop rapidly until about February 29th, and then maintain on the level close to zero; $\alpha_2$ for CUCC decreases until about February 22th,
and increases until about March 9th, and thereafter levels near zero;
$\alpha_5$ for POP decreases slowly until about February 9th, then increases until  about February 29th,
and decreases thereafter.

For the medium values of the normalized MPFW, the effect  increases as MPFW grows,
and some fluctuations appears for the large value (above 2), since large MPFW usually takes place in the
early stage and the period of work resumption.
The influence of CUCC increases as it grows,
and the rate of increases become slower above 2; while
the effect of CCFC ascends as it increases, and levels out above 3.
The effect of MDT drops under 0\textcelsius, and increases until 10\textcelsius,
and almost no influence between 10\textcelsius and 20\textcelsius, then
ascends rapidly above 20\textcelsius.
The trend of effect of POP grows as the population size ascends, especially when log-POP is larger than 7.5.
\end{example}

\begin{example}
 We revisit a CD4 data from the Multicenter AIDS Cohort Study, which contains 1817 observations
 from 283 homosexual men infected with HIV between 1984 and 1991.
\cite{chen2017,huang2002bspline}  have analyzed this data set using a VCM.
Now, we apply our method to this dataset.
 The response variable $Y_{ij}$ is the $i$-th subject's CD4 percentage
 at time $T_{ij}$. Following the covariates of \cite{huang2002bspline},
 we let $X_{1i}$ be the $i$-th subject's smoke status, a dichotomous variable,
 $X_{2i}$ the $i$-th subject's centred age,
 and $X_{3i}$ the $i$-th subject's centred pre-infection CD4 percentage.
 The relationship between response and covariates are modeled by a Semi-VCAM as below
 \begin{equation}
 Y_{ij}=\alpha_0\left(T_{ij}\right)+\alpha_{1}\left(T_{ij}\right)X_{1i}+\alpha_2\left(T_{ij}\right)\beta_{1}\left(X_{2i}\right)
 +\alpha_3\left(T_{ij}\right)\beta_{2}\left(X_{3i}\right),\label{CD4M}
 \end{equation}
 where the covariates are all time-invariant.

 Based on 500 bootstrap sampling, we do the time-varying testing \eqref{alp-test-all}
 and the linearity testing \eqref{test-beta},
obtaining the $p$ values 0.028 and 0.457, respectively.
That means, at significant level 0.05,  VCM is a reasonable choice, which verifies
that the model used in \cite{huang2002bspline} is appropriate.
\end{example}

\section{Concluding Remarks}
\label{sec:remarks}

In this paper, we have considered a Semi-VCAM for the functional/longitudinal data with different sampling plan. The Semi-VCAM is an extension of the existing VCAM. We have developed a pilot estimation based local linear estimation for the Semi-VCAM and have presented asymptotic distribution on a unified platform for sparse, dense and ultra dense cases of the data. The virtue of unified asymptotic results is to help us avoid deciding the types of data in advance, which is a subjective choice and may lead to wrong conclusions. From the viewpoint of model parsimony, we also have developed consistent testing procedures to justify whether a VCM or PLAM, especially an AM is sufficient for the real-life data. These test methods also avoid the subjective choice between the sparse, dense and ultra dense cases of the data.

Our model and inference methods may be extended in various directions. We close the paper by outlining some of them. In many application areas, data may be collected on a count or binary response. For example, daily death toll, suspected and confirmed cases of COVID-19. As a result, it is useful to extend our proposed model and inference to the generalized Semi-VCAM to accommodate the discrete functional/longitudinal responses.
Data in the form of samples of densities or distributions are increasingly encountered in practice and same as
\cite{han2019} there is a need for flexible regression models that accommodate random densities as responses.
We believe our proposed model could also be used to model the data in which the responses are random densities.
In addition, due to the fact that the proposed test method in our paper is based on the local smoothing,
 it may suffer the curse of dimensionality,
struggle to maintain the significance level and lose its power to an extent as the dimension of explanatory variables increases. Same as \cite{La2008} and \cite{li2017}, we may use projection technique, or bridging between local smoothing and global smoothing methods to avoid this. Due to the complication of our model,
extending the methods in \cite{La2008}
and \cite{li2017} to our scenario is not simple.

\appendix

\section{Appendix section}\label{app}

\subsection{Appendix subsection}

A function $m$ defined on the interval $[a,b]$ is called to be Lipschitz-continuous,
if there exists a fixed constant $C>0$, such that $|m\left(x\right)-m\left(x'\right)|\leq C|x-x'|$
for any $x,x'\in[a,b]$.
Denote $C_r[a,b]$ as the space of all functions $m\left(x\right)$ defined on $[a,b]$, such that
$m$ is differentiable of  $r-1$ order, and $m^{(r-1)}$ is Lipschitz-continuous,
where $m^{\left(l\right)}$ means the $l$-th order derivative of $m$.

The necessary conditions to validate asymptotic properties are  as follows.

\begin{itemize}
\item [(A1)] The observation time points $T_{ij}$'s are drawn from an unknown distribution, which has a density $f_{T}(t)$ with the support $\mathcal{T}$, and is continuously differentiable in a neighbourhood of $t$ and is uniformly bounded away from 0 and infinity.

\item[(A2)]  $\mathbf{X}_{i}$'s are independent realizations of  stochastic process $\mathbf{X}(t)$,   and $\mathbf{X}_{i}$'s are independent of $T_{ij}$'s. The marginal density function $f_{X_{k}}(\cdot)$  of covariates $X_{k}$
is continuously differentiable in a neighbourhood of $x$ and is uniformly bounded away from 0 and infinity.

\item [(A3)] $\mathbf{Z}_{i}$'s are independent realizations of  stochastic process $\mathbf{Z}\left(T\right)$,   and $\mathbf{Z}_{i}$'s are independent of $T_{ij}$'s. The eigenvalues of $\mathrm{E}\left[\mathbf{Z}\left(T\right)\mathbf{Z}\left(T\right)^{\tau}\right]$ are bounded from 0 and infinity uniformly in $T\in\mathcal{T}$. In addition, there exists a positive constant $M$ such that $|Z_{k}\left(T\right)|\leq M$ uniformly for $T\in\mathcal{T}$
    and $k=1,...,q$, where $Z_k$ i.i.d. with $Z_{ij,k}$, a random sample of $k$-th covariate.

\item [(A4)] $\alpha_{k}\in C_r[a,b]$ for $k=0,...,p$ and $\beta_{k}\in C_r[a_k,b_k]$ for $k=1,...,p$.

\item [(A5)]  $\{\nu_{i}(\cdot)\}_{i}$, $\{T_{ij}\}_{ij}$, $\{\varepsilon_{ij}\}_{ij}$ are independent and identically
distributed and mutually independent.  $\{\mathbf{x}_{ij}\}_{i}$ are independent and identically distributed.
Moreover, $\{\nu_{i}(\cdot)\}_{i}$, $\{\mathbf{x}_{ij}\}_{ij}$ and  $\{\varepsilon_{ij}\}_{ij}$ are mutually independent.

\item [(A6)] $\sigma^{2}(\cdot)<\infty$ is continuously differentiable. $\gamma(t,t')$ is continuously differentiable
and $\gamma(t,t)=\lim_{t'\to t}\gamma(t,t')<\infty$.

\item [(A7)] $\mathrm{E}\{|\nu_{i}(\cdot)+\sigma(\cdot)\varepsilon_{ij}|^{\nu}\}$ is continuous and bounded from infinity
for $\nu\leq 4$.

\item[(A8)] $k(\cdot)$ is bounded and symmetric probability density function with a bounded support and a bounded derivative.

\item [(A9)]  $\sqrt{K_{\mathrm{A}}}\{K_{\mathrm{A}}^{-r}+K_{\mathrm{C}}^{-r}\}=o(1)$ and
$K_{\mathrm{A}}^{2}K_{\mathrm{C}}/n =o(1)$.

\end{itemize}

\begin{remark}
{\rm
Assumptions~A1 and A2 involve the distributions of time points $T_{ij}$ and $k$-th covariate
$X_{k}$. Assumption A3 relates to covariates $\mathbf{Z}$, a similar conditions with \cite{chen2017}.
 Assumption~A4 specifies the degree of smoothness of varying-coefficient component functions and additive component functions.
Assumptions~A5--A7 are necessary for constructing asymptotic distribution, a common conditions with \cite{chen2017}.
Assumption A8 is a standard condition of kernel function in local polynomial smoothing,
and A9 is about the  number of interior knots in pilot spline estimation.
}
\end{remark}

\section*{Acknowledgements}

\begin{supplement}
\sname{Supplement A}\label{suppA}
\stitle{Preliminary Results}
\slink[url]{http://www.e-publications.org/ims/support/dowload/imsart-ims.zip}
\sdescription{\begin{proposition}\label{pro}
Under Assumption (A1) -- (A6) and (A9), it follows that
\[\sup_{x\in[a_k,b_k]}|\hat{\beta}_{k,\mathrm{I}}(x)-\beta_k(x)|
=O_p\left(\sqrt{K_{\mathrm{A}}}(K_{\mathrm{A}}^{-r}+K_{\mathrm{C}}^{-r})
+\sqrt{\frac{K_{\mathrm{C}}K_{\mathrm{A}}^2}{n\bar{N}_{\mathrm{H}}}
+\frac{K_{\mathrm{A}}}{n}}\right).\]
\end{proposition}
}
\end{supplement}



\end{document}